\pdfoutput=1

\documentclass[11pt,twoside,a4paper,cmspaper,final,collab]{cms-tdr}
\def\svnVersion{a14d444-D}\def\svnDate{2019/12/03}\def\cmsCernNoTag{CERN-EP-2019-174}\def\cmsCernDate{\today}\def\cmsMessage{"Published in the Journal of High Energy Physics as \href{http://dx.doi.org/10.1007/JHEP01(2020)036}{\doi{10.1007/JHEP01(2020)036}.}"}

\begin{document}\cmsNoteHeader{B2G-18-003}

\hyphenation{had-ron-i-za-tion}
\hyphenation{cal-or-i-me-ter}
\hyphenation{de-vices}
\RCS$HeadURL$
\RCS$Id$

\newlength\cmsFigWidth
\ifthenelse{\boolean{cms@external}}{\setlength\cmsFigWidth{0.49\textwidth}}{\setlength\cmsFigWidth{0.65\textwidth}}
\ifthenelse{\boolean{cms@external}}{\providecommand{\cmsLeft}{upper\xspace}}{\providecommand{\cmsLeft}{left\xspace}}
\ifthenelse{\boolean{cms@external}}{\providecommand{\cmsRight}{lower\xspace}}{\providecommand{\cmsRight}{right\xspace}}

\newlength\cmsTabSkip\setlength{\cmsTabSkip}{1ex}
\providecommand{\NA}{\ensuremath{\text{---}}}
\providecommand{\cmsTable}[1]{\resizebox{\textwidth}{!}{#1}}
\providecommand{\CL}{CL\xspace}

\newcommand{\rA}{\ensuremath{\mathrm{Q}_{\PH}}\xspace}
\newcommand{\rB}{\ensuremath{\mathrm{T}_{\PH}}\xspace}
\newcommand{\rC}{\ensuremath{\mathrm{R}_{\PH}}\xspace}
\newcommand{\rH}{\ensuremath{\mathrm{S}_{\PH}}\xspace}
\newcommand{\rW}{\ensuremath{\mathrm{Q}_{\PZ}}\xspace}
\newcommand{\rX}{\ensuremath{\mathrm{L}_{\PZ}}\xspace}
\newcommand{\rY}{\ensuremath{\mathrm{R}_{\PZ}}\xspace}
\newcommand{\rZ}{\ensuremath{\mathrm{S}_{\PZ}}\xspace}
\newcommand{\ThreeT}{\ensuremath{3T}\xspace}
\newcommand{\ThreeM}{\ensuremath{3M}\xspace}
\newcommand{\TwoMOneL}{\ensuremath{2M1L}\xspace}
\newcommand{\TwoTOneL}{\ensuremath{2T1L}\xspace}
\newcommand{\Tight}{\ensuremath{T}\xspace}
\newcommand{\Medium}{\ensuremath{M}\xspace}
\newcommand{\TbW}{\ensuremath{\PQT\PQb\PW}\xspace}
\newcommand{\TtZ}{\ensuremath{\PQT\PQt\PZ}\xspace}
\newcommand{\HZ}{\ensuremath{\PH/\PZ}\xspace}
\newcommand{\bW}{\ensuremath{\PQb\PW}\xspace}
\newcommand{\tH}{\ensuremath{\PQt\PH}\xspace}
\newcommand{\tZ}{\ensuremath{\PQt\PZ}\xspace}
\newcommand{\tX}{\ensuremath{\PQt\mathrm{X}}\xspace}
\newcommand{\tbw}{\ensuremath{\PQt\to\PQb\PW}\xspace}
\newcommand{\Hbb}{\ensuremath{\PH\to\PQb\PAQb}\xspace}
\newcommand{\Zbb}{\ensuremath{\PZ\to\PQb\PAQb}\xspace}
\newcommand{\Xbb}{\ensuremath{\mathrm{X}\to\PQb\PAQb}\xspace}
\newcommand{\Zqq}{\ensuremath{\PZ\to\PQq\cPaq (\PQq \neq \PQb)}\xspace}
\newcommand{\tprimetotH}{\ensuremath{\PQT\to\PQt\PH}\xspace}
\newcommand{\tprimetotZ}{\ensuremath{\PQT\to\PQt\PZ}\xspace}
\newcommand{\tprimetobW}{\ensuremath{\PQT\to\PQb\PW}\xspace}
\newcommand{\ttjets}{\ensuremath{\PQt\PAQt}+\text{jets}\xspace}
\newcommand{\mc}{Monte Carlo\xspace}
\newcommand{\intL}{35.9\fbinv\xspace}
\newcommand{\mtprime}{\ensuremath{m_{\PQT}}\xspace}
\newcommand{\GoM}{\ensuremath{\Gamma/{\mtprime}}\xspace}
\newcommand{\Mtilde}{\ensuremath{\widetilde{m}_{\PQT}}\xspace}
\newcommand{\qpr}{\ensuremath{\PQq'}\xspace}
\newcommand{\qapr}{\ensuremath{\PAQq'}\xspace}
\newcommand{\qgTbq}{\ensuremath{\PQq\Pg\to\PQT\PAQb\qpr}\xspace}
\newcommand{\qgTtq}{\ensuremath{\PQq\Pg\to\PQT\PAQt\PQq}\xspace}
\newcommand{\tXbq}{\ensuremath{\PQt\mathrm{X}\PAQb\qpr}\xspace}
\newcommand{\tXtq}{\ensuremath{\PQt\mathrm{X}\PAQt\PQq}\xspace}
\newcommand{\Tbq}{\ensuremath{\Pp\Pp\to\PQT\PQb\PQq}\xspace}
\newcommand{\Tbqonly}{\ensuremath{\PQT\PQb\PQq}\xspace}
\newcommand{\Ttqonly}{\ensuremath{\PQT\PQt\PQq}\xspace}
\newcommand{\Ttq}{\ensuremath{\Pp\Pp\to\PQT\PQt\PQq}\xspace}
\newcommand{\ttH}{\ensuremath{\ttbar\PH}\xspace}
\newcommand{\tHbq}{\ensuremath{\PQt\PH\PQb\PQq}\xspace}
\newcommand{\tZbq}{\ensuremath{\PQt\PZ\PQb\PQq}\xspace}
\newcommand{\tHZbq}{\ensuremath{\PQt\PH\PQb\PQq+\PQt\PZ\PQb\PQq}\xspace}
\newcommand{\tHZtq}{\ensuremath{\PQt\PH\PQt\PQq+\PQt\PZ\PQt\PQq}\xspace}
\newcommand{\tHtq}{\ensuremath{\PQt\PH\PQt\PQq}\xspace}
\newcommand{\tZtq}{\ensuremath{\PQt\PZ\PQt\PQq}\xspace}
\newcommand{\tHq}{\ensuremath{\PQt\PH\PQq}\xspace}
\newcommand{\wjets}{\PW{+}\text{jets}\xspace}
\newcommand{\zjets}{\PZ{+}\text{jets}\xspace}
\newcommand{\DRbtW}{\ensuremath{\Delta R({\PQb}_{\PQt},\PW)}\xspace}
\newcommand{\DRjj}{\ensuremath{\Delta R(j_{\PW}, j_{\PW})}\xspace}
\newcommand{\DRbbHH}{\ensuremath{\Delta R({\PQb}_{\PH}, {\PQb}_{\PH})}\xspace}
\newcommand{\DRbbZZ}{\ensuremath{\Delta R({\PQb}_{\PZ}, {\PQb}_{\PZ})}\xspace}
\newcommand{\DRbbHZ}{\ensuremath{\Delta R({\PQb}_{\PH/\PZ}, {\PQb}_{\PH/\PZ})}\xspace}
\newcommand{\mwmc}{\ensuremath{m_{\PW}^\mathrm{MC}}\xspace}
\newcommand{\mtmc}{\ensuremath{m_{\PQt}^\mathrm{MC}}\xspace}
\newcommand{\mzmc}{\ensuremath{m_{\PZ}^\mathrm{MC}}\xspace}
\newcommand{\mhmc}{\ensuremath{m_{\PH}^\mathrm{MC}}\xspace}
\newcommand{\mhzmc}{\ensuremath{m_{\PH/\PZ}^\mathrm{MC}}\xspace}
\newcommand{\swmc}{\ensuremath{\sigma_{\PW}^\mathrm{MC}}\xspace}
\newcommand{\stmc}{\ensuremath{\sigma_{\PQt}^\mathrm{MC}}\xspace}
\newcommand{\szmc}{\ensuremath{\sigma_{\PZ}^\mathrm{MC}}\xspace}
\newcommand{\shmc}{\ensuremath{\sigma_{\PH}^\mathrm{MC}}\xspace}
\newcommand{\shzmc}{\ensuremath{\sigma_{\PH/\PZ}^\mathrm{MC}}\xspace}
\newcommand{\mwm}{\ensuremath{m_{\PW}^\text{meas}}\xspace}
\newcommand{\mtm}{\ensuremath{m_{\PQt}^\text{meas}}\xspace}
\newcommand{\mhzm}{\ensuremath{m_{\PH/\PZ}^\mathrm{meas}}\xspace}
\newcommand{\kw}{\ensuremath{\kappa_{\PW}}\xspace}
\newcommand{\kh}{\ensuremath{\kappa_{\PH}}\xspace}
\newcommand{\kz}{\ensuremath{\kappa_{\PZ}}\xspace}
\newcommand{\TB}{\ensuremath{\PQT\PQB}\xspace}

\cmsNoteHeader{B2G-18-003}
\title{Search for electroweak production of a vector-like \PQT quark using fully hadronic final states}

\date{\today}

\abstract{
A search is performed for electroweak production of a vector-like top
quark partner \PQT of charge $2/3$ in association with a
top or bottom quark, using proton-proton collision
data at $\sqrt{s} = 13$\TeV collected
by the CMS experiment at the LHC in 2016.
The data sample corresponds to an integrated luminosity of 35.9\fbinv.
The search targets \PQT quarks over a wide range of masses and
fractional widths, decaying to a top quark and
either a Higgs boson or a \PZ boson in fully hadronic final states.
The search is performed using
two experimentally distinct signatures that depend on whether
or not each quark from the decays of the top quark, Higgs boson, or \PZ boson
produces an individual resolved jet.
Jet substructure, {\PQb} tagging, and kinematic variables
are used to identify the top quark and boson jets, and also to suppress
the standard model backgrounds.
The data are found to be
consistent with the expected backgrounds.
Upper limits at $95\%$ confidence level are set on the cross sections
for \PQT quark-mediated production of
$\PQt\PH\mathrm{Q}\PQq$, $\PQt\PZ\mathrm{Q}\PQq$, and their sum,
where Q is the associated top or bottom heavy quark
and \PQq is another associated quark.
The limits are given for each search signature for various \PQT quark
widths up to 30\% of the \PQT quark mass,
and are between 2\unit{pb} and 20\unit{fb}
for \PQT quark masses in the range 0.6--2.6\TeV.
These results are significantly more sensitive than prior
searches for electroweak single production
of \tprimetotH and represent the first constraints on \tprimetotZ
using hadronic decays of the \PZ boson with this production mode.
}

\hypersetup{
pdfauthor={CMS Collaboration},
pdftitle={Search for electroweak production of a vector-like T quark using fully hadronic final states},
pdfsubject={CMS},
pdfkeywords={CMS, physics, vector-like quarks}}

\maketitle

\section{Introduction}\label{sec:intro}
We report on a search for electroweak production of a new heavy quark of
charge 2/3 with nonchiral couplings, referred to as a vector-like quark.
Unlike the standard model (SM) chiral fermions, such particles do not
acquire their mass from a Yukawa coupling to the Higgs boson (\PH). Many
proposed extensions of the SM contain vector-like quarks, which usually
mix with the top quark (\PQt). Such particles could have a role in
stabilizing the Higgs boson mass, and thus offer a potential
solution to the hierarchy problem. Vector-like quarks are discussed in
detail in Refs.~\cite{Okada:2012gy, Aguilar-Saavedra:2013qpa,
DeSimone:2012fs, Buchkremer:2013bha} and have been the subject of
phenomenological studies in various frameworks including those of
Refs.~\cite{AguilarSaavedra:2009es, Dobrescu:2009vz, Vignaroli:2012nf,
Han:2003wu}.

\begin{figure}[!htbp]
  \centering
    \includegraphics[width=0.49\textwidth,height=0.25\textheight]{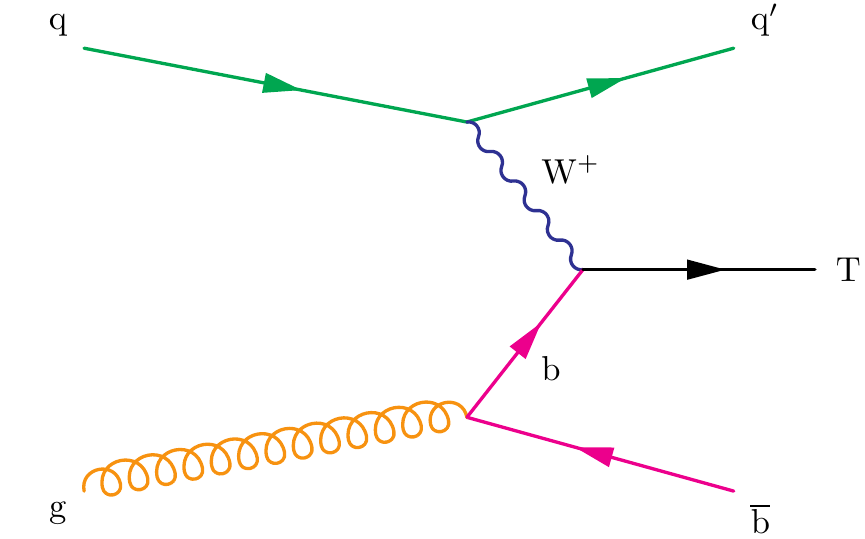}
    \includegraphics[width=0.49\textwidth,height=0.25\textheight]{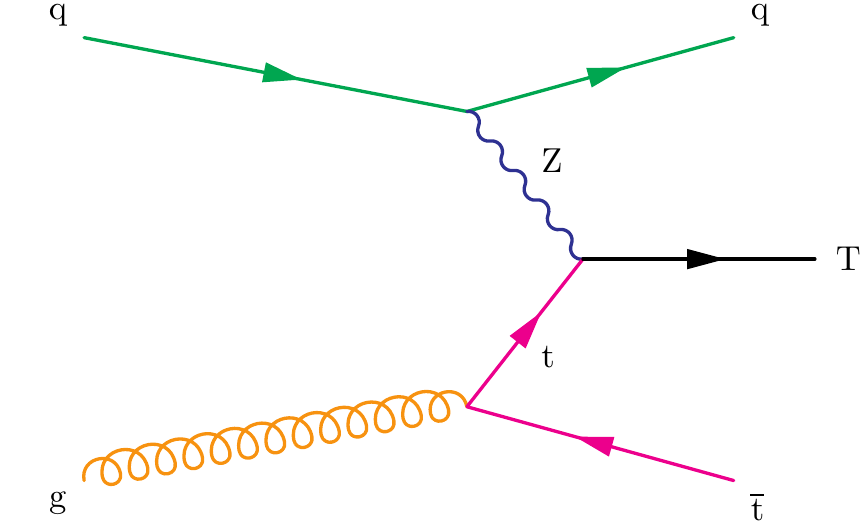}
    \caption{Example Feynman diagrams for electroweak production of vector-like \PQT quarks.
Charged-current (left) and neutral current (right).}
    \label{fig:diagrams}

\end{figure}

Much like the top quark, a vector-like top quark partner (\PQT) could be produced
either in pairs, dominantly through the strong interaction, or singly,
in association with additional quarks through the electroweak interaction.
The \PQT quark could couple to \bW, \tZ, or \tH; this leads to the
corresponding \PQT quark decays and to the associated electroweak production from processes such
as those depicted in Fig.~\ref{fig:diagrams}.
The branching fractions and dominant electroweak production processes
depend on the particular model;
many models have substantial branching fractions to \tZ or \tH resulting in
signatures that are of primary relevance to this paper.
Neglecting the corrections due to decay particle masses, the branching fractions for
the \PQT singlet model of Ref.~\cite{AguilarSaavedra:2009es}
are 50\% (\bW), 25\% (\tZ), 25\% (\tH), while for
the (\TB) doublet model of Ref.~\cite{AguilarSaavedra:2009es},
the \tZ and \tH branching fractions tend to be approximately equal
and depend on two mixing angles, $\theta^u_R$ and $\theta^d_R$, with
each branching fraction ranging from zero to 50\%.
Therefore specific models can have branching fractions
as large as 50\% for \tZ and 50\% for \tH.

We perform a search targeting the electroweak production of a
vector-like top quark partner \PQT
in fully hadronic final states in proton-proton ($\Pp\Pp$) collisions
at $\sqrt{s}=13$\TeV with the CMS detector at the CERN LHC.
We use two searches that target separately lower and higher
mass values for the \PQT quark.
Both searches are designed to be sensitive to the decay to a top quark
and a Higgs boson (\tprimetotH),
and to the decay to a
top quark and a \PZ boson (\tprimetotZ)
with subsequent hadronic decays of
X ($\mathrm{X}=\PH,\PZ$).
Both also consider a wide range of widths of the \PQT quark, ranging from
narrow, defined as small compared to the experimental mass resolution,
to as much as 30\% of the \PQT quark mass.
The event selections primarily require {\PQb} tagging for the Higgs
and \PZ boson candidates and so are most sensitive to $\Xbb$.
The experimental signature is a resonant peak in
the \tX invariant mass spectrum.

The searches are designed to seek evidence of \PQT quarks produced in
association with a bottom quark, dominated by the
\qgTbq process,
and, separately, associated production with a top quark dominated by
the \qgTtq process, where the
charge conjugate processes are also implied.
These are electroweak production modes, with the production of only a
single \PQT, that rely on a nonzero $\PQT\PQb\PW$ coupling for
the charged-current production, and a nonzero $\PQT\PQt\PZ$
coupling for the neutral-current production.
In order to be produced with an observable cross section, one needs a
substantial partial width for the coupling to the initial state.
As a consequence, currently accessible production cross sections in
electroweak single production are associated with particle widths
exceeding about 5\%, which would affect the experimentally observable
invariant mass distributions. The total width could also be enlarged
if additional decay modes were present.

As a result of the lower requirement on the constituent
center-of-mass energy and the larger available phase space,
single production via the electroweak mechanism allows a search for
vector-like top quarks with masses beyond those already tested with
pair production. The $\qgTbq \to \tXbq$ process, with
the top quark from the \PQT quark decaying hadronically and X
decaying to two \PQb\,quarks,
results in up to seven jets, four of which are {\PQb} jets. The seven
jets are associated with the production of seven fermions, namely
$\PQq \Pg \to ( \PQq \qapr \PQb ) (\bbbar ) \PAQb \qpr$.
Similarly the $\qgTtq \to \tXtq$
process results in at least nine fermions. In each case the other
associated quark (\qpr or \PQq) often results in a forward jet at
high absolute pseudorapidity. The \qgTbq process
is expected to have a higher cross section than \qgTtq from
kinematic and coupling considerations.

Recent searches at the LHC for pair production of vector-like quarks
have severely constrained the possible existence
of lower-mass vector-like quarks that couple to heavy
quarks~\cite{Aaboud:2017qpr,Sirunyan:2017usq,Aaboud:2017zfn,
Sirunyan:2017pks,Aaboud:2018xuw,Sirunyan:2018omb,Aaboud:2018saj,
Aaboud:2018pii,Sirunyan:2018qau,Sirunyan:2019sza}.
These searches use several final states arising from
the \bW, \tZ, and \tH decay channels
and usually model pair production under the assumption of a narrow width.
In particular, for the \PQT singlet model, the
most stringent expected lower mass limit from pair production to date
is 1.2\TeV~\cite{Aaboud:2018pii} at 95\% confidence level (\CL).
Pair production is based on the assumed universal strong coupling and so
the quantum chromodynamics (QCD) pair production cross sections are
known and model independent, and depend only on the \PQT
quark mass, \mtprime. On the other hand, electroweak production depends
on the strength of the \PQT quark coupling at the production vertex,
either \TbW or \TtZ, and
therefore the production cross sections are model dependent.
In some models, such as that of Ref.~\cite{Aguilar-Saavedra:2013qpa},
the couplings are constrained by precision observables to be quite
small. In other models, such as that of Ref.~\cite{Vignaroli:2012nf},
cross sections two orders of magnitude higher than
in Ref.~\cite{Aguilar-Saavedra:2013qpa} may be feasible.
The first direct experimental constraints on electroweak production
of vector-like quarks were published in Ref.~\cite{Aad:2014efa}.
Search results at $\sqrt{s}=13$\TeV include
the search already performed by CMS for electroweak production
of \PQT with \tprimetotH
for both semileptonically decaying top quarks and hadronically decaying top quarks using the
2015 data set~\cite{Khachatryan:2016vph,Sirunyan:2016ipo}. Other results
at $\sqrt{s}=13$\TeV targeting electroweak production of \PQT
are described in
Refs.~\cite{Sirunyan:2017tfc, Aaboud:2018ifs} for \tprimetobW,
and for \tprimetotZ with dielectron and dimuon decays
of the \PZ in Refs.~\cite{Sirunyan:2017ezy,Sirunyan:2017ynj,Aaboud:2018saj},
and using a missing energy signature for
$\PZ \to \PGn \PAGn$ decays in Ref.~\cite{Aaboud:2018zpr}.
The search reported here uses the 2016 data set to study fully hadronic
final states with both merged and resolved jets resulting
from electroweak production of a \PQT quark
with \tprimetotH and \tprimetotZ. This search represents
a significant advance  over prior electroweak production
searches for \tprimetotH, with expected 95\% \CL cross section upper limits
typically 5--10 times lower than those reported in Ref.~\cite{Sirunyan:2016ipo},
and is competitive with other searches for \tprimetotZ in this production mode.

For low values of the \PQT quark mass, the quarks resulting from the top quark decay
and from the Higgs or \PZ boson decay can be resolved as individual jets.
However, as the \PQT quark mass increases, the
larger Lorentz boost from the decay will lead to the decay products of the top
quark and the quarks from the Higgs or \PZ boson becoming progressively
less and less resolved as separate jets.
The jet multiplicity, correspondingly, is reduced.
Furthermore, one can reconstruct both the top quark and
the Higgs or \PZ boson by using large-area jet and substructure techniques, where
area refers to the jet's extent in $\eta$-$\phi$ space.
Consequently, two search signatures are defined as follows:
\begin{itemize}
\item{\textit{Low-mass search:}} Reconstruction of a five-jet invariant
mass signature for the \tprimetotH and \tprimetotZ decay modes.
This search signature is
based on multijet triggers with {\PQb} tagging
and is effective
for low \PQT masses (0.6--1.2\TeV) where
the individual jets from the decays can be resolved.
\item{\textit{High-mass search:}} Reconstruction of an invariant mass signature
from two large-area jets for both the \tprimetotH and \tprimetotZ
decay modes. This search signature is based on triggers
using high transverse momentum jets and is effective
for high \PQT mass  ($>$1.0\TeV).
In this mass range, the final state particles
from the decays of each of the two daughter particles resulting
from the \PQT quark decay (the \PQt and the \PH or \PZ) produce a
single large-area jet. This leads to events with two large-area jets.
\end{itemize}
Each search is designed to be sensitive to
T quark production in association
with either a bottom quark or a top quark.
Besides the primary motivation of exploring the possibility
of electroweak production of a vector-like quark,
this analysis can be viewed more broadly
as two independent searches for high mass signatures
of physics beyond the SM at the LHC.
As such they provide potential for discovery
of new physics, independent of the specific models discussed here.

The paper is organized as follows: this section
has given the motivation to search for the singly produced \PQT quark
with two distinct signatures and two decay modes.
The CMS detector and event reconstruction are described in Section~\ref{sec:detector}.
The data set and the modeling of signal and background processes are described in
Section~\ref{sec:sbmodeling}.
Reconstruction methods common to the two searches are discussed in Section~\ref{sec:reconstruction}.
The event selection criteria, background estimation, and
results are described for the low-mass search
in Section~\ref{sec:fh-resolved}, and for the high-mass search
in Section~\ref{sec:fh-boosted}.
Systematic uncertainties for both signatures are discussed in Section~\ref{sec:systematics}.
The overall results are presented in Section~\ref{sec:combination} and summarized in Section~\ref{sec:summary}.
\section{The CMS detector and event reconstruction}\label{sec:detector}
The central feature of the CMS apparatus is a superconducting solenoid
of 6\unit{m} internal diameter, providing a magnetic field of
3.8\unit{T}. Within the solenoid volume are a silicon pixel and strip
tracker, a lead tungstate crystal electromagnetic calorimeter (ECAL),
and a brass and scintillator hadron calorimeter (HCAL), each composed
of a barrel and two endcap sections. Forward calorimeters extend the
pseudorapidity coverage provided by the barrel and endcap detectors.
Muons are detected in gas-ionization chambers embedded in the steel
flux-return yoke outside the solenoid.

The silicon tracker measures charged
particles within the pseudorapidity range $\abs{\eta} < 2.5$.
For nonisolated particles with transverse momentum, $\pt$, in the
range $1 < \pt < 10\GeV$ and $\abs{\eta} < 1.4$, the track resolutions
are typically 1.5\% in \pt and 25--90 (45--150)\mum in the
transverse (longitudinal) impact parameter~\cite{TRK-11-001}.

The ECAL consists of 75\,848 crystals covering $\abs{\eta} < 3.00$.
The HCAL cells have widths of 0.087 in pseudorapidity and 0.087 in azimuth ($\phi$) in
the region $\abs{\eta} < 1.74$.
In the $\eta$-$\phi$ plane, and for $\abs{\eta} < 1.48$, the HCAL cells map
onto $5{\times}5$ arrays of ECAL crystals
to form calorimeter towers projecting radially outwards from
close to the nominal interaction point. For $\abs{\eta} > 1.74$,
the coverage of the towers increases progressively with $\abs{\eta}$
to a maximum of 0.174 in $\Delta \eta$ and $\Delta \phi$.
The forward calorimeters extend the calorimetric coverage
for hadronic jets to $\abs{\eta} = 5.0$.

Events of interest are selected using a two-tiered trigger system~\cite{Khachatryan:2016bia}. The first level,
composed of custom hardware processors, uses information from the calorimeters and muon detectors to select events at a rate of around 100\unit{kHz} within a time interval of less than 4\mus. The second level, known as the high-level trigger, consists of a farm of processors running a version of the full event reconstruction software optimized for fast processing, and reduces the event rate to around 1\unit{kHz} before data storage.

In the reconstruction, the vertex with
the largest value of summed physics-object $\pt^2$ is taken to be the primary $\Pp\Pp$ interaction vertex. The physics objects are the jets, clustered using the anti-\kt
jet finding algorithm~\cite{Cacciari:2008gp,Cacciari:2011ma}, with the tracks
assigned to the vertex as inputs.

A particle-flow algorithm~\cite{CMS-PRF-14-001} aims to reconstruct and identify each individual particle in an event, with an optimized combination of information from the various elements of the CMS detector. The energy of photons is directly obtained from the ECAL measurement, corrected for zero-suppression effects. The energy of electrons is determined from a combination of the electron momentum at the primary interaction vertex as determined by the tracker, the energy of the corresponding ECAL cluster, and the energy sum of all bremsstrahlung photons spatially compatible with originating from the electron track. The momentum of muons is obtained from the fitted trajectory of the corresponding track reconstructed from the tracker and the muon detectors.
The energy of charged hadrons is determined from a combination of their momentum measured in the tracker and the matching ECAL and HCAL energy deposits, corrected for zero-suppression effects and for the response function of the calorimeters to hadronic showers. Finally, the energy of neutral hadrons is obtained from the corresponding
corrected ECAL and HCAL energies.

Jet momentum is determined as the vectorial sum of all particle momenta in the jet, and is found from simulation to be within 5 to 10\% of the true momentum over the whole \pt spectrum and detector acceptance. Additional $\Pp\Pp$ interactions
within the same or nearby bunch crossings (pileup) can contribute additional tracks and calorimetric energy depositions to the jet momentum. To mitigate this effect, tracks identified to be originating from pileup vertices are discarded, and an offset correction is applied to correct for remaining contributions. Jet energy corrections are derived from simulation to bring the measured response of jets to that of
particle-level jets on average. In situ measurements of the momentum balance in dijet, photon+jet, \PZ{+}jet, and multijet events are used to estimate any residual differences in jet energy scale in data and simulation~\cite{Khachatryan:2016kdb}. Additional selection criteria are applied to each jet to remove jets potentially dominated by anomalous contributions from various subdetector components or reconstruction failures.
The jet energy resolution amounts typically to 15\% at 10\GeV, 8\% at 100\GeV, and 4\% at 1\TeV, to be compared to about 40, 12, and 5\% obtained when the calorimeters alone are used for
jet clustering~\cite{CMS-PRF-14-001}.

A more detailed description of the CMS detector, together with a definition of the coordinate system used and the relevant kinematic variables, can be found in Ref.~\cite{Chatrchyan:2008zzk}.

\section{Data and modeling of signals and backgrounds}\label{sec:sbmodeling}
This analysis uses proton-proton collision data collected at a
center-of-mass energy of $\sqrt{s}=13$\TeV, recorded in 2016,
amounting to a total integrated luminosity of \intL.

Simulated samples
for the $2 \to 3$ signal processes, \Tbq and \Ttq, were
generated at leading order
using the \mc (MC) event generator
\MGvATNLO 2.3.3~\cite{MG5_aMCNLO}
for various masses of the \PQT quark and for the
decays \tprimetotH and \tprimetotZ.
The signal generation for these ``narrow width'' $2 \to 3$ process
samples has the width set to 10\GeV, which is small on the
scale of the experimental resolution of about 6\%.
Separate samples were generated for both
left- and right-handed chiralities of the \PQT quark for each decay mode,
for these narrow-width cases.
In addition, \MGvATNLO 2.4.2 at leading order was used to simulate
the large width $2 \to 4$
processes, $\Pp\Pp\to\tHbq$, $\Pp\Pp\to\tHtq$,
$\Pp\Pp\to\tZbq$ and $\Pp\Pp\to\tZtq$,
with fractional widths $\GoM$ of 10, 20, and 30\%.
All of the large-width samples assume left- \mbox{(right-)handed} \PQT
chiralities for the \Tbq (\Ttq) case, as expected
in the singlet (doublet) model.

The benchmark \PQT quark masses used
for the results range from 0.6 to 2.6\TeV.
The \textsc{NNPDF3.0} parton distribution function (PDF)
set~\cite{Ball:2014uwa} was used. The samples are generated with both
the top quark and the Higgs boson decaying inclusively.
The masses of the Higgs boson and top quark are set
to 125 and 172.5\GeV, respectively.
In the samples, the SM Higgs boson
branching fraction $\mathcal{B}(\Hbb)$ of 58\% is assumed.
Similarly the $\PQt$ and $\PZ$ are decayed inclusively in
the \tprimetotZ samples.

The SM background simulation samples include \ttjets, \wjets, \zjets,
single top quark, \tHq, \ttH,
$\PQt\PAQt\PW$, $\PQt\PAQt\PZ$, $\PW\PW$, $\PZ\PZ$, $\PW\PZ$,
$\PW\PH$, and $\PZ\PH$. These processes are generated at
next-to-leading order with \MGvATNLO 2.3.3 unless otherwise specified.
The parton-level MC simulations for SM backgrounds and signal are
interfaced with \PYTHIA~8.212~\cite{Pythia8}.
The large-width signal samples and the main backgrounds involving top
quarks use the CUETP8M2T4 tune~\cite{CMS-PAS-TOP-16-021}. The other
samples use the earlier CUETP8M1 tune~\cite{Khachatryan:2015pea}.

The \ttjets events are inclusive, and are simulated
using \POWHEG~2.0~\cite{1126-6708-2004-11-040,1126-6708-2007-11-070,
Alioli2010,1126-6708-2007-09-126}.
The \wjets and \zjets samples include only
hadronic \PW or \PZ boson decays and contain an $\HT>600$\GeV requirement,
where $\HT$ is the scalar sum of jet transverse momenta.
The single top quark $\PQt\PW$ process was generated using
the \POWHEG~2.0+\PYTHIA~8 generator combination.
The SM \tHq process simulation included all
decay modes for the top quark and Higgs boson.
The \ttH sample was generated with the decay \Hbb
with a Higgs boson mass of 125\GeV using \POWHEG~2.0.
The $\PW\PW$ sample was generated with hadronic decays
using \POWHEG~2.0.
The $\PZ\PZ$ sample was generated with hadronic decays
using \textsc{Madspin}~\cite{MadSpin} and applying the \textsc{FxFx}
merging procedure~\cite{FxFx} for matching jets from
the matrix element calculation with those from the parton shower.
The $\PW\PZ$ sample was inclusive and generated with \PYTHIA~8.
The $\PZ\PH$ sample was generated with \Zbb with \POWHEG~2.0.

The SM background events
comprised uniquely of jets produced through the strong interaction,
referred to as QCD multijet events, were also considered
in the design of the analyses.
Two sets of simulated samples were used: a sample
generated using \PYTHIA~8 that is binned in the
invariant $\pt$ associated
with the hard process
and an $\HT$-binned sample using
{\MGvATNLO} at leading order with
up to four partons in the matrix element calculations, using the MLM jet
matching scheme~\cite{MLM} with \PYTHIA fragmentation and showering.
All simulated event samples were generated using
the {NNPDF3.0} PDF set except that based on \PYTHIA~8 which used
the {NNPDF2.3} PDF set~\cite{Ball:2012cx}.

\section{Reconstruction methods and primary selection}
\label{sec:reconstruction}

Particle-flow anti-\kt jets are used.
Small-area jets, denoted AK4 jets, are defined using
a distance parameter of 0.4, whereas large-area jets are defined
with a distance parameter of 0.8 and are denoted as AK8 jets.

For both searches, {\PQb} tagging is used to identify jets
that contain \PQb-flavored hadrons ({\PQb} jets).
The {\PQb} tagging is applied
to AK4 jets and also subjets reconstructed as part of an AK8 jet.
Depending on the search, {\PQb} tagging involves several
secondary vertexing algorithms:
the online and offline CSVv2 discriminators and the DeepCSV
discriminators~\cite{CMS-BTV-16-002}. Offline {\PQb} tagging working points
are defined with light-flavor jet mistag rates of
approximately 0.1\% (tight), 1\% (medium), and 10\% (loose).
For AK4 jets, the tight CSVv2 {\PQb} tagging working point has an efficiency
of 41\% while the medium {\PQb} tagging working points have
CSVv2 and DeepCSV efficiencies of 63 and 68\%, respectively.
In the case of AK8 jets, a grooming algorithm~\cite{softdrop}
looks for jet substructure and {\PQb} tagging is applied to the resulting
subjets. For AK8 jets with \pt of around 400\GeV,
the medium {\PQb} tagging working point has an
efficiency per subjet of about 51\%, whereas the loose {\PQb} tagging
working point has an efficiency per subjet of about 75\%.
The mistag rates of jets originating from \PQc-flavored hadrons for the tight,
medium, and loose {\PQb} tagging working points are about 2, 12, and 37\%, respectively.

The event selection requires that the events have at least one
satisfied trigger condition among a set of unprescaled high-level
trigger algorithms. The set is specific to each search strategy.
The trigger conditions for the low-mass search strategy
rely on online jet information and, for some trigger conditions,
also on {\PQb} tagging information.
The event selection is dominated by the trigger condition with the
lowest jet multiplicity and the lowest \pt threshold.
This condition requires at least six jets with $\pt>30$\GeV with
two of them passing the {\PQb} tagging online criteria.
The trigger conditions of the high-mass search strategy consist
of a scalar \PT \ sum trigger formed from all jets
with a summed \PT \ threshold of 900\GeV and an inclusive
large-area single jet trigger with a \PT \ threshold of 450\GeV.
The high-mass search uses three other trigger conditions
that include some loose jet
substructure requirements and lower thresholds on either
the scalar \PT \ sum (700\GeV), the inclusive single-jet
\PT \ (360\GeV), or on two jets, with the higher (lower) jet \PT \
required to exceed 300 (200)\GeV.

At least one primary vertex must be found within 24\cm longitudinally
and within 2\cm radially of the center of the luminous region.

\section{Low-mass search}\label{sec:fh-resolved}
The low-mass search strategy
uses the invariant mass reconstructed from five AK4 jets as the main
discriminating variable. The event selection requires at least six jets
to conform with the trigger requirements. The selection criteria are
based on the properties of the signal final state, in the cases of \tbw
and $\HZ \to \bbbar$.
The final state is composed of two jets coming from the \PW decay and
three {\PQb} jets (two coming from the \HZ and one from the top quark decay).
The main background processes consist
of QCD multijet production and top quark pair production.
These backgrounds are not expected to result in a resonance in
the five-jet invariant mass variable.
A reduction of the QCD multijet background is achieved by
imposing {\PQb} tagging, and by requiring events to be consistent with
the presence of all of the relevant states (\PW, \HZ, and top quark).
The presence of only one top quark
candidate from the
selected jets is used to reduce the \ttbar background.

\subsection{Event selection}
The following criteria define the first part of the selection:

\begin{itemize}
\item {\textit{Small-area jet multiplicity.}}
The event should have at least six AK4 jets with $\pt>40$\GeV
within $\abs{\eta}<4.5$.

\item{\textit{Leading jets.}}
The jets with the highest \pt (leading jets) have
larger \pt in the signal than in most of the backgrounds.
Therefore, the leading jets must have $\pt > 170$, 130, and 80\GeV,
for the leading, second-leading, and third-leading jets, respectively.

\item{\textit{\PQb-tagged jets.}}
The considered \PQT quark decay leads to three {\PQb} quarks while most
backgrounds have at most two {\PQb} quarks. In the signal region
labeled as \ThreeT, at least three \PQb-tagged jets using the tight DeepCSV working point are required
for jets with $\abs{\eta}<2.4$.
Other {\PQb} tagging working points are used to estimate
the background using control samples in data.

\item{\textit{\PQT candidate identification.}}
The correct identification of a \HZ boson and a top quark in
the five-jet final state relies on a $\chi^{2}$ sorting algorithm.
Here the presence of all three states, namely the \HZ,
the \PW, and the top quark, is exploited.
The algorithm loops over jet combinations and
considers two {\PQb}-tagged jets for the \HZ candidate,
two jets (potentially {\PQb}-tagged) for the {\PW} candidate, and
a combination of the dijet {\PW} candidate and a {\PQb}-tagged jet
for the top quark candidate. These combinations of jets are used
to construct the variables defined in
Eqs.~(\ref{eq:fh-chi2H})--(\ref{eq:fh-chi2Tot}):
\begin{equation}
\chi^{2}_{\HZ}=\left( \frac{\mhzm-\mhzmc}{\shzmc} \right)^2 ,
\label{eq:fh-chi2H}
\end{equation}
\begin{equation}
\chi^{2}_{\PW}=\left( \frac{\mwm-\mwmc}{\swmc} \right)^2 ,
\label{eq:fh-chi2W}
\end{equation}
\begin{equation}
 \chi^{2}_{\PQt} =  \left( \frac{\mtm-\mtmc}{\stmc} \right)^2 ,
\label{eq:fh-chi2top}
\end{equation}
\begin{equation}
\chi^{2} = \chi^{2}_{\HZ} + \chi^{2}_{\PW} + \chi^{2}_{\PQt} ,
\label{eq:fh-chi2Tot}
\end{equation}
where $m^\text{meas}$ denotes the measured mass quantities reconstructed
from the considered combination of jets,
and $m^\mathrm{MC}$ and $\sigma^\mathrm{MC}$ denote
the expected mass values and standard deviations
from Gaussian fits to simulated signal samples.
The mass values fitted for each particle are:
$\mhmc=121.9$\GeV, $\mzmc=90.9$\GeV, $\mwmc=83.8$\GeV and $\mtmc=173.8$\GeV;
these values differ only slightly from
the input world-average values~\cite{PDG2018}.
For the \bbbar decays of the Higgs and \PZ bosons, the fitted standard
deviations are $\shmc=13.5$\GeV and $\szmc=11.4$\GeV, and for the
fully hadronic decays of the \PW and top quark,
they are $\swmc=10.0$\GeV and $\stmc=16.0$\GeV.
One first chooses the lowest $\chi^{2}_{\HZ}$ {\PQb}-tagged jet pair
as the \HZ candidate and then selects the other jets making up the {\PW}
and top quark candidates by minimizing the total $\chi^{2}$.
This procedure is found to improve the signal-to-background
ratio by 30\% compared to simply choosing the combination with
the best total $\chi^2$.
Finally, the total $\chi^2$ must not exceed 15 in order to ensure good
quality of the \HZ, \PW, and top quark candidates.
It is found that the five jets are correctly identified
about 73 and 64\% of the time for
the narrow width \tHbq and \tZbq cases, respectively.

\item{\textit{Second top quark mass.}}
A large fraction of \ttbar events
survive the requirement on at least three {\PQb}-tagged jets.
These originate from incorrect {\PQb} tagging of the jet arising
from a charm quark from the \PW boson part of one of the top quark decays.
In order to reduce this background, we define the second top quark mass
as the invariant mass formed by the \HZ candidate and the
remaining highest \pt jet not used in the $\chi^{2}$ calculation.
For \ttbar events, the second top quark mass has a peak around 172\GeV
and there are nearly no signal events expected in that region. Therefore
we require that the second top quark mass is greater than 250\GeV. This
leads to about a factor of two reduction in the \ttbar background.

\item{\textit{\bbbar mass.}}
Finally, the reconstructed boson from the \PQT candidate must have a mass
larger than 100\GeV, if looking for a \PH, and the mass must be smaller
than 100\GeV, if looking for a \PZ.
This ensures that there is no overlap between the two channels.
\end{itemize}

The second part of the selection uses the presence
of a top quark and a Higgs or \PZ boson in the event.
The variables are chosen to be as model
independent as possible and the selection criteria are optimized using the
figure of merit described in~\cite{Punzi:2003bu}.
The selection criteria are described below.

\begin{itemize}
   \item {\textit{Relative \HT.}}
The relative \HT variable is defined as $\big(\pt(\HZ_\text{cand})+\pt({\PQt}_\text{cand})\big)/\HT$.
In single \PQT quark production, most of the momentum should be carried by
the top quark and \HZ candidates, therefore the relative \HT is
an extremely good discriminator against \ttbar and multijet events.
The \HZ and \PQt candidates from the \PQT candidate decay
must have a relative \HT greater than 0.40.

   \item{\textit{Max($\chi^{2}$).}}
The maximum among the $\chi^{2}$ values defined in
Eqs.~(\ref{eq:fh-chi2H})--(\ref{eq:fh-chi2top}) is examined and is required to be less than 3.0.
This criterion is highly correlated to the $\chi^2$ criteria but represents a tighter condition that
ensures that each mass is identified with high quality.
It is equivalent
to requiring a mass window of at most $\pm \sqrt{3}\sigma$ for each candidate.

   \item{\textit{$\Delta R$ of jets from \HZ decay.}}
Because of the large mass of the \PQT quark (above 0.6\TeV),
the \HZ decay tends to be boosted (but the {\PQb} jets not completely merged).
A small spatial separation
of $\Delta R(\PQb_{\HZ}, \PQb_{\HZ}) <1.1$ between
the two {\PQb}-tagged jets is required, leading to a reduction of the background.
The $\Delta R$ is defined as the
inter-jet separation in $\eta$-$\phi$ space
($\Delta R = \sqrt{\smash[b]{ (\Delta \eta)^2 + (\Delta \phi)^2}}$),
where $\Delta \eta$ and $\Delta \phi$ are the
corresponding inter-jet separations in pseudorapidity and azimuth (in radians).

  \item{\textit{\HZ $\chi^{2}$.}} As most of the backgrounds do not
contain a genuine Higgs boson, $\chi^{2}_{\HZ}$
is a very discriminating criterion for the Higgs boson decay channel.
We require $\chi^{2}_{\PH }<1.5$ for the \PH case
and $\chi^{2}_{\PZ}<1.0$ for the \PZ case. It is equivalent to a mass
window of $\pm 16.5$\GeV for the Higgs boson and
$\pm 11.4$\GeV for the \PZ boson. A tighter $\chi^{2}$ requirement
is made for \PZ candidates to avoid background contamination from lower masses and to
reduce overlap with \PH candidates.

  \item{\textit{$\Delta R$ of jets from \PW decay.}} Given the Lorentz
boost of the \PW in signal events, a requirement of $\Delta R(j_{\PW}, j_{\PW}) <1.75$
reduces the QCD multijet background, while retaining most of the signal.

  \item{\textit{$\Delta R$ of jets from top quark decay.}}
The top quark decay products tend to be Lorentz-boosted (but the jets do not completely merge) for the signal.
A spatial separation between the {\PQb}-tagged jet and
  the \PW~candidate that is used to make the top quark candidate of $\Delta R({\PQb}_{\PQt},\PW) <1.2$ is required. This
further reduces the QCD multijet background.

\end{itemize}

The total number of events selected from the data
sample in the \ThreeT signal region is 615 (290 for the \tZ selection and 325 for the \tH selection).
The number of expected signal events is 7.6 for a \PQT quark
mass of 0.7\TeV, $\GoM = 0.01$, left-handed chirality, and a \PQT quark produced
in association with a bottom quark with a product of cross section and branching fraction
of 89\unit{fb} for each channel.
For this signal process, the selection efficiency is presented
in Table~\ref{tab:fh-effcutflow} together with various simulated
background processes for the Higgs and \PZ boson decay channels.

\begin{table}[htb]
\centering
\topcaption{Cumulative efficiencies for the low-mass search
after applying event selections for the signal and main backgrounds
in the Higgs boson decay channel (upper half) and the \PZ boson
decay channel (lower half).
The first and last rows of each section give the expected numbers
of events normalized to the integrated luminosity of 35.9\fbinv.
Uncertainties are statistical only. The signal values are for
a mass of 0.7\TeV, $\GoM = 0.01$, left-handed chirality, and a \PQT quark produced
in association with a bottom quark with a product of cross section and branching fraction
of 89\unit{fb} for each channel.
The ``Other backgrounds'' column includes \wjets, \zjets,
single top quark, and \ttH processes.
It has been checked that the \ttH process does not
present a resonance in the \tH channel.
The number of expected \ttH events is comparable to the
expected \PQT signal.}
\label{tab:fh-effcutflow}
\cmsTable{
\begin{tabular}{lcccc}
 Selection for \tH & Signal & QCD Multijet & \ttbar & Other backgrounds\\
\hline
Basic selection  ($m_{\bbbar}>100\GeV$)& 23.1 $\pm$ 0.9 & 9360 $\pm$ 810 & 2612 $\pm$ 28  & 353 $\pm$ 23 \\
 Relative $\HT > 0.4$&  81.4\%  & 42.8\% & 51.9\% & 52.9\%\\
 $\mathrm{Max}(\chi^2) < 3.0$ & 54.3\%   & 14.1\% & 25.1\% & 21.8\%\\
 $\DRbbHH < 1.1 $ & 44.4\% & 7.5\% & 11.9\% & 8.9\% \\
 $\chi^{2} _{\PH } < 1.5$ & 39.8\% & 4.9\% & 9.3\% & 7.1\% \\
 $\DRjj < 1.75$& 33.7\% & 3.2\% & 7.2\% & 5.6\%\\
 $\DRbtW < 1.2$ & 25.7\% & 1.9\% &  4.5\% & 2.5\%\\
Full selection & 5.9 $\pm$ 0.4 & 181 $\pm$ 52 & 116.5 $\pm$ 6.1 & 9.3 $\pm$ 0.6\\[\cmsTabSkip]
 Selection for \tZ & Signal & QCD Multijet & \ttbar & Other backgrounds\\
\hline
Basic selection ($m_{\bbbar}<100\GeV$)& 5.7 $\pm$ 0.2 & 6810 $\pm$ 630 & 1270 $\pm$ 17   &  223 $\pm$ 24 \\
 Relative $\HT > 0.4$&  86.9\%  & 48.5\% & 47.2\% & 57.5\% \\
 $\mathrm{Max}(\chi^2) < 3.0$ & 53.3\%   & 15.9\% & 24.1\% & 28.8\% \\
 $\DRbbZZ < 1.1 $ & 51.1\% & 11.7\% & 16.4\% & 22.7\%\\
 $\chi^{2} _{\PZ} < 1.0$ & 45.0\% & 7.3\% & 11.5\% & 18.4\% \\
 $\DRjj < 1.75$& 37.6\% & 5.2\% & 9.6\% & 9.9\%\\
 $\DRbtW < 1.2$ & 28.8\% & 1.5\% &  5.7\% & 5.5\%\\
Full selection & 1.6 $\pm$ 0.1 & 103$\pm$ 38 & 72.7 $\pm$ 4.7 & 8.1 $\pm$3.9\\
\end{tabular}}
\end{table}

\subsection{Background estimation and validation}\label{sec:fh-BkgEst}
None of the SM backgrounds are expected to result in a resonance in
the five-jet invariant mass, therefore the spectrum of the background
five-jet invariant mass distribution should have a monotonically decreasing
shape. However, the second part of the selection
criteria tends to shape the five-jet invariant mass distribution.
In order to evaluate the shape of the five-jet invariant mass distribution
for the background in data,
two regions that are independent from the main \ThreeT signal region are defined using looser {\PQb} tagging criteria.
In these two regions
it is important to ensure
that no bias with respect to the selection criteria is present, as all
backgrounds are estimated from data.
The extraction of signal
is done by fitting the signal and background simultaneously in all three regions.

The {\PQb} tagging does not strongly influence the kinematic distributions
of objects used to construct the five-jet invariant mass.
Therefore we relax the {\PQb} tagging criteria required for three of the
five jets forming the \PQT candidate.
The first new region is called the \ThreeM signal region; it
requires three medium {\PQb}-tagged jets but excludes events with three tight {\PQb}-tagged jets, bringing information on the background and possible signal shapes.
The second new region is denoted as the \TwoMOneL signal region; in order to
have significant numbers of events in this background-dominated region
and to keep events with similar kinematics to the \ThreeT signal region,
the {\PQb} tagging criteria are relaxed to two medium and
one loose {\PQb}-tagged jets but excluding three medium {\PQb}-tagged jets.

Two additional samples are defined and used to validate the
method; one is enriched in QCD multijet events, the other in \ttbar events.
In order to define the QCD multijet enriched control sample,
the $\chi^2$ criterion is relaxed to 50, the
Max($\chi^{2}_{\PH/\PZ}$, $\chi^{2}_{\PW}$, $\chi^{2}_{\PQt}$) criterion is inverted,
and $\chi^2_{\PQt} > 1$ is required to reduce the fraction of \ttbar events.
The QCD multijet sample is subdivided into a \ThreeT region
and a \TwoMOneL region (excluding \ThreeM) based on the {\PQb}-tagged jet configurations.
For the \ttbar control sample,
the $\chi^2$ criterion is relaxed to 50,
the Max($\chi^{2}_{\PH/\PZ}$, $\chi^{2}_{\PW}$, $\chi^{2}_{\PQt}$) and
$\chi^2_{\PH/\PZ}$ criteria are inverted, the
\DRbbHZ criterion is relaxed to 1.5, and $\chi^2_{\PQt} <1.5$ is required.
The \ttbar sample is subdivided into a \TwoTOneL (two tight and one loose {\PQb}-tagged jets) region and a \TwoMOneL region (excluding \TwoTOneL) based on the {\PQb}-tagged jet configurations.
For the \TwoTOneL (\TwoMOneL) region, one of the tight (medium) {\PQb}-tagged jets must be from the top
quark candidate.
A summary of the criteria changed to define each region is presented in Table~\ref{tab:fh-Regions}.
The fraction of expected signal events is of the order of 3\% in the QCD multijet region and 1\% in the \ttbar \TwoTOneL region (for a cross section times branching fraction of 600\unit{fb}).

\begin{table}[htb]
\centering
\topcaption{Criteria defining the various signal and control regions.
The first line of each section gives the \PQb tagging requirements.
The criteria that differ are preceded by an asterisk "*".
The {\bbbar} mass requirements are different for
the \PH ($m_{\bbbar}>100\GeV$) and \PZ~channels ($m_{\bbbar}<100\GeV$).
\label{tab:fh-Regions}
}
{
\begin{tabular}{lll}
  \ThreeT region & \ThreeM region & \TwoMOneL region \\ \hline
  \ThreeT    & *{\ThreeM but not \ThreeT} &  *{\TwoMOneL but not \ThreeM} \\
  $\chi^{2}<15$ & $\chi^{2}<15$ & $\chi^{2}<15$ \\
  Relative $\HT > 0.4$  & Relative $\HT > 0.4$ & Relative $\HT > 0.4$ \\
  Max($\chi^{2}) < 3.0$ &  Max($\chi^{2}) < 3.0$ & Max($\chi^{2}) < 3.0$ \\
  $\DRbbHZ < 1.1 $ &  $\DRbbHZ < 1.1 $ & $\DRbbHZ < 1.1 $  \\
  $\chi^{2}_{\HZ} < 1.5 / 1.0$   & $\chi^{2}_{\HZ} < 1.5 / 1.0$  & $\chi^{2}_{\HZ} < 1.5 / 1.0$\\
  $\DRjj < 1.75$ & $\DRjj < 1.75$ & $\DRjj < 1.75$ \\
  $\DRbtW < 1.2$ & $\DRbtW < 1.2$ & $\DRbtW < 1.2$ \\[\cmsTabSkip]

\end{tabular}
}
{
\begin{tabular}{ll}
  QCD \ThreeT region & QCD \TwoMOneL region \\ \hline
  \ThreeT &   *{\TwoMOneL but not \ThreeM} \\
  *{$\chi^{2}<50$} & *{$\chi^{2}<50$} \\
  Relative $\HT > 0.4$ & Relative $\HT > 0.4$  \\
  *{$5 < \mathrm{Max}(\chi^{2}) < 20$ and $\chi^{2}_{\PQt} > 1.0$ } &
  *{$5 < \mathrm{Max}(\chi^{2}) < 20$ and $\chi^{2}_{\PQt} > 1.0$ }  \\
  $\DRbbHZ < 1.1 $ &  $\DRbbHZ < 1.1 $ \\
  $\chi^{2}_{\HZ} < 1.5 / 1.0$ & $\chi^{2}_{\HZ} < 1.5 / 1.0 $ \\
  $\DRjj < 1.75$ & $\DRjj < 1.75$ \\
  $\DRbtW < 1.2$ & $\DRbtW < 1.2$ \\[\cmsTabSkip]

  \ttbar \TwoTOneL region & \ttbar \TwoMOneL region \\ \hline
  *{\TwoTOneL}  & *{\TwoMOneL but not \TwoTOneL}   \\
  *{Top \PQb-tag \Tight } & *{Top \PQb-tag \Medium}\\
  *{$\chi^{2}<50$}     & *{$\chi^{2}<50$} \\
  Relative $\HT > 0.4$ & Relative $\HT > 0.4$  \\
  *{$3 < \mathrm{Max}(\chi^{2}) < 5$}  &  *{$3 < \mathrm{Max}(\chi^{2}) < 5$}  \\
  *{$\DRbbHZ < 1.5 $} &  *{$\DRbbHZ < 1.5 $} \\
  *{$\chi^{2}_{\PQt} < 1.5$ and $\chi^{2}_{\HZ} > 3$} &
  *{$\chi^{2}_{\PQt} < 1.5$ and $\chi^{2}_{\HZ} > 3$} \\
  $\DRjj < 1.75$ & $\DRjj < 1.75$ \\
  $\DRbtW < 1.2$ & $\DRbtW < 1.2$ \\

\end{tabular}
}
\end{table}

Relaxing the {\PQb} tagging requirement induces a
change in the {\PQb} tagging
efficiency depending on the $\pt$ and $\eta$ of the jet.
As $\pt$ and $\eta$ are two highly correlated variables,
a reweighting procedure using $\eta$ and momentum is used.
Weights are derived jet-by-jet for each channel and for
each {\PQb} tagging working point using the QCD multijet control region.
These weights reflect the differences in the efficiency between
loose and medium and between medium and tight {\PQb}-tagged jets.
For each event, the product of the weights for all three {\PQb}-tagged jets
is applied to correct for the change in the {\PQb} tagging efficiency
going from \ThreeT to \ThreeM and from \ThreeM to \TwoMOneL.

The validation of the method is done with the QCD multijet and \ttbar control regions.
The shape of the five-jet invariant mass distribution is compared
between the QCD \ThreeT region
and the QCD \TwoMOneL region reweighted as \ThreeT; it is found to be satisfactory
using the Kolmogorov--Smirnov test.
Similarly, when comparing the shape of the five-jet invariant mass for the
the \ttbar \TwoTOneL region and the \ttbar \TwoMOneL region reweighted as \ThreeT, acceptable consistency is found.

\begin{figure}[!hbtp]
  \centering
   \includegraphics[width=0.535\textwidth]{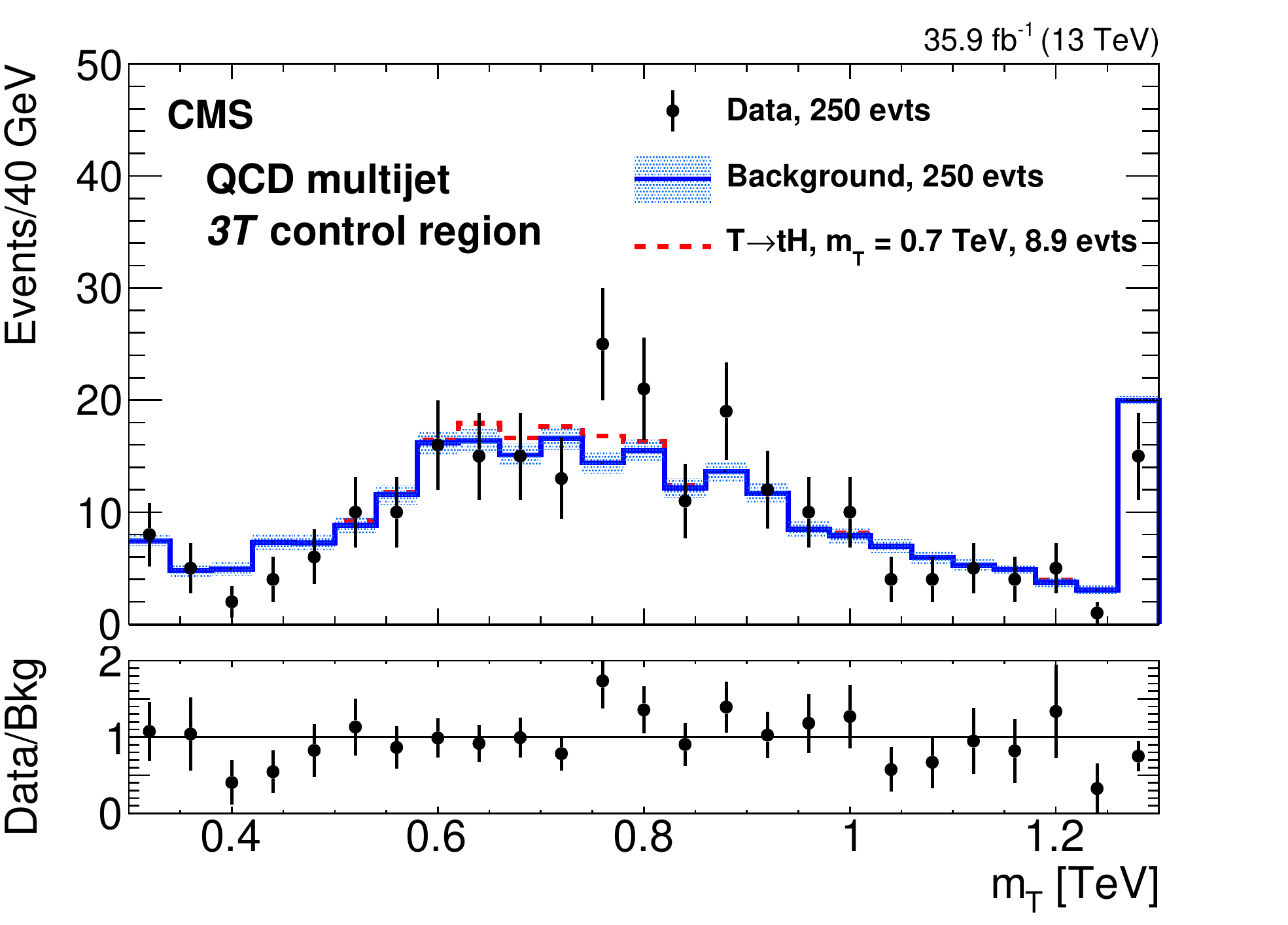}
   \includegraphics[width=0.535\textwidth]{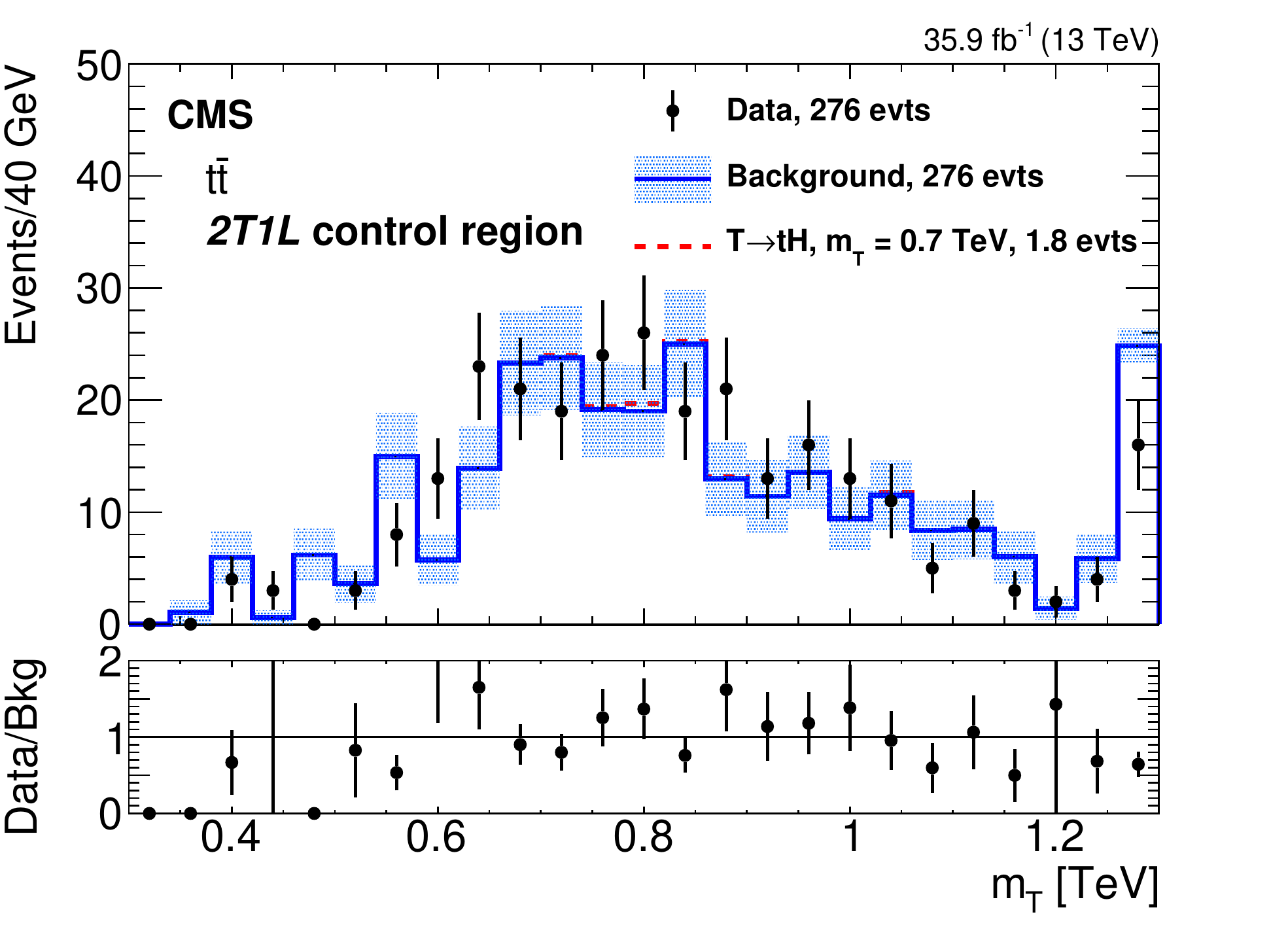}
   \includegraphics[width=0.535\textwidth]{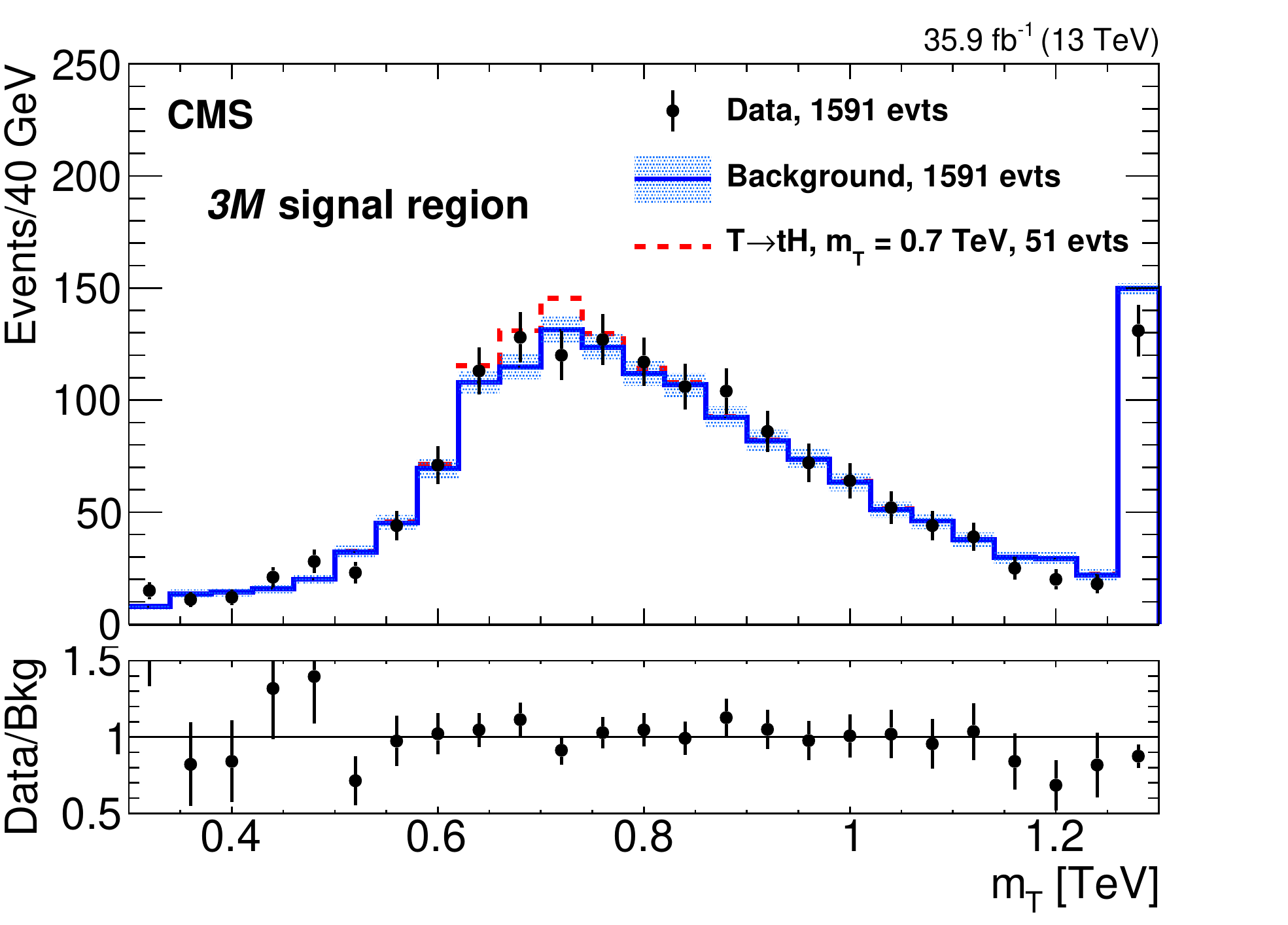}
   \caption{The five-jet invariant mass distribution (black points with error bars) for the \tHbq channel after the full selection
      in the QCD multijet \ThreeT control region (upper), the \ttbar \TwoTOneL control region (middle), and the \ThreeM signal region (lower).
The superimposed blue histogram, labeled ``background'', is the reweighted \TwoMOneL region distribution, used as
an estimate of the background shape, with its normalization adjusted to match the number of
entries observed in each region.
A potential narrow-width signal (dashed red histogram) is added on top of
the blue histogram for ${\mtprime}=0.7$\TeV and $\GoM=0.01$, for a product of signal
cross section and branching fraction of 600\unit{fb}.
The light blue shaded area corresponds to the statistical uncertainties in the
corresponding \TwoMOneL region. The last bin in each distribution also contains events with masses exceeding 1.3\TeV.}
    \label{fig:fh-FinalCR}

\end{figure}

The five-jet invariant mass distributions for
the QCD multijet \ThreeT region, the \ttbar \TwoTOneL region, and
the \ThreeM signal region are presented in Fig.~\ref{fig:fh-FinalCR} together
with a potential signal for ${\mtprime}=0.7$\TeV
corresponding to a product of
cross section times branching fraction of 600\unit{fb}.
The \TwoMOneL region distribution is overlaid after applying the {\PQb} tagging weight
computed with respect to the \ThreeT, \TwoTOneL, and \ThreeM regions, respectively.
An acceptable agreement is observed in each sample.

\subsection{Low-mass search results}
\begin{figure}[!hbt]
  \centering
   \includegraphics[width=0.48\textwidth]{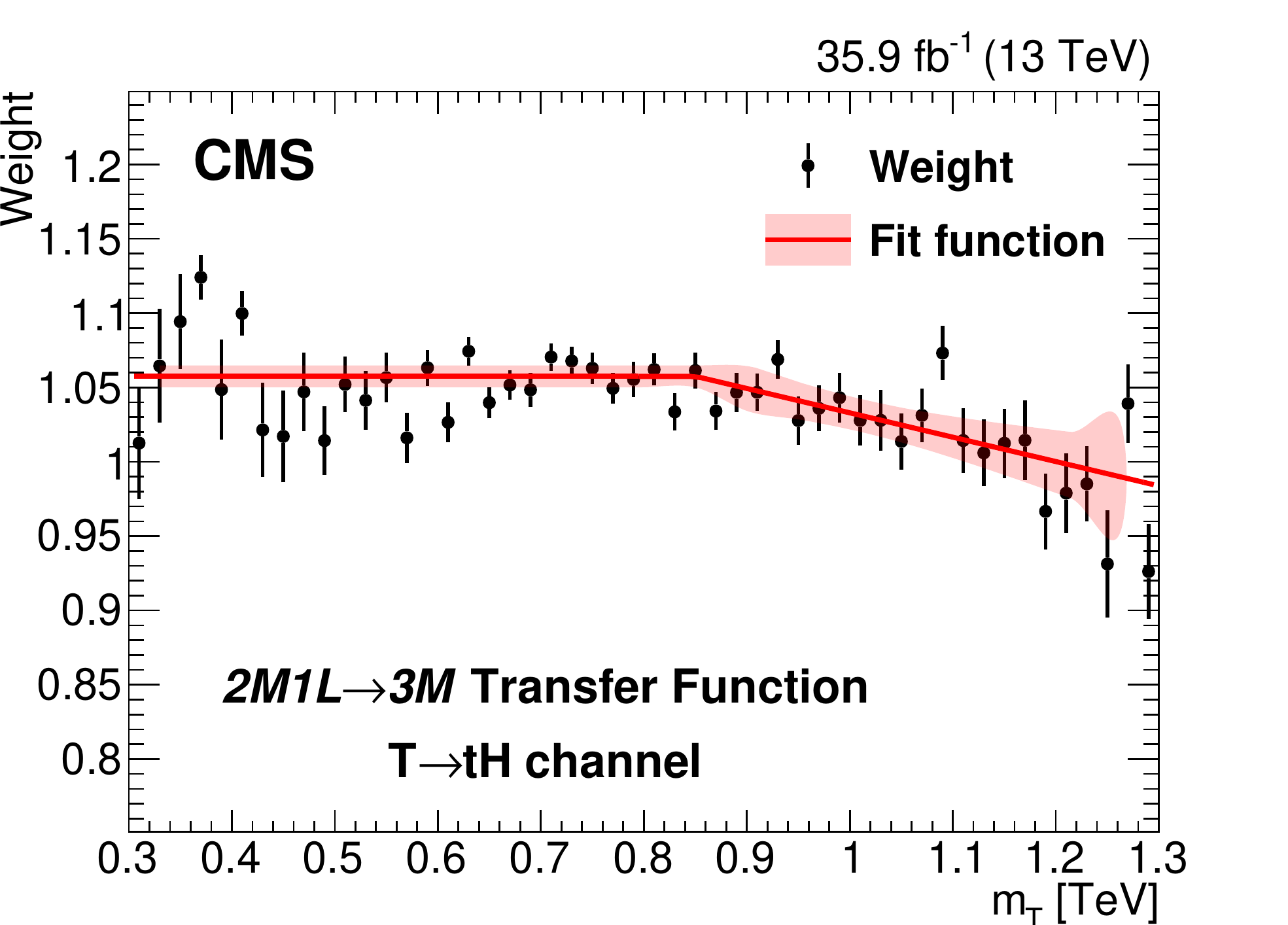}
   \includegraphics[width=0.48\textwidth]{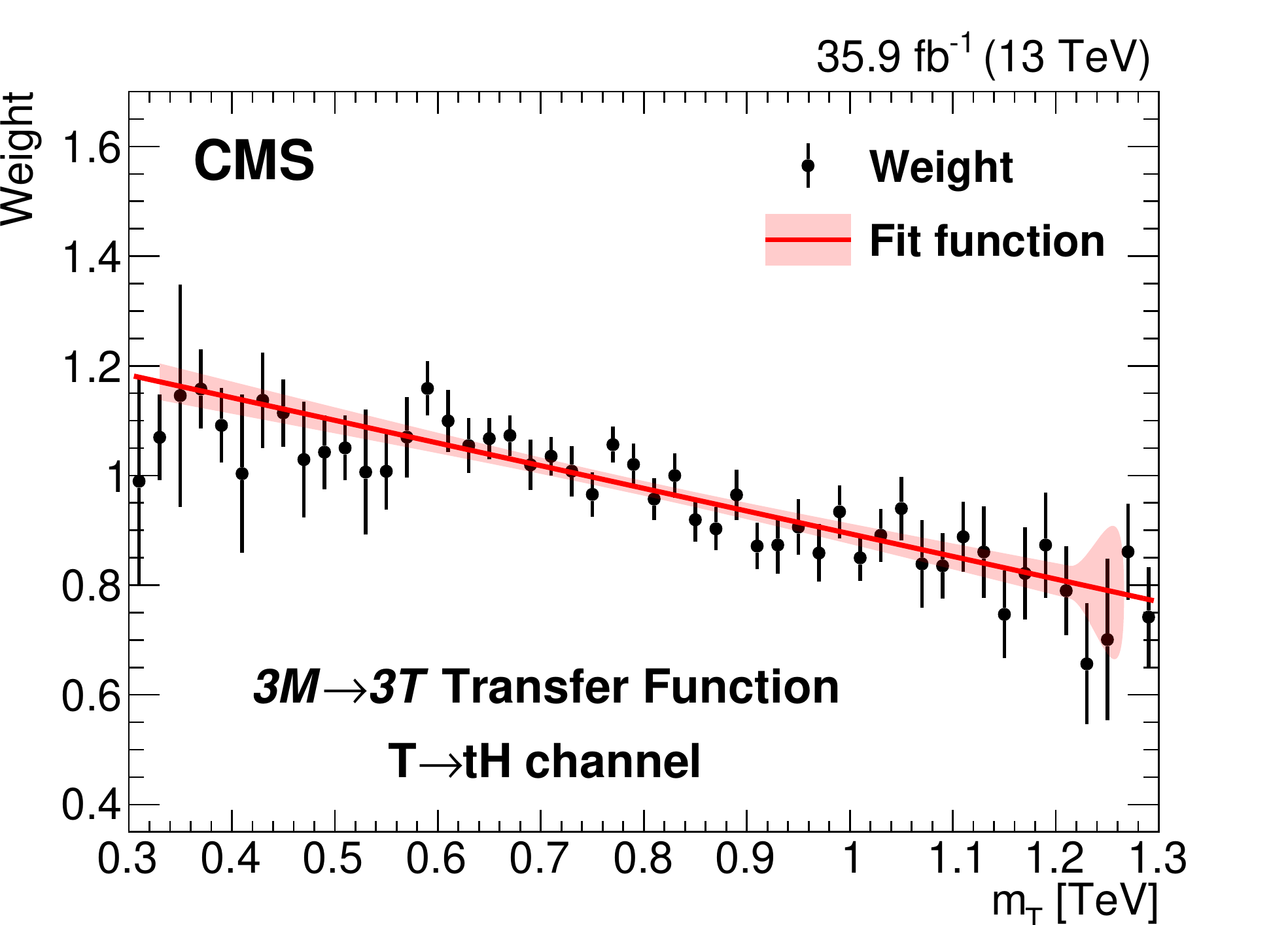}
   \includegraphics[width=0.48\textwidth]{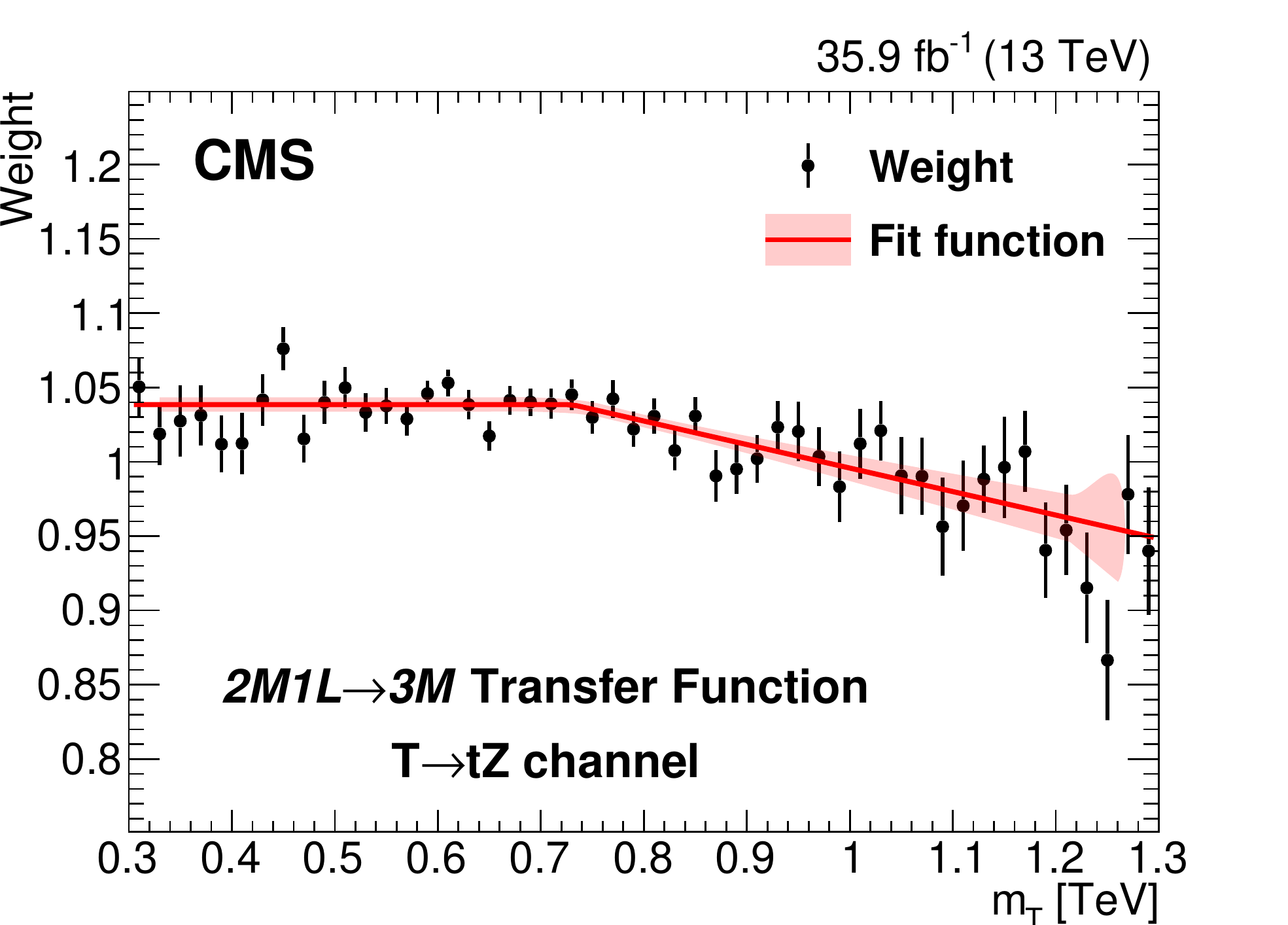}
   \includegraphics[width=0.48\textwidth]{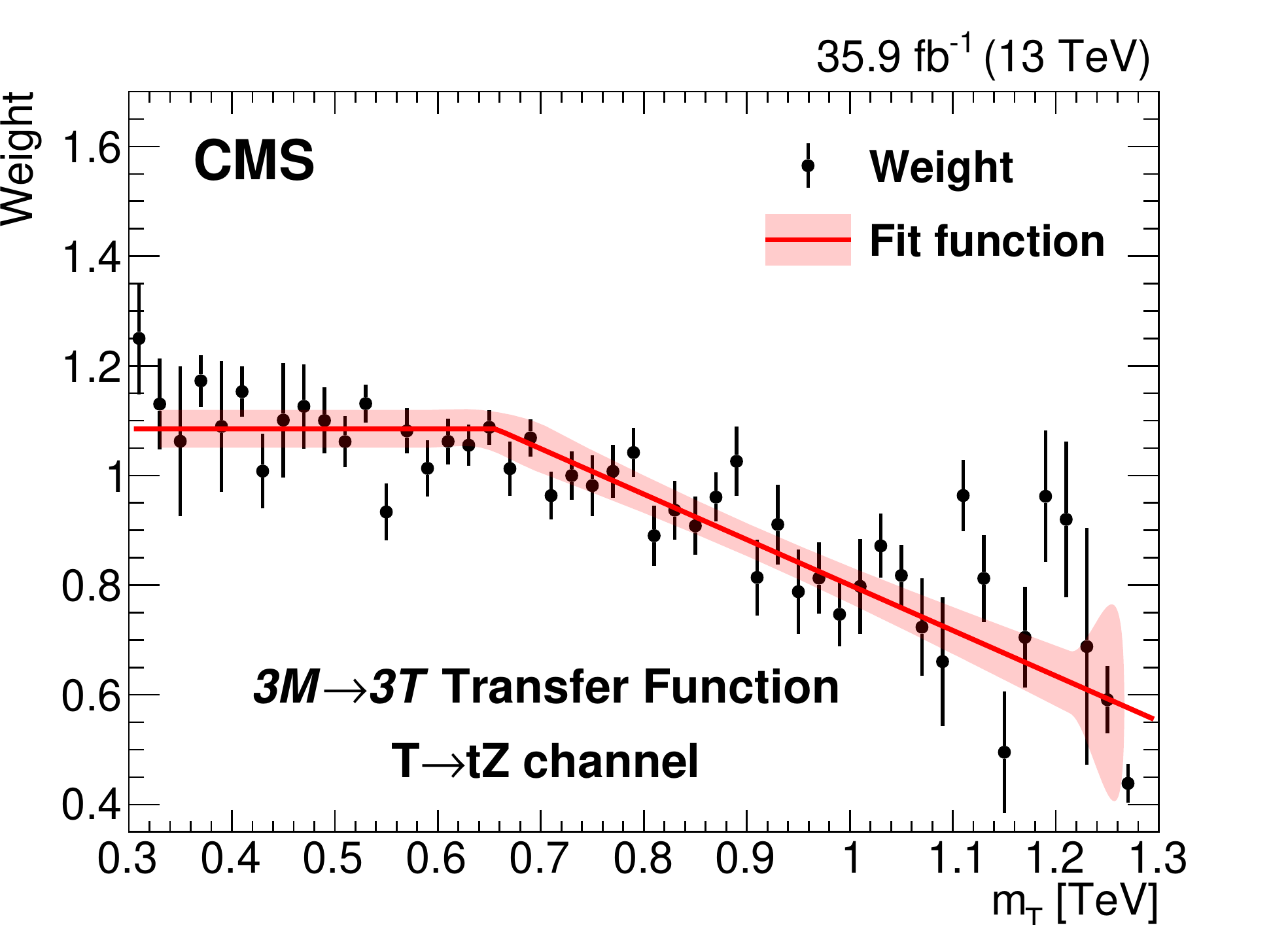}
    \caption{Dependence of the reweighting values (product of all {\PQb} tagging weights) on
the five-jet invariant mass for the \TwoMOneL region (left) and for
the \ThreeM region (right), in the case of the \tH channel (upper) and the \tZ channel (lower).
These variations are fitted to obtain the transfer functions (in red)
using either a 3-parameter function with a constant term and a slope, or a
2-parameter straight line. The light red shaded regions represent
the central 68\%~\CL interval for each fit when taking into
account only the statistical uncertainties.
}
    \label{fig:fh-TF}
\end{figure}

For each decay channel, three independent regions based on the {\PQb}-tagged jet
requirements are examined: \ThreeT (largest signal over background ratio),
\ThreeM, and \TwoMOneL (background dominated).
The overall background shape and normalization is driven by the
observations in the \TwoMOneL region.
The background shape is linked between the regions by two transfer
functions;
these are derived from the {\PQb} tagging weights to correct for {\PQb} tagging
differences between the regions.
One transfer function links the \ThreeT region to the \ThreeM region and the other links the \ThreeM region to the \TwoMOneL region.
The transfer functions, based on simple parametrizations of the dependence of the
reweighting values on the five-jet invariant mass, are displayed in Fig.~\ref{fig:fh-TF}.

The signal is parametrized as a Gaussian shape following the fit of the \PQT
quark reconstructed mass for each of the simulation
samples for each region. The variations
of the Gaussian fit parameters (mean and standard deviation)
with \PQT quark mass
are fitted for each region.
The parametrizations for the \tH and \tZ channels are found to be compatible.
The systematic uncertainties are discussed in Section~{\ref{sec:systematics}}.
Here we note simply that they are all taken as correlated
between the channels, except the ones related to the transfer
functions and the normalization between regions.
For a given channel, the fit procedure adjusts
the shape of the background bin by bin based on the data
in each of the regions, taking into account the transfer function between regions.
The overall fit uses 40\GeV wide bins; it includes three bin-independent fit parameters, namely the
signal strength and two relative normalization factors between each region, and fit parameters for the
background contribution in each bin of the \TwoMOneL region.
\begin{figure}[!hbtp]
  \centering
   \includegraphics[width=0.49\textwidth]{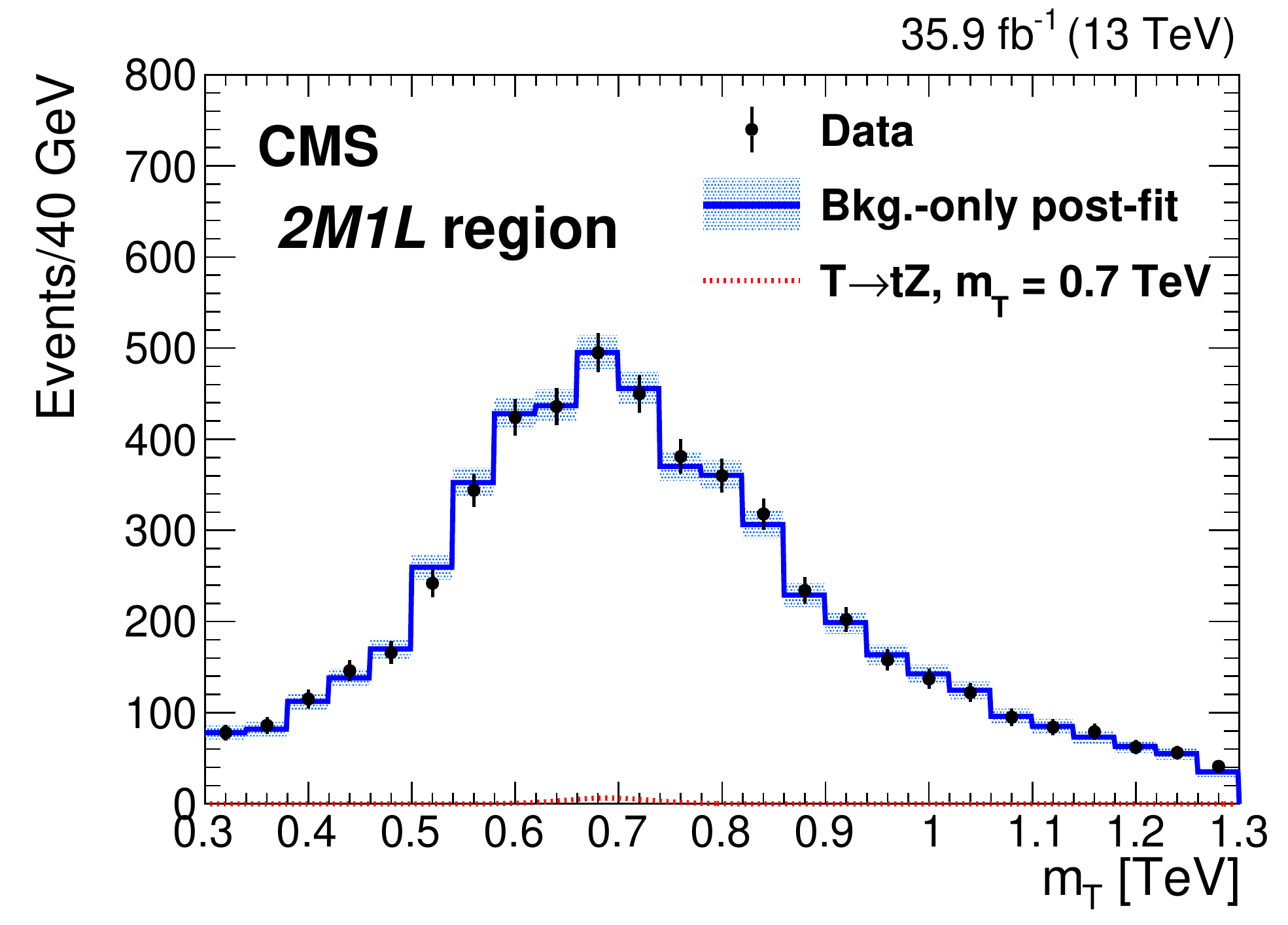}
   \includegraphics[width=0.49\textwidth]{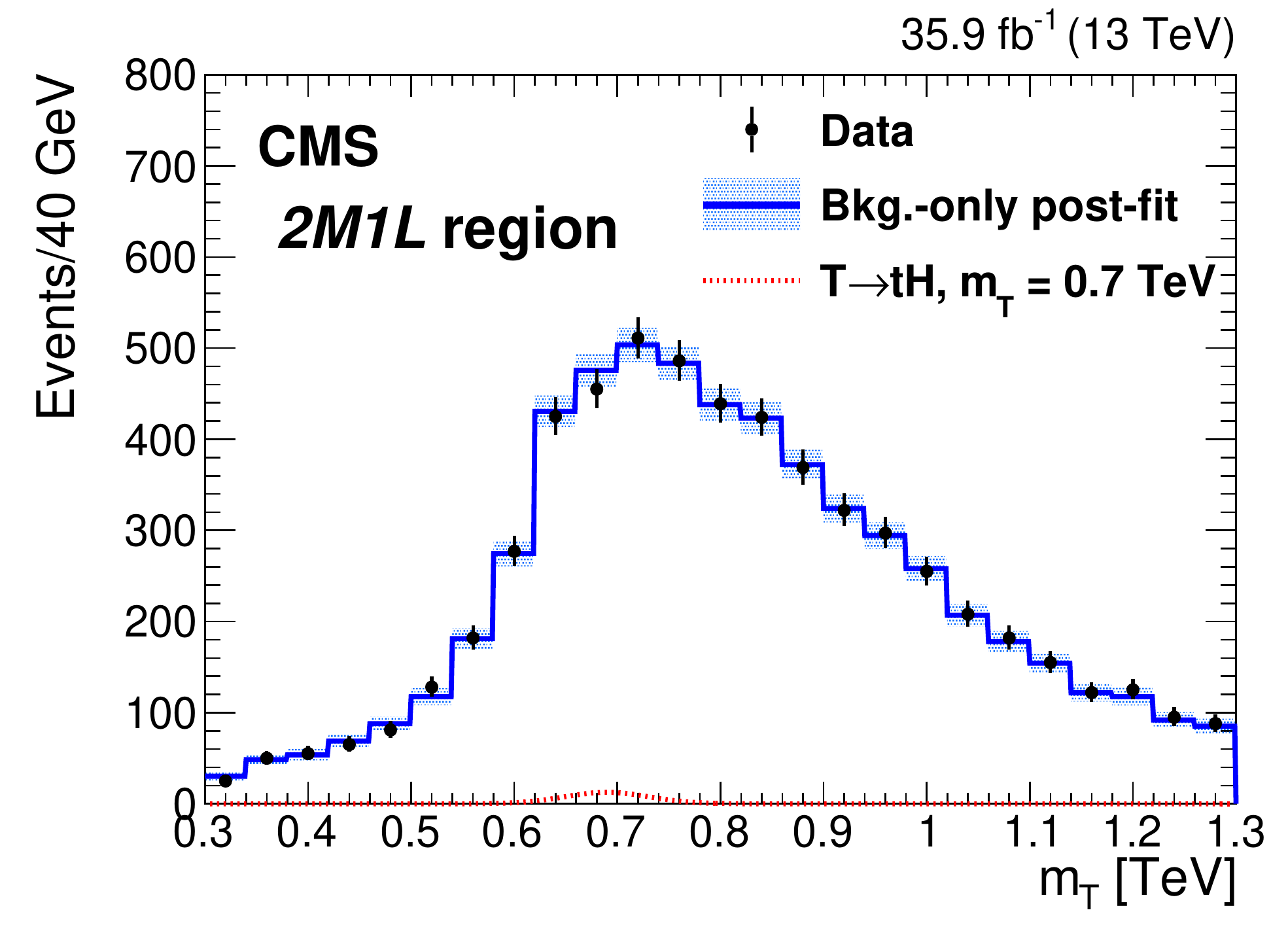}
   \includegraphics[width=0.49\textwidth]{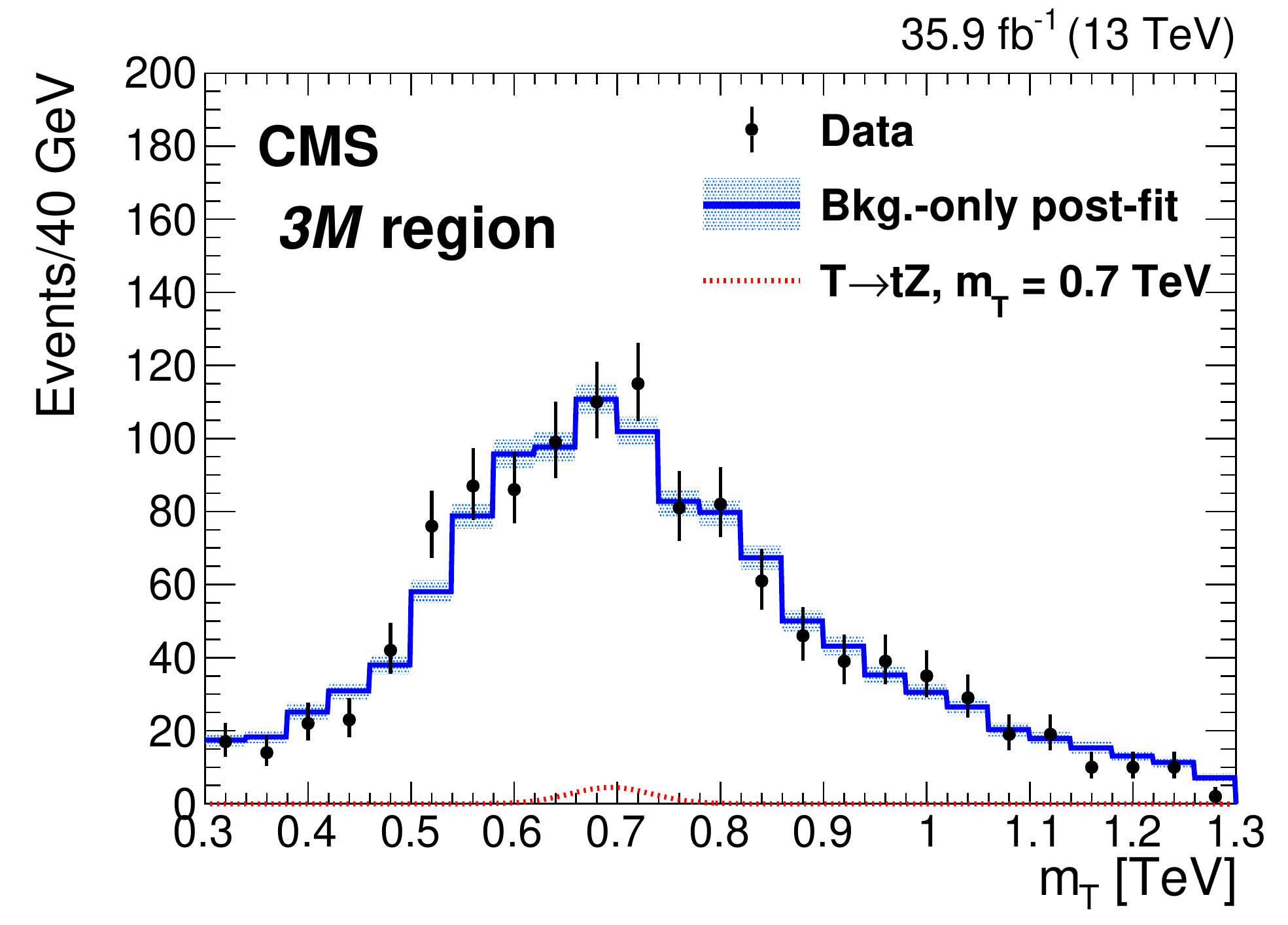}
   \includegraphics[width=0.49\textwidth]{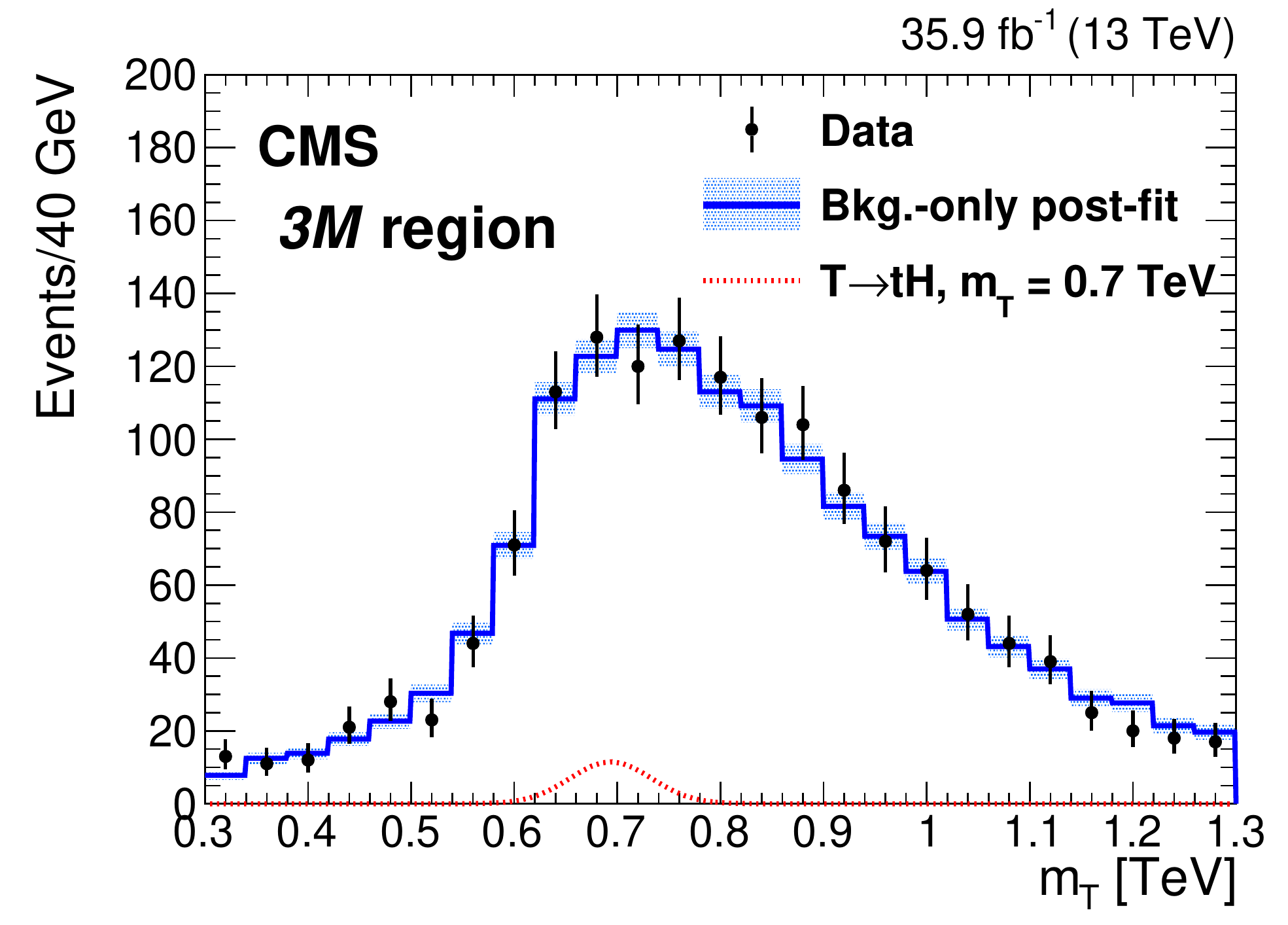}
   \includegraphics[width=0.49\textwidth]{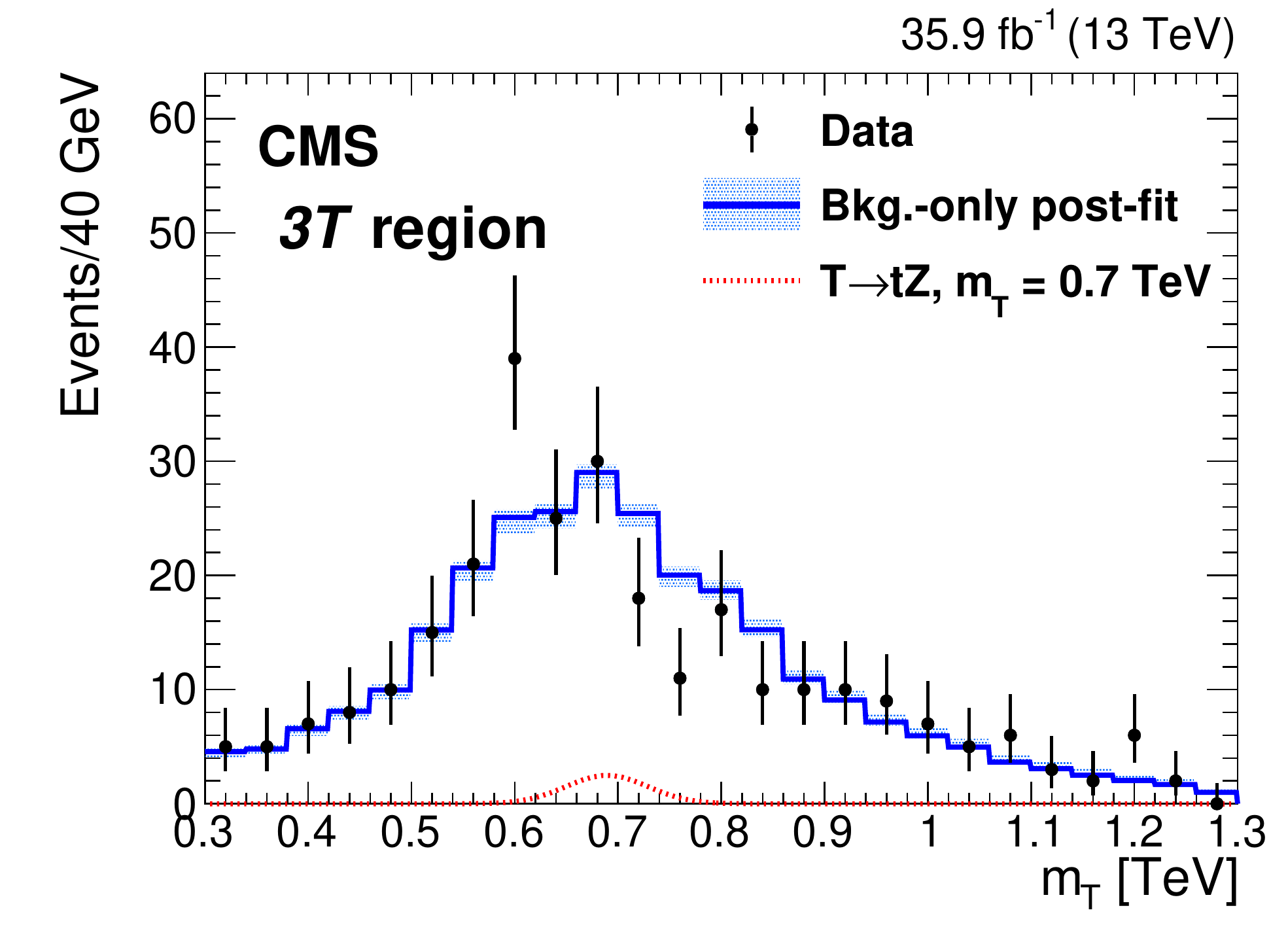}
   \includegraphics[width=0.49\textwidth]{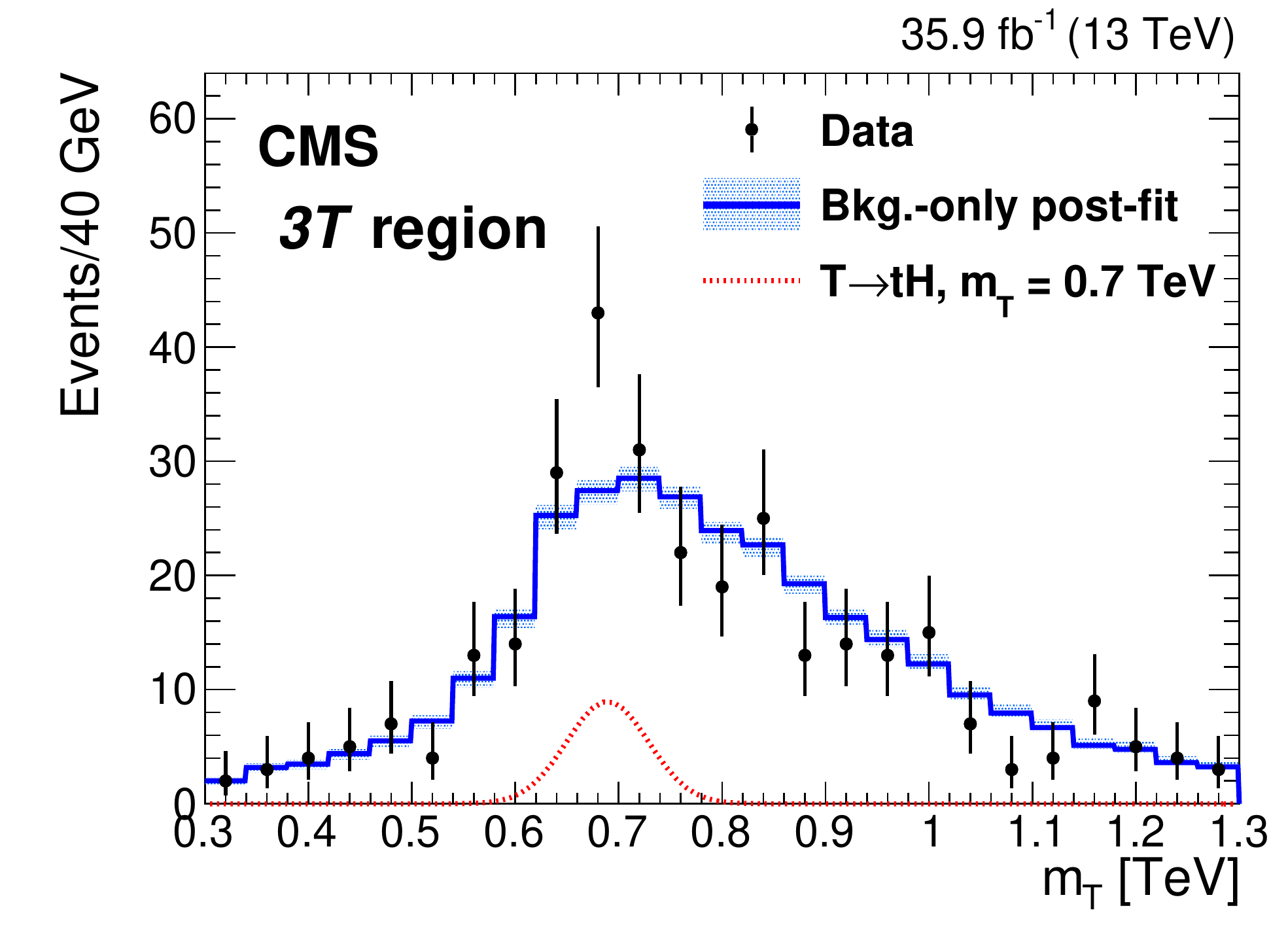}
    \caption{The background-only post-fit invariant mass distributions for the
\tZ candidates (left) and \tH candidates (right) for each region fitted: \TwoMOneL (upper  row),
\ThreeM (middle row), and \ThreeT (lower row). The signal hypothesis shown is
a \PQT with a mass of 0.7\TeV, narrow width,
and a product of the cross section and branching fraction of 600\unit{fb} for the \tZbq and \tHbq channels.
The data are represented by the black points with error bars, the signal hypothesis is represented by the red dashed line, the blue histogram gives the fitted background, and the light blue band represents the uncertainty in the
background fit.
}
    \label{fig:fh-PostFit}

\end{figure}

The background-only post-fit invariant mass distributions for each of the
regions (\TwoMOneL, \ThreeM, and \ThreeT) as well as for each channel (\tZ and \tH) are
displayed in Fig.~\ref{fig:fh-PostFit}.
A signal with a mass of ${\mtprime}=0.7$\TeV and product of the cross section and branching fraction
of 600\unit{fb} is superimposed.
An excess is observed when fitting the three regions for the \tH channel.
The local
significance is 3.0 standard deviations for a \PQT quark mass of 0.68\TeV.
For the same \PQT quark mass the local significance is 0.2 standard
deviations in the \tZ channel.
In a search for a vector-like quark, one expects similar branching
fractions for the \tH and \tZ channels.
No overall excess is measured when considering the fit of all six distributions.

The result of the median value for the limit in terms of the
appropriate cross section
is calculated
using the asymptotic \CLs framework~\cite{CLS2,CLS1,AsympCLs}.

Similarly, results are obtained in the cases where the \PQT quark
has 10, 20, and 30\% widths and in
the case of production in association with a top quark.
In all cases, the observed results show no evidence for a signal.
The studies have also been performed for the case of right-handed chirality.
No effect of the chirality is observed, indicating that the
low-mass search is insensitive to this property.
The resulting limits are reported in Section~{\ref{sec:combination}}.
\section{High-mass search}\label{sec:fh-boosted}
This search strategy focuses on reconstructing the invariant mass of
\tprimetotH and \tprimetotZ candidates formed
from two large-area jets in the fully hadronic channel.
The large-area jets are associated with
events in which the top quark and the Higgs or $\PZ$ boson are each
highly Lorentz-boosted, and correspondingly the search
targets \PQT masses of 1\TeV and above.
The background consists mostly of top quark pair production and QCD multijet production.

\subsection{Particle tagging}\label{ss:fh-particletagging}
To identify boosted $\tbw\to\PQb\qqbar'$, $\Hbb$, and $\Zbb$ decays,
jet substructure  techniques~\cite{Salam:2009jx} are used, which remove soft and collinear radiation from the clustered jet constituents. Clusters of the remaining constituents are identified with each of the quarks from the decay of the original particle.
The soft-drop algorithm~\cite{Dasgupta:2013ihk,softdrop} is used to groom the jets, using the soft radiation fraction parameter $z=0.1$ and the angular exponent parameter $\beta=0$. The algorithm yields two soft-drop subjets. The jet mass after applying the soft-drop algorithm will be referred to as the soft-drop mass.
The pruning grooming algorithm~\cite{Ellis:2009me} is also used, leading to the pruned jet mass.
For pruning, the minimum subjet \pt as a fraction of the parent jet \pt is required to exceed 0.1 and
the separation angle in $\eta$-$\phi$ space between the two subjets must exceed 0.5.
Furthermore, the \textit{N}-subjettiness algorithm~\cite{Thaler:2010tr} is used to further select jets with three or two substructures for the top quark jets, and the $\PH$ and $\PZ$ boson jets, respectively.
Flavor tagging is applied to identify $\PQb$ quark subjets using the CSVv2 multivariate discriminator,
in order to further enhance the signal purity and suppress backgrounds from non-\ttjets multijet processes.
The particle tagging criteria for boosted top quark, and $\PH$ and $\PZ$ boson
jets are as follows:
\begin{itemize}
\item{\textit{$\PH$ jet:}} An AK8 jet with $\pt>300\GeV$ must have a pruned jet mass within the range 105--135\GeV.
The ratio of the \textit{N}-subjettiness variables $\tau_2/\tau_1$ of
the jet is required to be $<$0.6. At least one of the two soft-drop subjets must pass the medium $\PQb$ tag criterion and the other subjet should pass at
least the loose $\PQb$ tag criterion.
\item{\textit{$\PZ$ jet:}} An AK8 jet with $\pt>200\GeV$ must have a pruned jet mass within the range 65--105\GeV.
  The requirements on the \textit{N}-subjettiness ratio $\tau_2/\tau_1$ and subjet $\PQb$ tagging are the same as those for the $\PH$ jet.
\item{\textit{$\PQt$ jet:}} An AK8 jet with $\pt>400\GeV$ and a soft-drop mass within the range 105--220\GeV is required.
The jet should have an \textit{N}-subjettiness ratio $\tau_3/\tau_2< 0.57$,
indicating that the large-area jet is likely to have three subjets. The soft-drop subjet with the highest CSVv2 discriminator value should pass the medium $\PQb$ tag criterion~\cite{CMS-PAS-JME-15-002}.
\end{itemize}

In addition, reversed-$\PH$-tagged, reversed-$\PZ$-tagged,
and reversed-$\PQt$-tagged jets are defined, with the same kinematic and
\textit{N}-subjettiness requirements as those for their tagged counterparts, but with complementary $\PQb$ tagging criteria, as follows:

\begin{itemize}
\item{\textit{Reversed-$\PH$-tagged jet:}} Same criteria as for an $\PH$-tagged jet but with one subjet
passing the medium $\PQb$ tag criterion and the other subjet failing the loose \PQb tag criterion.
\item{\textit{Reversed-$\PZ$-tagged jet:}} Same criteria as for a $\PZ$-tagged jet but with both subjets failing the loose $\PQb$ tag criterion.
\item{\textit{Reversed-$\PQt$-tagged jet:}} Same criteria as for a $\PQt$-tagged jet but with the highest soft-drop subjet \PQb discriminant failing the medium $\PQb$ tag criterion.
\end{itemize}

The reversed-$\PZ$ tag is defined differently from the reversed-$\PH$ tag so
that sensitivity to a potential \tZ signal, including
efficiency from $\Zqq$ decays, can be retained.

\subsection{Event selection}\label{sec:fh-boosted-evsel}

Only events satisfying the following primary selection
criteria are considered further, either as candidates for the signal
or for the
associated background control regions:

\begin{itemize}
  \item{\textit{Small-area jet multiplicity:}} At least four AK4 jets with $\pt > 30\GeV$ and $\abs{\eta} < 5$.
  \item{\textit{Large-area jet multiplicity:}} At least two AK8 jets with $\pt > 200\GeV$ and $\abs{\eta} < 2.4$.
  \item{\textit{Leading jet:}} The highest-\pt AK8 jet should have $\pt > 400\GeV$ and a pruned mass greater than 50\GeV.
  \item{\textit{Scalar \pt sum:}} The scalar sum of the transverse
momenta of the two highest-\pt AK8 jets should exceed 850\GeV.
  \item{\textit{Extra jet multiplicity:}} At least two of the AK4 jets should be separated by a distance $\Delta R > 1.2$ from the two leading AK8 jets.
  \item{\textit{Forward extra jet:}} At least one of the extra AK4 jets defined above should have $\abs{\eta} > 2.4$.
\end{itemize}

The last two criteria are imposed in order to ensure evidence that the
selected events contain a diquark ($\PQb\PQq$ or $\PQt\PQq$) system
that is produced in association with the \PQT quark, where the
quark ($\PQq$) associated with the vector boson tends to be forward.
Events passing the selection requirements above
are by design expected to be almost fully efficient for the trigger requirements.

Signal candidates passing the primary event selection are
further categorized according to the following criteria.
At this stage, the focus is on choosing the best pair of AK8 jets for
constructing the \PQT candidate mass.

\begin{itemize}
   \item{\textit{Double-tag:}} Each of the two highest-\pt selected AK8 jets must have either a $\PQt$ tag or an \HZ tag.
                      Furthermore, one of these jets must have a $\PQt$ tag and the other an \HZ tag. In the ambiguous case,
                      where both jets are $\PQt$-tagged and \HZ-tagged,
                      the higher-\pt jet is assigned as the top quark candidate, and the lower-\pt jet as the Higgs/$\PZ$ boson candidate.
   \item{\textit{\HZ tag isolation:}} Motivated by reducing \ttbar
background, events are rejected if any AK4 jet is separated from
the $\HZ$ candidate jet by $0.55 < \Delta R (j,\HZ) < 0.9$.
\end{itemize}

Signal regions, \rH and \rZ, are defined using these criteria for
the \tH and \tZ searches respectively. Related background control
regions are defined and are described in
Section~\ref{sec:fh-boosted-bkgd}.

We denote the measured AK8 jet four-vectors corresponding
to the top quark, Higgs boson and \PZ candidates
as $P_{\PQt} = (E_{\PQt}, \vec{p}_{\PQt})$,
$P_{\PH} = (E_{\PH}, \vec{p}_{\PH})$ and
$P_{\PZ} = (E_{\PZ}, \vec{p}_{\PZ})$ and construct a corrected
\PQT mass-sensitive observable, $\Mtilde$. This observable
takes advantage of the knowledge of the top quark and Higgs/\PZ boson
masses to correct the masses of the AK8 jets; it is inspired by
a similar variable used in Ref.~\cite{Sirunyan:2018iwt} that
was based on a suggestion in Ref.~\cite{Kumar:2014bca}.
The reconstructed mass of the \PQT candidate
from the \tX dijet system ($m_{\tX}^{j}$), with $\mathrm{X}=\HZ$,
is adjusted for deviations of the reconstructed top
quark and Higgs/$\PZ$ boson AK8 masses ($m_{\PQt}^{j}$, $m_{\mathrm{X}}^{j}$)
from the known $\PQt$ and \HZ masses~\cite{PDG2018},
as follows:
\begin{linenomath}
\begin{equation}
\Mtilde =
     \sqrt{(P_{\PQt}+P_{\mathrm{X}})^{2}} -
  \sqrt{P_{\PQt}^{2}} - \sqrt{P_{\mathrm{X}}^{2}} + m_{\PQt} + m_{\mathrm{X}}
     = m_{\tX}^{j} - (m_{\PQt}^{j} - m_{\PQt}) - (m_{\mathrm{X}}^{j} - m_{\mathrm{X}}) ,
\end{equation}
\end{linenomath}
where the $j$ superscripts denote jet-based measured mass quantities.
This estimator is found to have better performance
in terms of mass resolution by about 10\% compared to the uncorrected mass estimator.
It has also been verified that it is accompanied by a commensurate
reduction in background acceptance.

\begin{figure}[!htb]
  \centering
    \includegraphics[width=0.49\textwidth]{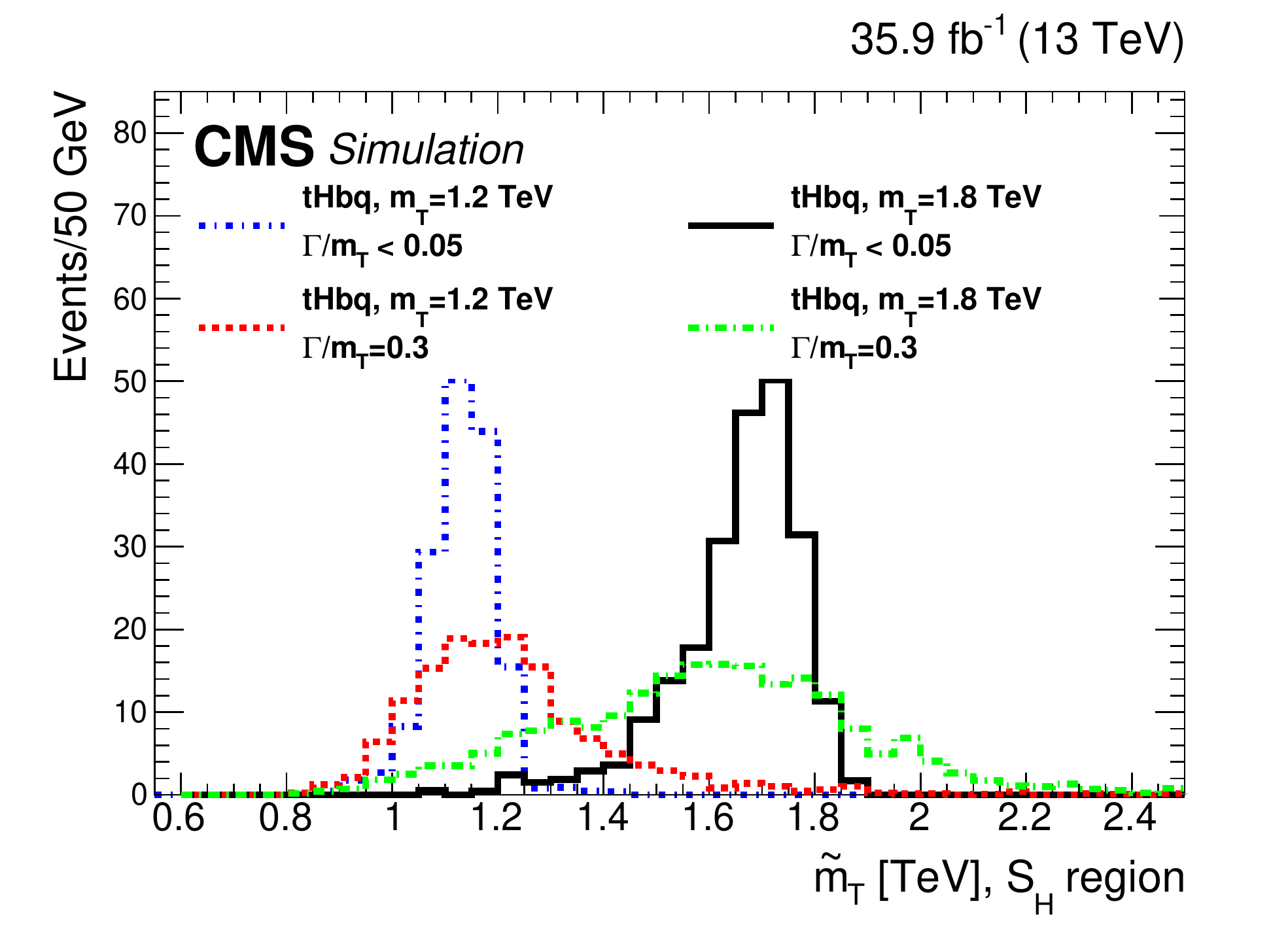}
    \includegraphics[width=0.49\textwidth]{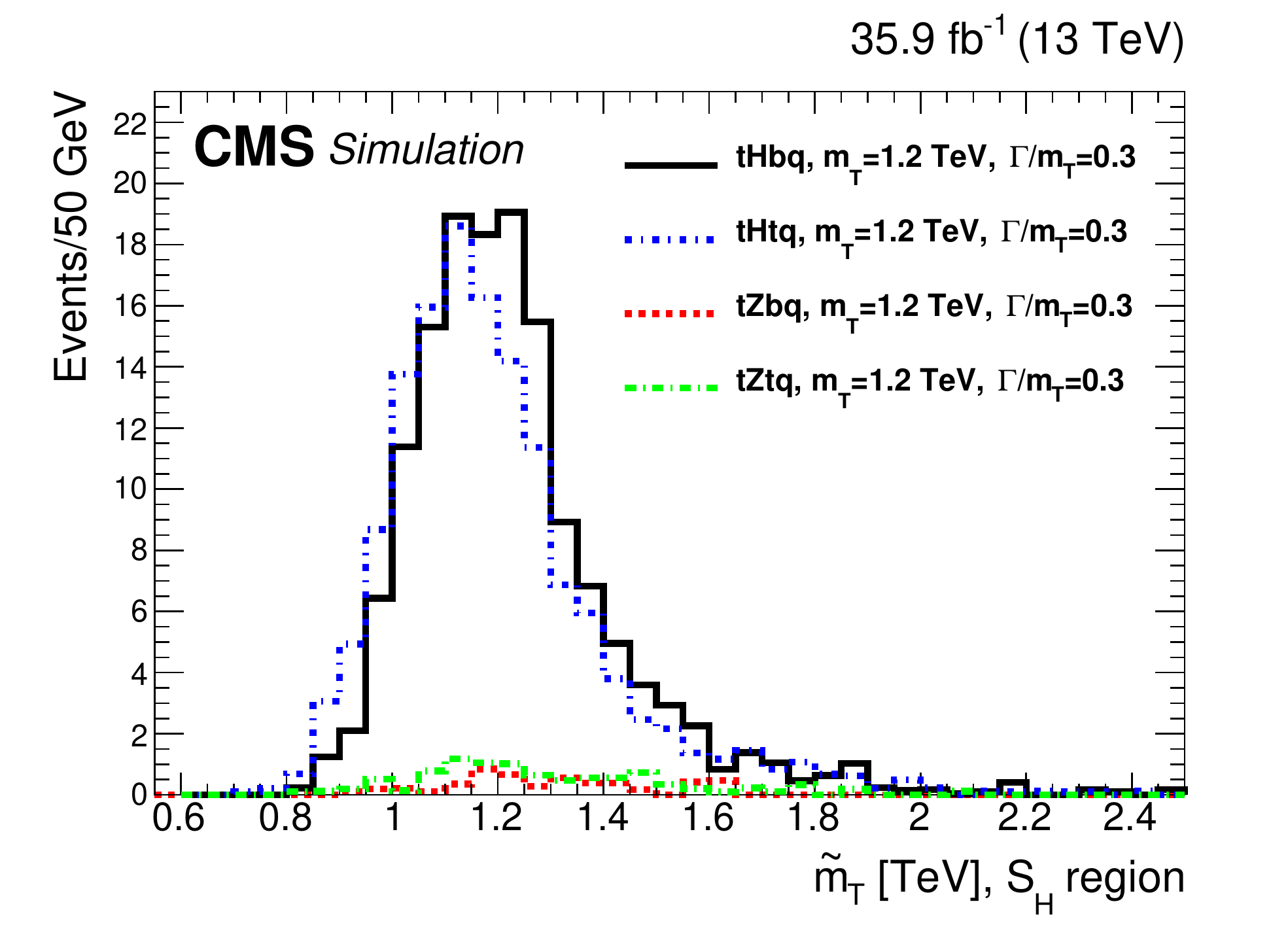}
    \includegraphics[width=0.49\textwidth]{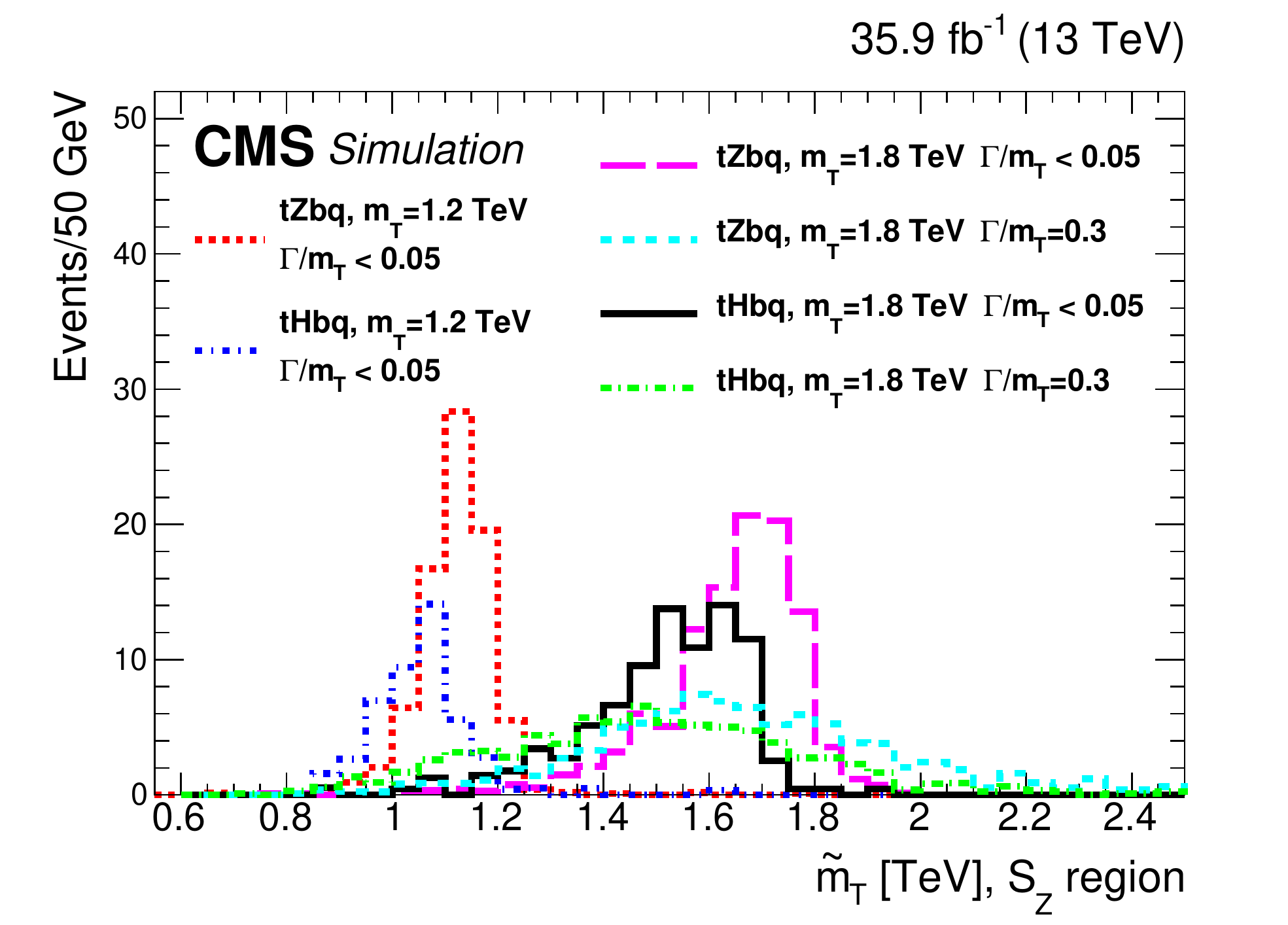}
    \includegraphics[width=0.49\textwidth]{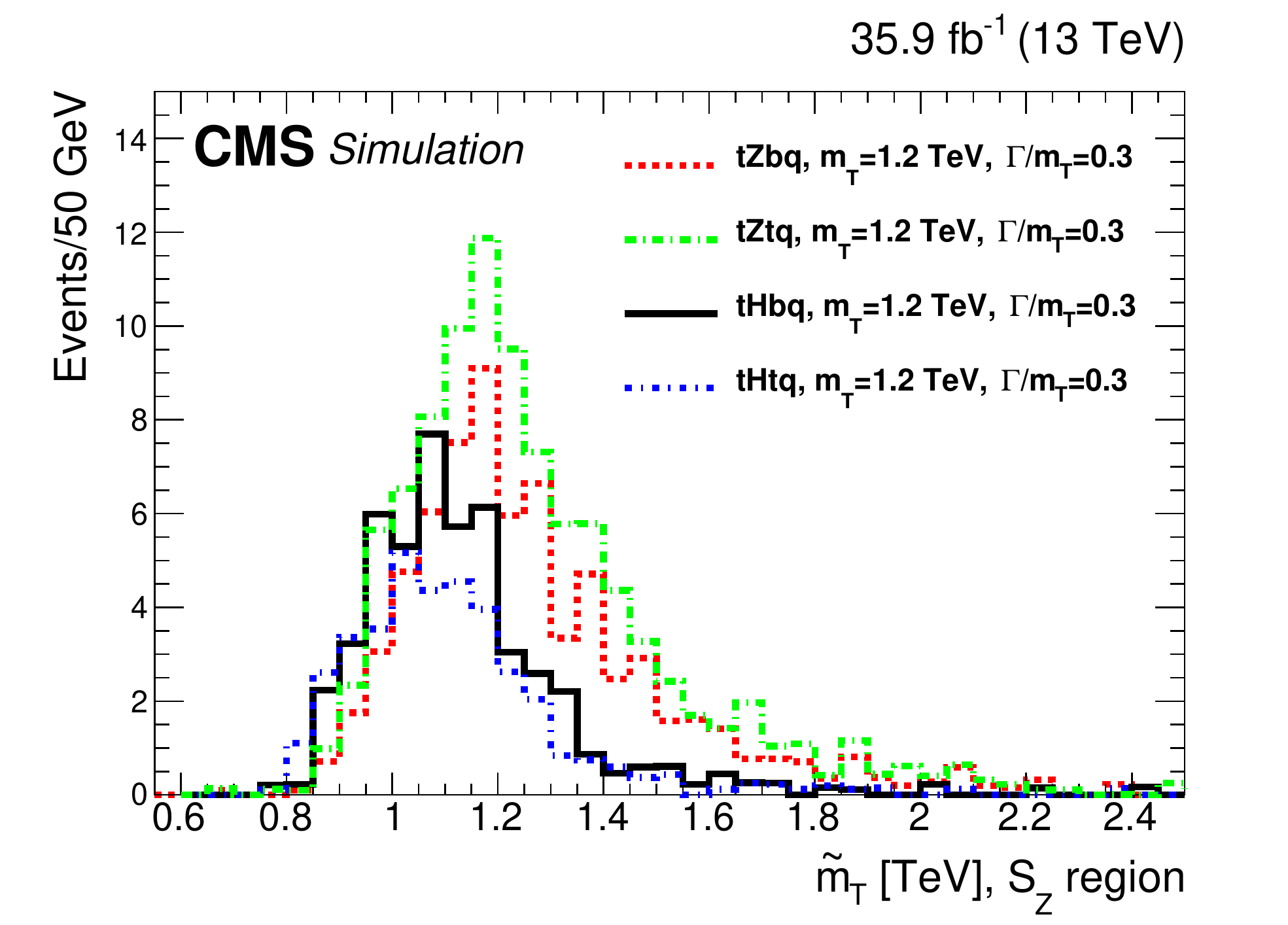}
    \caption{Example \Mtilde distributions in the signal regions, \rH (upper row), and \rZ (lower row).
    For presentation purposes, the cross sections for \tHbq, \tZbq, \tHtq and \tZtq are set equal to 1\unit{pb}
for all masses and fractional widths and normalized to the integrated luminosity of the data set.
The left column illustrates potential signals with a
range of masses and widths for the \tHbq and \tZbq channels.
The right column illustrates potential signals for
one mass and a large width for all four processes including also \tHtq and \tZtq.
}
    \label{fig:lineshapes}

\end{figure}

Example distributions of \Mtilde in the signal regions, \rH and \rZ, are shown
in Fig.~\ref{fig:lineshapes} for \PQT masses of 1.2 and 1.8\TeV for both narrow and large widths.
It can be seen that in the \rZ region, in addition to the expected
efficiency for the \tZbq process, there is also substantial efficiency
for the \tHbq process; the reverse is not true for the \rH region.

\subsection{Background estimation}\label{sec:fh-boosted-bkgd}

The \Mtilde distribution is
used to determine the amount of signal potentially present in the data.
A fit is performed that takes advantage of
the relatively narrow signal shape in \Mtilde compared to the broader
shape expected from the backgrounds.
After the primary event selection criteria, the
main backgrounds are \ttbar and QCD multijet events.
The \ttbar background is estimated using simulated events
and the QCD multijet background is estimated using control regions in data.
Other smaller background sources, designated as ``other'' and
consisting of \wjets, \zjets, single top quark,
\tHq, \ttH, $\PQt\PAQt\PW$,
$\PQt\PAQt\PZ$, $\PW\PW$, $\PZ\PZ$, $\PW\PZ$, $\PW\PH$, and $\PZ\PH$,
are all estimated from simulation.
Background templates are constructed using a smoothing procedure that
fits the \Mtilde distributions with an empirical functional form.
A simultaneous fit is then performed in eight regions: four regions designed
to test for tH signal contributions
and four regions designed to test for $\tZ$ signal contributions.
The fit examines all eight regions and fits the \Mtilde distributions for the
amounts of signal and QCD multijet background contributions
using \ttbar and other backgrounds predicted from simulation.
Fits are performed under three hypothetical signal
scenarios, $\tH$ only, $\tZ$ only, and $\tH$+$\tZ$.
In the latter case, the small difference in cross section
for \tHbq relative to \tZbq is taken from the singlet model
calculation.
For \tHtq relative to \tZtq, the difference from
the (\TB) doublet model calculation is used.

The criteria described in Section~\ref{sec:fh-boosted-evsel} define
the main signal regions (\rH and \rZ) using
the $\PQt$-, $\PH$-, and $\PZ$-tagged jets.
The additional six mutually exclusive regions are used
as control regions in the fit and to predict the shapes and normalization of
the QCD multijet background from data; these are denoted \rA, \rB, \rC for
the tH signal and \rW, \rX, \rY for the $\tZ$ signal.
Regions \rA and \rW are control regions for the QCD multijet background.
Region \rB is a $\ttbar$ enriched control region, while region \rX has
sensitivity to non-$\bbbar$ \PZ decays.
Regions \rC and \rY serve as control regions that test
the rejection of QCD multijet events
by the $\PH$ tag and $\PZ$ tag criteria.

For the definition of the regions, only the two highest $\pt$ AK8
jets are examined and each jet must be either tagged or reversed-tagged.
The signal region \rH requires a $\PQt$-tagged jet and an $\PH$-tagged jet.
Region \rA requires a reversed-$\PQt$-tagged jet and a reversed-$\PH$-tagged jet.
Region \rB requires a $\PQt$-tagged jet and a reversed-$\PH$-tagged jet.
Region \rC requires a reversed-$\PQt$-tagged jet and an $\PH$-tagged jet.
Regions \rC and \rH include the isolation requirement
on the $\PH$-tagged jet,
while regions \rA and \rB do not include an isolation requirement
around the reversed-$\PH$-tagged jet.
These choices define a \rB region that is enhanced in \ttbar events, thus
providing a suitable control region.
The order of assigning events starts with region \rH, then proceeds with region \rB, \rC, and then \rA,
where each subsequent region
is not allowed to contain any of the events assigned to the previous region.
In ambiguous cases, as is done for the signal region, the higher-$\pt$ jet is
assigned as the $\PQt$-tagged or reversed-$\PQt$-tagged jet.

There are four analogous regions designated by \rW, \rX, \rY, and \rZ for events
with $\PQt$- and $\PZ$-tagged jets for events that do not satisfy
the \rA, \rB, \rC, or \rH region definitions.
The \rW region has a reversed-$\PQt$-tagged jet and a reversed-$\PZ$-tagged jet.
The \rX region has a $\PQt$-tagged jet and a reversed-$\PZ$-tagged jet.
The \rY region has a reversed-$\PQt$-tagged jet and a $\PZ$-tagged jet.
These criteria make region \rX sensitive to hadronic decays of $\PZ$ bosons other
than $\bbbar$ from both
the $\tprimetotZ$ signal and the background.
Regions \rY and \rZ include the isolation requirement on
the $\PZ$-tagged jet, while regions \rW and \rX
do not include an isolation requirement around the reversed-\PZ-tagged jet.
Events can only be assigned to one of the eight regions. These criteria lead to a well-defined
\Mtilde value for each event corresponding to the mass assignments implicit in the tagging criteria.
Table~\ref{tab:control} summarizes the criteria for the eight regions.
\begin{table} [htb!]
  \centering
  \topcaption{Overview of the criteria used to define
the mutually exclusive \rA, \rB, \rC, \rH, \rW, \rX, \rY, and \rZ regions.
These are based on the particle tagging criteria for \PQt, \PH,
and \PZ jets and for the reversed-\PQt-tagged, reversed-\PH-tagged, and
reversed-\PZ-tagged jets using the two highest \pt AK8 jets.}
  \label{tab:control}
  \begin{tabular}{lcccc}
  Region & Channel & First jet & Second jet & \HZ tag isolation  \\
  \hline
  \rA & \tH & reversed-\PQt-tagged & reversed-\PH-tagged & \NA \\
  \rB & \tH &          \PQt tag & reversed-\PH-tagged & \NA \\
  \rC & \tH & reversed-\PQt-tagged &          \PH tag & required \\
  \rH & \tH &          \PQt tag &          \PH tag & required \\[\cmsTabSkip]
  \rW & \tZ & reversed-\PQt-tagged & reversed-\PZ-tagged & \NA \\
  \rX & \tZ &          \PQt tag & reversed-\PZ-tagged & \NA \\
  \rY & \tZ & reversed-\PQt-tagged &          \PZ tag & required \\
  \rZ & \tZ &          \PQt tag &          \PZ tag & required \\
  \end{tabular}
\end{table}

A simultaneous fit is performed to the \Mtilde distributions
in each of the eight regions to determine the amount of signal present.
The signal templates are taken directly from the simulated signal
samples. The \ttbar and other background contributions are found using
the smoothed templates. The smoothed QCD multijet background shape is
determined from the data in region \rA for the regions \rA, \rB, \rC,
and \rH and in region \rW for the \rW, \rX, \rY, and \rZ regions.
A binned likelihood fit is performed using the \Mtilde variable,
with 50 bins of 50\GeV width over the range 0.6--3.1\TeV.
All background components, except QCD multijet, are constrained within
uncertainties using predictions from MC simulations.
The numbers of QCD multijet events in regions \rH and \rZ are estimated
using the control regions.
The amount of QCD multijet background in each of the control
regions \rA, \rB, \rC, \rW, \rX, and \rY is found from the following.
If the expected numbers of QCD multijet events in the four regions are
$N_{\rA},N_{\rB},N_{\rC},N_{\rH}$ then, if the Higgs boson
and reversed-Higgs boson tagging are independent of the top quark
tagging and reversed-top quark tagging criteria, one may write:
\begin{linenomath}
\begin{equation}\label{eq:abcdformula1}
\frac{N_{\rH}}{N_{\rB}} = \frac{N_{\rC}}{N_{\rA}} \Rightarrow N_{\rH} = N_{\rB} \, \frac{N_{\rC}}{N_{\rA}} .
\end{equation}
\end{linenomath}
Using Eq.~(\ref{eq:abcdformula1}),
one may then make a data-based prediction of
the number of QCD multijet events in the signal region \rH. Similarly,
the number of QCD multijet background events in region \rZ can be
estimated from $N_{\rZ} = N_{\rX} \, (N_{\rY}/N_{\rW})$. In our
fitting method,
$N_{\rA}$, $N_{\rB}$, $N_{\rC}$, $N_{\rW}$, $N_{\rX}$,
and $N_{\rY}$ are free parameters determined by the fit. The fit assumes
that the double ratios $(N_{\rA}/N_{\rC})/(N_{\rB}/N_{\rH})$ and
$(N_{\rW}/N_{\rY})/(N_{\rX}/N_{\rZ})$ are consistent with unity
in order to constrain the number of QCD multijet events
in regions \rH and \rZ.

The double ratios measured from the QCD multijet simulation
are
$(N_{\rA}/N_{\rC})/(N_{\rB}/N_{\rH}) = 0.77\pm 0.39\pm 0.21$
and
$(N_{\rW}/N_{\rY})/(N_{\rX}/N_{\rZ}) =0.67\pm 0.35\pm 0.24$,
which are both consistent with the
expected value of unity within their combined statistical
and systematic uncertainties. The systematic uncertainties
are discussed in Section~\ref{sec:systematics}.
The predictions of the QCD multijet contributions, which include the
overall effect of the shape and the normalization, have been validated using a
fit that uses a different set of eight control regions in data.
These control regions are mutually exclusive to the previously defined eight regions,
contain larger numbers of events, and are defined using loose {\PQb} tagging criteria on
the \PQt tag and the reversed-\PQt tag in a sample that excludes events
with forward ($\abs{\eta} > 2.4$) jets.
The double ratio is taken to be fully correlated between the $\tH$ and $\tZ$ regions,
with a central value of 1.0, and an assigned
uncertainty of 0.6.
This uncertainty is assessed based on
the measured double ratios
from the relatively low number of events
in the QCD multijet simulation.
The fits to data are observed to be insensitive to the
exact uncertainty used in that the preferred value for the double ratio
tends to be close to unity.

Limits on the signal strength are extracted by fitting the signal and
backgrounds to the data. The fit finds the amount of signal as well as the
amounts of QCD multijet background in each of the eight regions.
The QCD multijet event yields in the regions \rA, \rB, and \rC are allowed
to float freely. The QCD multijet event yield in
the region \rH is constrained using Eq.~(\ref{eq:abcdformula1}) with
the double ratio being modeled with a Gaussian prior.
The same procedure is used for
regions \rW, \rX, \rY, and \rZ. All other uncertainties are treated
either using log-normal priors (for those that change the event yields only),
or as Gaussian priors (with shape variations corresponding to the $\pm 1$
standard deviation change in those uncertainties that affect the \Mtilde
distributions as well as the yields).
The fitting method is validated with a data sample based on simulation.
The fit uses the modified frequentist approach for confidence levels, taking
the profile likelihood ratio as the test statistic~\cite{CLS2,CLS1} and using
the asymptotic approximation for limit setting~\cite{AsympCLs}.

\subsection{High-mass search results}\label{ss:boosted_results}
Table~\ref{tab:dataqcd} gives the total number of events
in regions \rA, \rB, \rC, and \rH, while Table~\ref{tab:datawxyz} gives the total
number of events in regions \rW, \rX, \rY, and \rZ.
These tables also show the fitted contributions from each background source for the background-only hypothesis fit when fitting
the observed \Mtilde distributions in the eight regions.
Also included in the tables are the expected numbers of events and efficiencies for various signals.
The efficiencies are inclusive; they include all decay modes of the \PH, \PZ, and \PQt quark.
\begin{table}[htbp]
 \centering
  \topcaption{Post-fit numbers of events for the \rA, \rB, \rC, and \rH regions for the data and specified background sources,
for the overall eight-region background-only fit.
The uncertainties include both the statistical and systematic components.
The fitted background sums depend on the data.
The expected event yields for various signal samples are also listed with
statistical uncertainties only, along
with the corresponding masses (\TeVns) and
cross sections (fb).
The fractional width considered is 30\%.
The percent efficiency in region \rH is also
noted in parentheses, alongside the event yield.
}
  \label{tab:dataqcd}
  \begin{tabular}{lcccccc}
  Data set & ${\mtprime}$ & $\sigma$ & \rA & \rB & \rC & \rH  \\
  \hline
    \ttjets    &  \NA &  \NA   & $ 140 \pm 20  $ & $ 230 \pm 30  $ & $  16.8 \pm 3.8  $ & $  21.7 \pm 4.8  $ \\
    Other background & \NA & \NA   & $  21.7 \pm 9.6  $ & $  20.5 \pm 7.0  $ & $   7.4 \pm 4.4  $ & $   3.0 \pm 1.6  $ \\
    QCD multijet    & \NA &  \NA  & $478 \pm 42 $ & $  91 \pm 35  $ & $ 125 \pm 12  $ & $  28.4 \pm 9.1  $ \\
    Total background   & \NA &  \NA   & $ 640 \pm 28  $ & $ 342 \pm 23  $ & $ 149 \pm 12  $ & $  53.1 \pm 7.7  $ \\
    Data             & \NA  &  \NA   & 640     &   345  &   151 &   52     \\[\cmsTabSkip]
    \tHbq  & 1.2 & 142 & $6.80 \pm 0.30$       & $10.7 \pm 0.4$      & $11.6 \pm 0.4$       & $20.6 \pm 0.6$ (0.40) \\
    \tZbq  & 1.2 & 131 & $1.56 \pm 0.15$       & $1.18 \pm 0.13$       & $0.49 \pm 0.08$       & $0.73 \pm 0.11$ (0.02) \\
    \tHtq  & 1.2 & 40.7 & $2.32 \pm 0.10$        & $3.70 \pm 0.13$        & $3.21 \pm 0.12$       & $5.63 \pm 0.16$ (0.39) \\
    \tZtq  & 1.2 & 32.9 & $0.47 \pm 0.03$       & $0.49 \pm 0.03$       & $0.13 \pm 0.01$       & $0.14 \pm 0.02$ (0.01) \\[\cmsTabSkip]
    \tHbq  & 1.8 & 13.6 & $1.08 \pm 0.04$       & $1.54 \pm 0.05$       & $1.83 \pm 0.05$       & $2.83 \pm 0.07$ (0.58) \\
    \tZbq  & 1.8 & 11.0 & $0.12 \pm 0.01$       & $0.13 \pm 0.01$       & $0.08 \pm 0.01$       & $0.07 \pm 0.01$ (0.02) \\
    \tHtq  & 1.8 &  4.0 & $0.33 \pm 0.01$       & $0.49 \pm 0.01$       & $0.50 \pm 0.01$        & $0.76 \pm 0.02$ (0.53) \\
    \tZtq  & 1.8 &  3.2 & $0.11 \pm 0.01$       & $0.11 \pm 0.01$       & $0.03 \pm 0.01$        & $0.04 \pm 0.01$ (0.03) \\
  \end{tabular}
\end{table}
\begin{table}[htbp]
 \centering
  \topcaption{Post-fit numbers of events for the \rW, \rX, \rY, and \rZ regions for the data and specified background sources, for the overall eight-region
background-only fit.
The uncertainties include both the statistical and systematic components.
The fitted background sums depend on the data.
The expected event yields for various signal samples are also listed with statistical uncertainties only, along with
   the corresponding masses (\TeVns) and cross sections (fb).
The fractional width considered is 30\%.
The percent efficiency in region \rZ is also noted in parentheses, alongside the event yield.
}
  \label{tab:datawxyz}
  \begin{tabular}{lcccccc}
  Data set & ${\mtprime}$ & $\sigma$ & \rW & \rX & \rY & \rZ \\
  \hline
    \ttjets    & \NA &  \NA   & $ 258 \pm 32  $ & $ 421 \pm 53 $ & $  16.4 \pm 4.4  $ & $  30.2 \pm 5.8  $ \\
    Other background & \NA & \NA   &  $ 271 \pm 64 $ & $ 223 \pm 94 $ & $  12.1 \pm 4.0  $ & $   2.4 \pm 1.5  $ \\
    QCD multijet    & \NA &  \NA  & $5710 \pm 150 $ & $ 830 \pm 230 $ & $ 259 \pm 19  $ & $  45.0 \pm 9.7  $ \\
    Total background   & \NA & \NA   & $6230 \pm 120 $ & $1480 \pm 180 $ & $ 288 \pm 17  $ & $  77.5 \pm 9.7  $ \\
    Data             & \NA & \NA   & 6253 & 1475 & 286  & 80 \\[\cmsTabSkip]
    \tHbq  & 1.2 & 142    & $6.44 \pm 0.30$       & $10.1 \pm 0.4$        & $3.46 \pm 0.22$       & $6.97 \pm 0.33$ (0.14) \\
    \tZbq  & 1.2 & 131    & $27.3 \pm 0.6$        & $45.6 \pm 0.8$        & $6.01 \pm 0.29$       & $9.87 \pm 0.39$ (0.21) \\
    \tHtq  & 1.2 & 40.7   & $2.22 \pm 0.09$       & $3.42 \pm 0.12$       & $0.93 \pm 0.06$       & $1.55 \pm 0.08$ (0.11) \\
    \tZtq  & 1.2 & 32.9   & $4.10 \pm 0.08$       & $6.71 \pm 0.10$       & $0.83 \pm 0.04$       & $1.41 \pm 0.05$ (0.12) \\[\cmsTabSkip]
    \tHbq  & 1.8 & 13.6   & $1.12 \pm 0.04$       & $1.48 \pm 0.05$       & $0.66 \pm 0.03$       & $1.09 \pm 0.04$ (0.22) \\
    \tZbq  & 1.8 & 11.0   & $3.98 \pm 0.07$       & $5.64 \pm 0.09$       & $0.89 \pm 0.04$       & $1.21 \pm 0.04$ (0.31) \\
    \tHtq  & 1.8 &  4.0   & $0.28 \pm 0.01$       & $0.44 \pm 0.01$       & $0.15 \pm 0.01$       & $0.23 \pm 0.01$ (0.16) \\
    \tZtq  & 1.8 &  3.2   & $1.27 \pm 0.02$       & $1.83 \pm 0.03$       & $0.24 \pm 0.01$       & $0.37 \pm 0.01$ (0.32) \\
  \end{tabular}
\end{table}

\begin{figure}[!htb]
  \centering
    \includegraphics[width=0.497\textwidth]{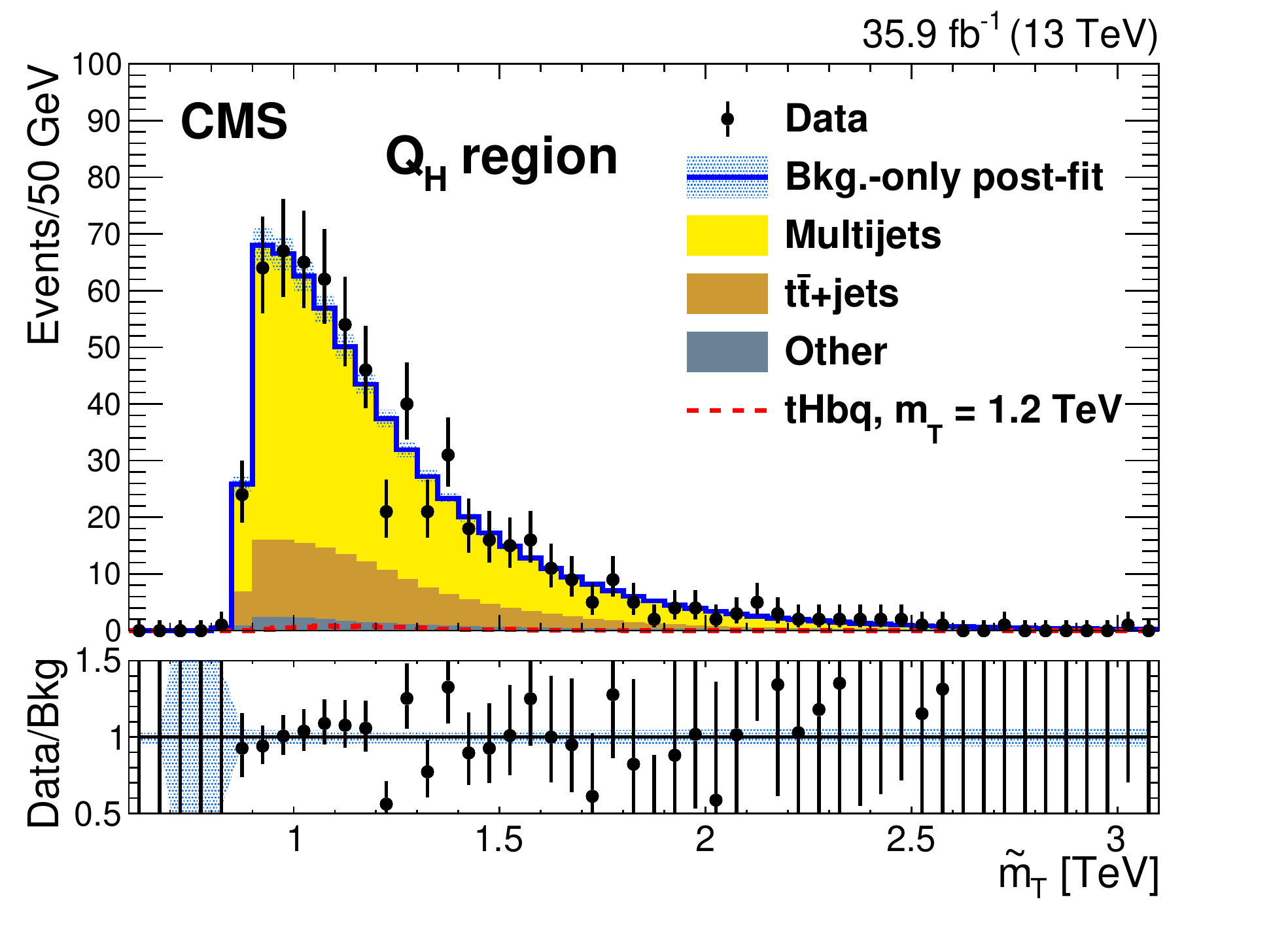}
    \includegraphics[width=0.497\textwidth]{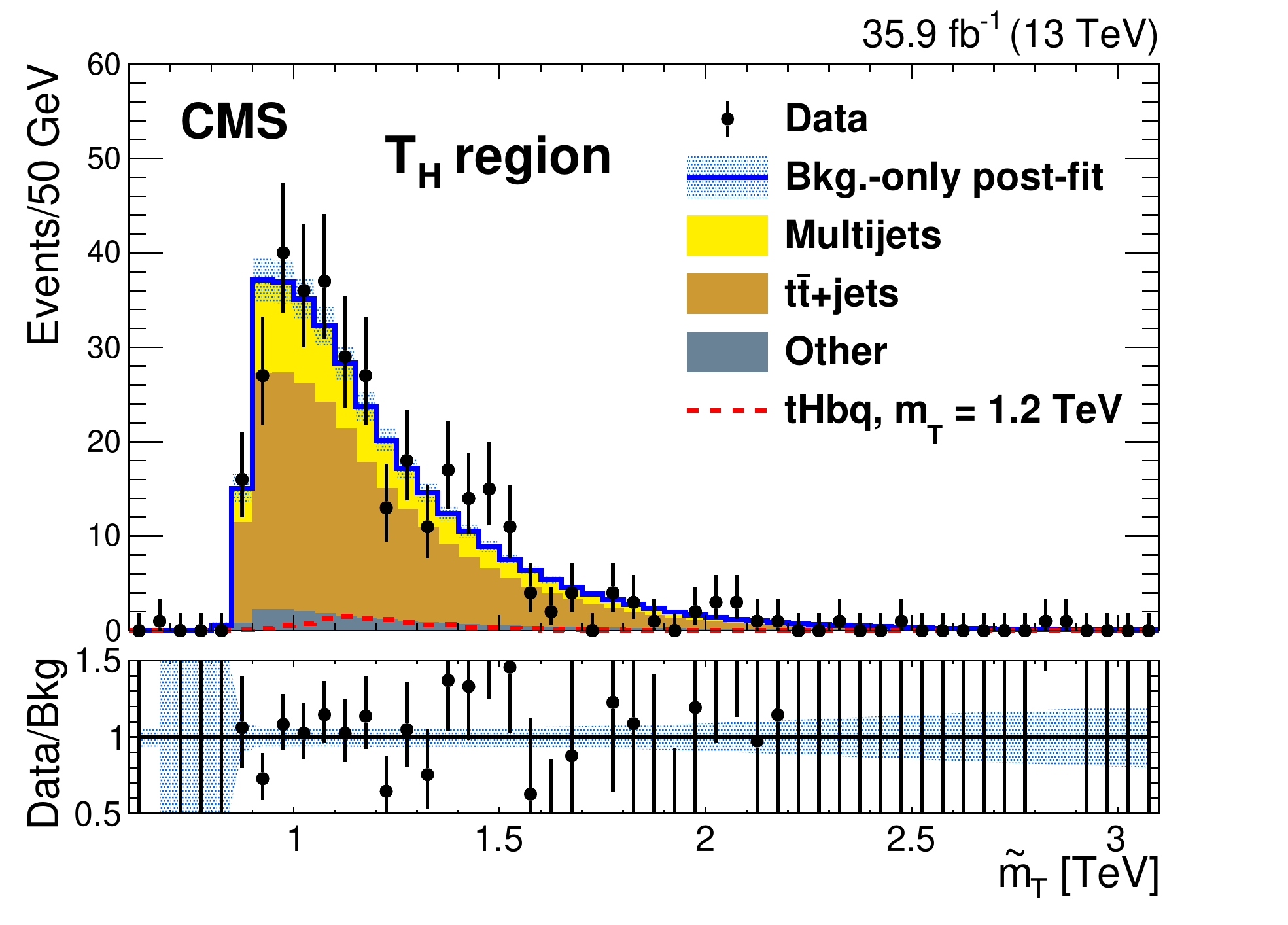}
    \includegraphics[width=0.497\textwidth]{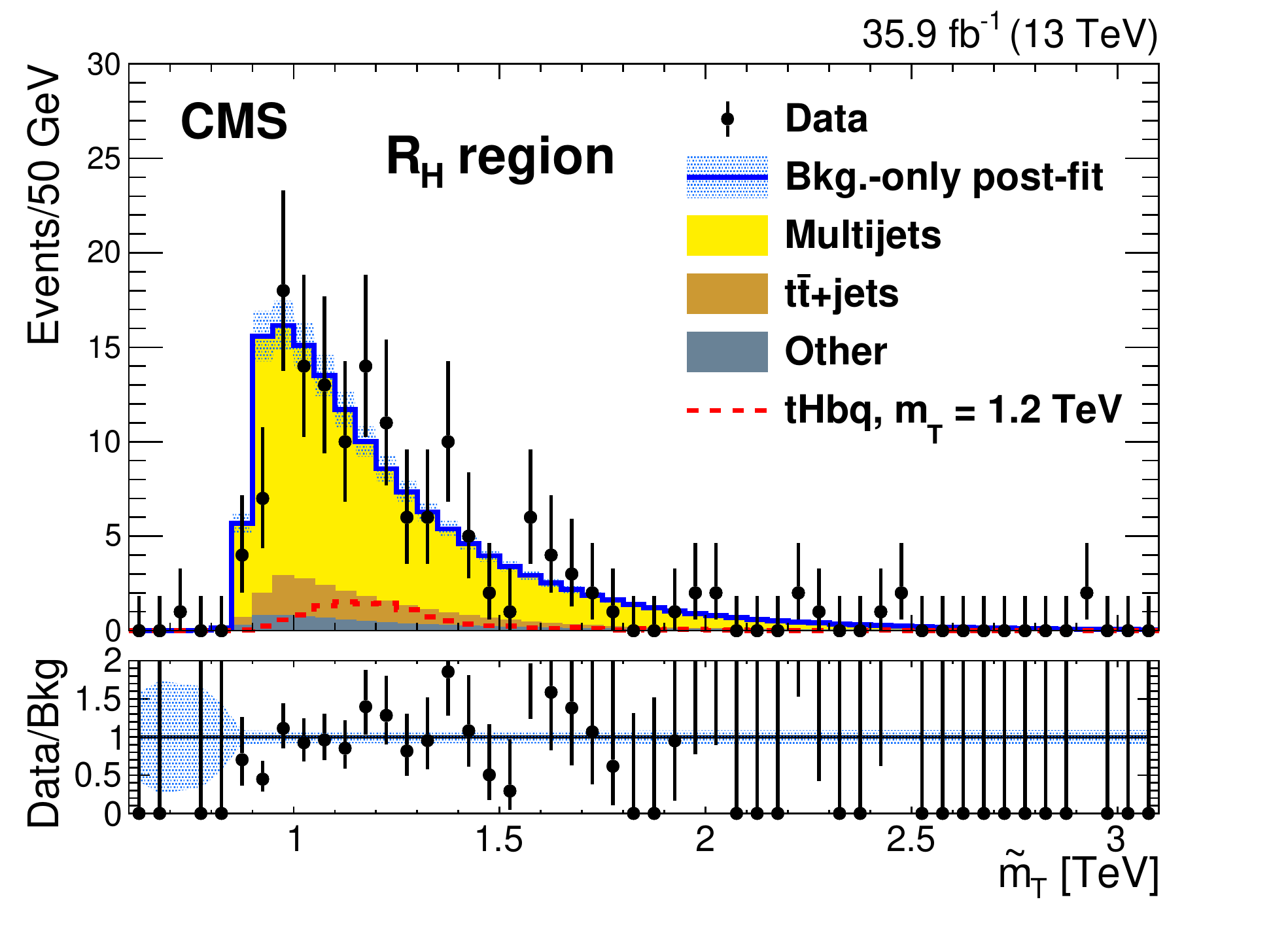}
    \includegraphics[width=0.497\textwidth]{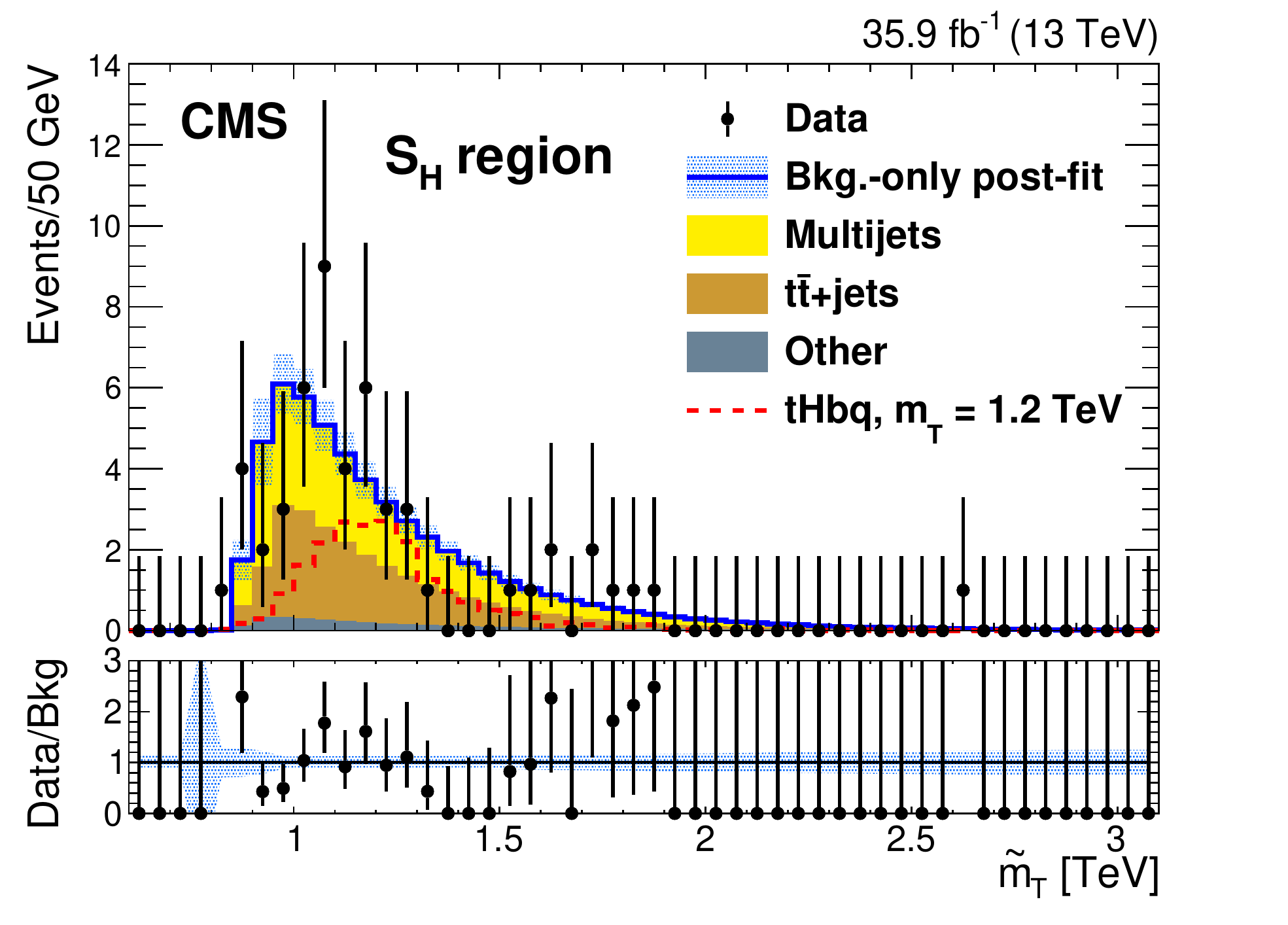}
    \caption{The background-only post-fit distributions in data
for the \rA, \rB, \rC, and \rH regions that are used as
signal and control regions primarily for the \tprimetotH channel.
The upper plots show regions \rA (left) and \rB (right), while the lower plots
show regions \rC (left) and \rH (right).
The dashed red histogram is an
example \tprimetotH signal for the \tHbq process with a 1.2\TeV \PQT
quark mass and a fractional width of 30\%
with a cross section from the singlet model of 142\unit{fb}.
The lower panels show the ratio of observed data
to fitted background per bin.
The error bars on the data
represent 68\% \CL Poisson intervals. The light blue band in each ratio panel
shows the fractional uncertainties in the fitted background.
}
    \label{fig:postfit_tH}

\end{figure}

\begin{figure}[!htb]
  \centering
    \includegraphics[width=0.497\textwidth]{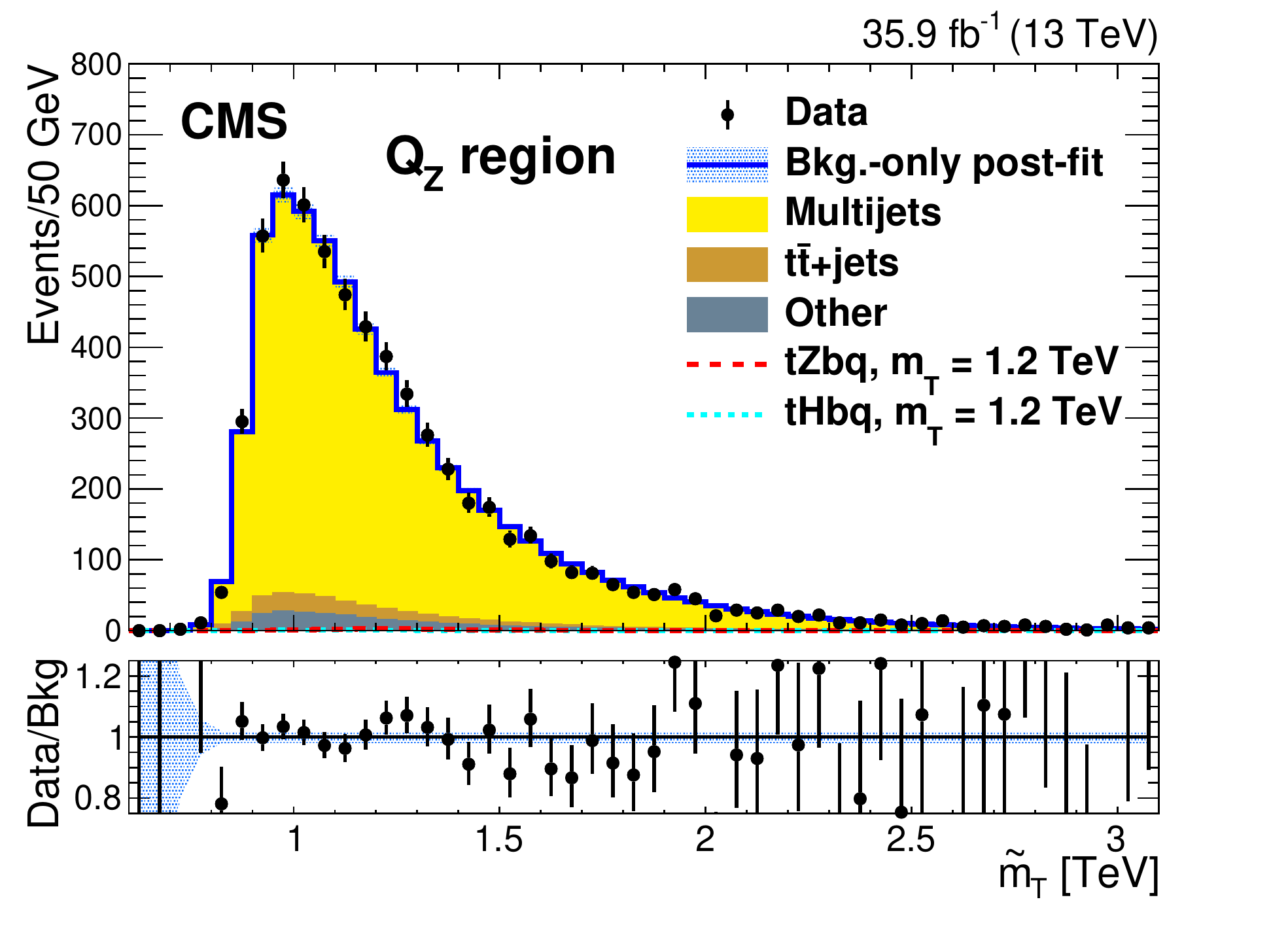}
    \includegraphics[width=0.497\textwidth]{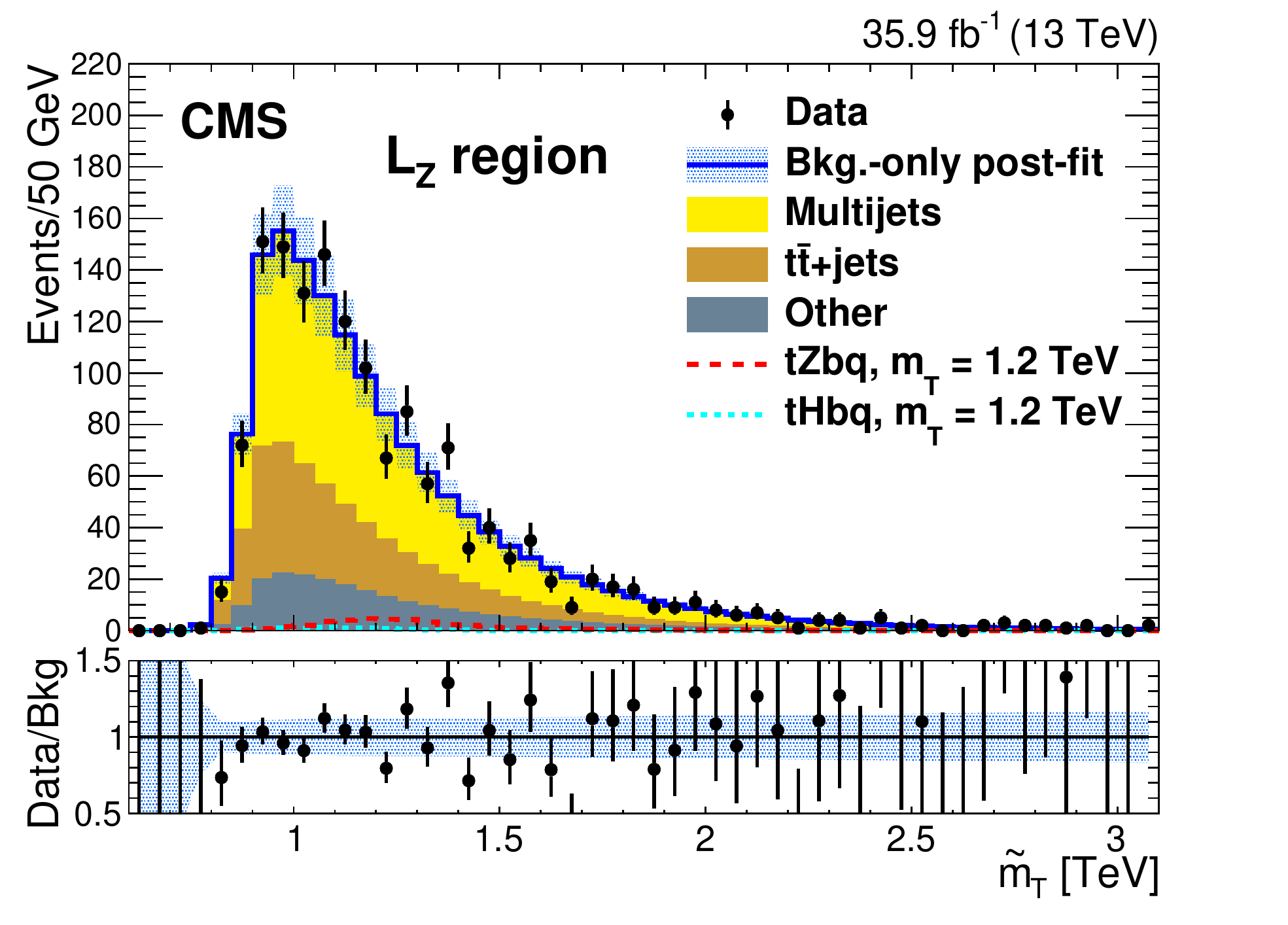}
    \includegraphics[width=0.497\textwidth]{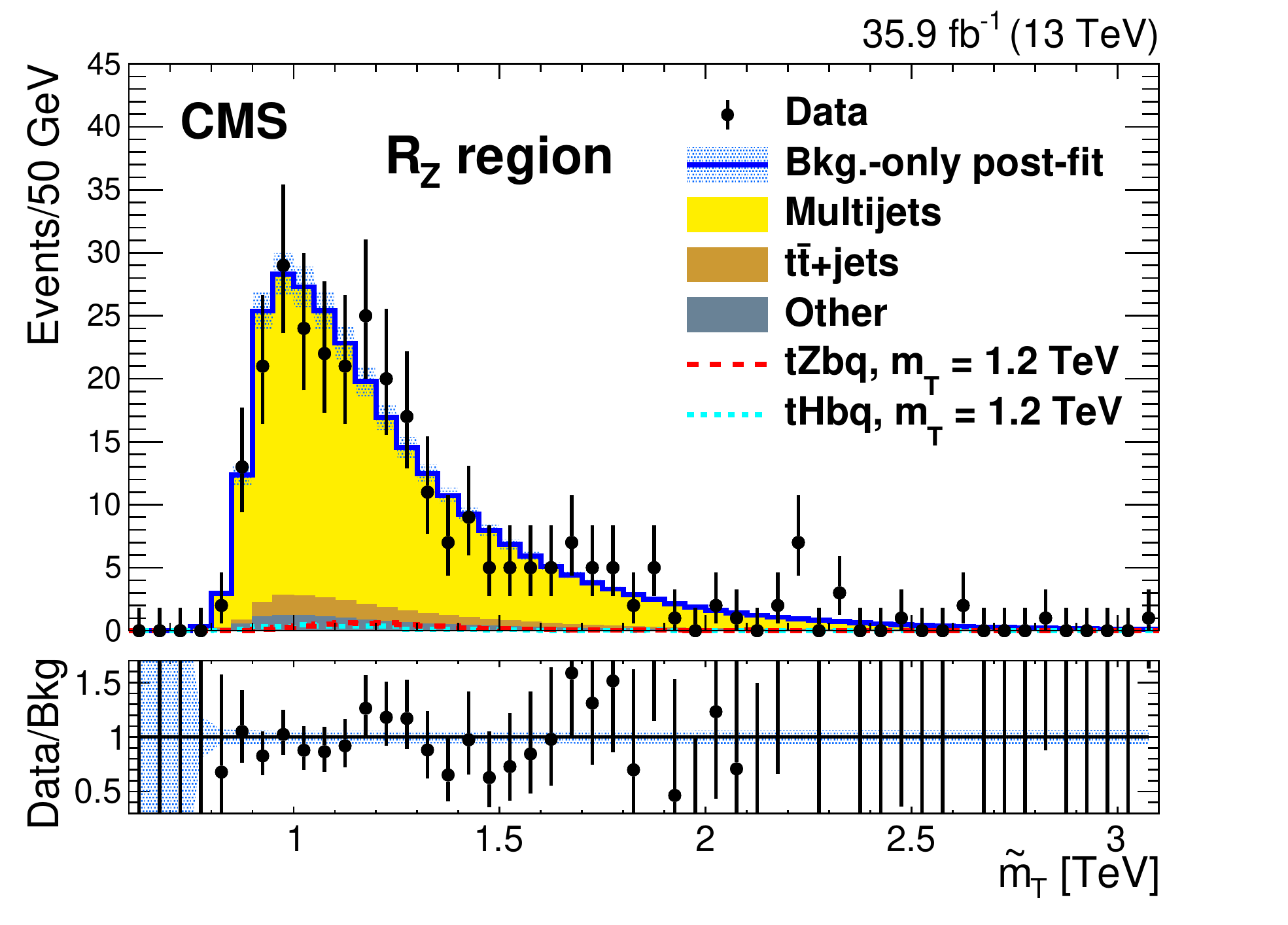}
    \includegraphics[width=0.497\textwidth]{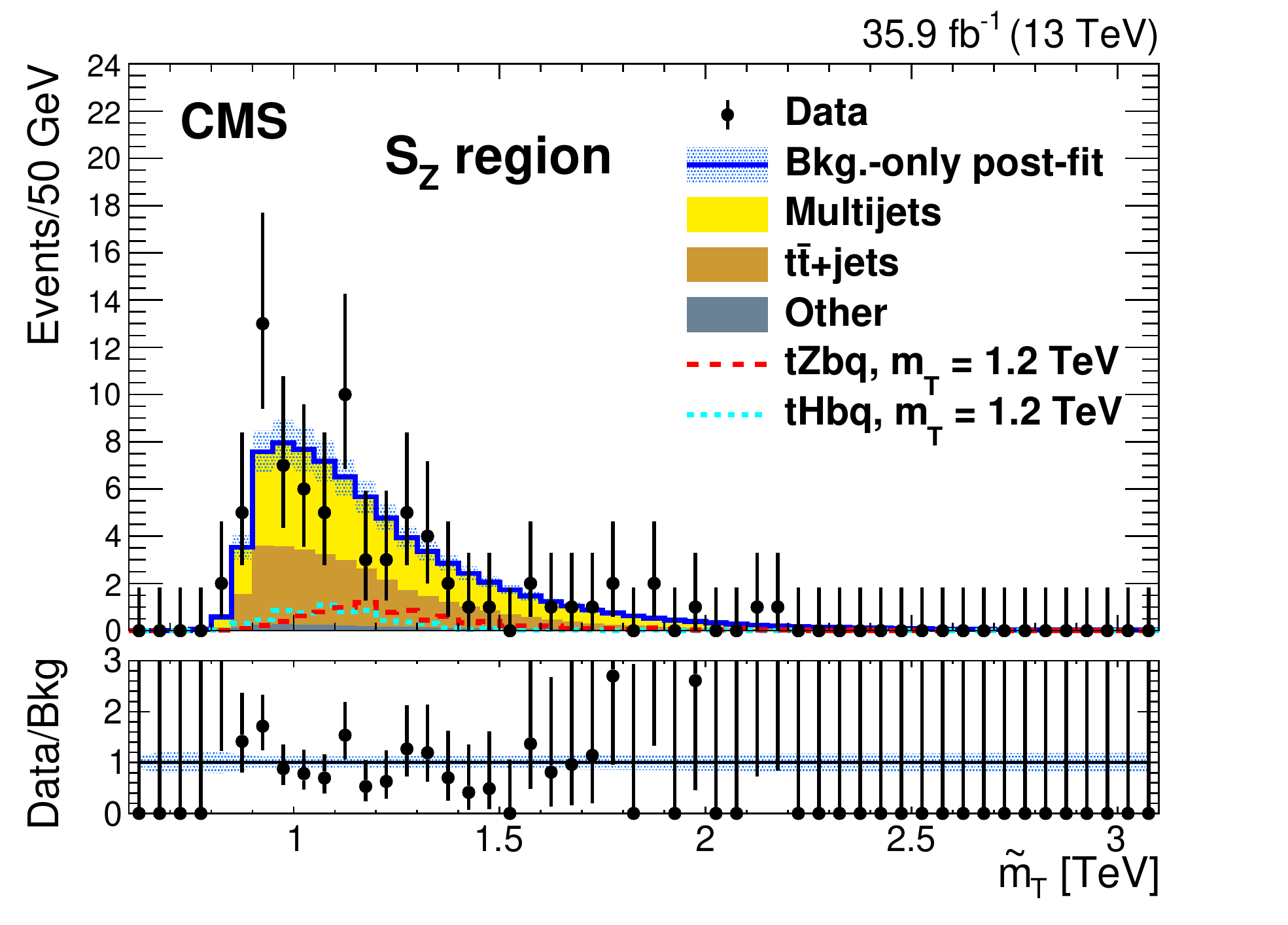}
    \caption{The background-only post-fit distributions in data
for the \rW, \rX, \rY, and \rZ regions that are used as signal and control regions primarily for the \tprimetotZ channel.
The upper plots show regions \rW (left) and \rX (right), while the lower plots
show regions \rY (left) and \rZ (right).
The dashed red histogram is an
example \tprimetotZ signal for the \tZbq process with a 1.2\TeV \PQT quark mass
and a fractional width of 30\%
with a cross section from the singlet model of 131\unit{fb}.
The shorter dashed cyan histogram is for \tprimetotH signal
for the \tHbq process with the model assumptions used in Fig.~\ref{fig:postfit_tH}.
The lower panels show the ratio of observed data
to fitted background per bin. The error bars on the data
represent 68\% \CL Poisson intervals. The light blue band in each ratio panel
shows the fractional uncertainties in the fitted background.
    }
    \label{fig:postfit_tZ}

\end{figure}

The resulting post-fit \Mtilde distributions in data based on the background-only hypothesis are shown for the \rA, \rB, \rC, \rH
and \rW, \rX, \rY, \rZ regions in Figs.~\ref{fig:postfit_tH} and \ref{fig:postfit_tZ}, respectively. It is found that these post-fit distributions are consistent with the background-only model
with an acceptable goodness-of-fit.
Upper limits are then set on the cross sections for the
two production modes ($\Tbq$ and $\Ttq$). These upper limits
are reported in Section~\ref{sec:combination} together with the
limits from the low-mass search
for four fractional width ($\GoM$) values and
the two decay modes ($\tH$ and $\tZ$) as well as their sum ($\tH$+$\tZ$).
\section{Systematic uncertainties\label{sec:systematics}}
The systematic uncertainties can be classified into those that affect the
overall yields of the signal and the background processes, and those that
affect the invariant mass distributions \mtprime (for the low-mass search)
and \Mtilde (for the high-mass search).
The sources of systematic uncertainties and their effects on the signal
and the background are summarized in Table~\ref{tab:syst}.

\begin{table}[!htb]
  \centering
  \topcaption{The systematic uncertainties in the signal and background yields for each search.
The uncertainties marked ``Shape'' affect both the event yields and the distributions.}
  \label{tab:syst}
    {\renewcommand{\arraystretch}{1.2}
    \begin{tabular}{lccc}
      Source & Low-mass & & High-mass \\
      \hline
                                        &                    & Signal yield (\%) & \\
      Trigger efficiency                & 3                  & & 3                                \\
      Jet energy scale and resolution   & Shape              & & Shape                            \\
      Jet mass scale and resolution     & \NA                & & 2                                \\
      \PH/\PZ tagging correction factor     & \NA            & & 10                               \\
      \PH/\PZ jet $\tau_{21}$ selection     & \NA            & & 7+jet \pt-dependence             \\
      Jet mass resolution               & \NA                & & Shape                            \\
      Top quark jet tagging             &   \NA              & & $^{+5}_{-3}$                        \\
      $\PQb$ tagging selection         & 5--7               & & Shape                            \\
      PDF                               & 0.5--1             & & 0.5--1                           \\
      Pileup modeling                  & 2--4                & & 2                                \\
      Integrated luminosity                        & 2.5     & & 2.5                              \\
      Renorm./fact. scales                        & 1--2     & & Shape                            \\[\cmsTabSkip]
                                        &                    & Background yield (\%) & \\
      Low-mass background         & Shape                    & &       \NA            \\
      \ttjets cross section             & \NA                & & $^{+6.1}_{-5.5}$                        \\
      \wjets cross section             & \NA                 & & 3.8                              \\
      QCD multijet background           & \NA                & & 23--25                           \\
      Renorm./fact. scales                        & \NA      & & Shape                            \\
    \end{tabular}
    }
\end{table}

The trigger efficiency for the low-mass category is measured in data and
is found to be about 97\% with an assigned uncertainty of 3\%.
The trigger efficiency for the high-mass analysis is measured using
hadronic triggers to be over 99.5\%.
There is a mild dependence on \Mtilde that is
evaluated using a muon-based monitor trigger. The maximum variation is 3\%
and this is taken as the uncertainty in the overall event yields.

The jet energy scale uncertainties depend on the \pt and $\eta$ of the
jets~\cite{CMS-PAS-JME-16-003}.
The jet energy resolution in data is found to be worse than
that in simulation, and the discrepancy is corrected by applying an
extra smearing to the energy of jets in simulated events.
Both the jet energy scale and resolution uncertainties affect the overall
scale and shapes of the invariant mass distributions.

The uncertainties in the $\PH$ and $\PZ$ jet mass scale and resolution and
the \textit{N}-subjettiness selection~\cite{CMS-PAS-JME-16-003} affect the
high-mass search. The jet mass scale, resolution, and the
\textit{N}-subjettiness selection efficiency were measured in a sample
of \ttjets events with one top quark decaying leptonically
and the other top quark decaying hadronically (semileptonic \ttjets events).
The hadronically decaying top quark is boosted enough to produce a merged $\PW \to \qqbar'$
jet separated from a $\PQb$ jet.
The jet mass scale was evaluated to be unity with an uncertainty of 2\%.
An additional uncertainty of about 10\% was derived using simulations to account
 for the difference between the jet showering for a $\PW\to\qqbar'$ jet and an
$\Hbb$ or $\Zbb$ jet, using parton shower models from \PYTHIA~8
and \HERWIGpp~2.7.1~\cite{HERWIGpp}. The EE5C tune~\cite{HerwigppEE5C}
is used for \HERWIGpp.
The jet mass resolution was found to be 23\% larger in the data than
in the simulation, with an uncertainty of 18\%.
The ratio of the \textit{N}-subjettiness selection efficiency in the data to the simulation (scale factor) was found to be $1.11 \pm 0.08$ with an
additional jet $\pt$-dependent uncertainty.
The jet mass scale and \textit{N}-subjettiness scale factor uncertainties affect
only the event yields whereas the jet mass resolution affects both the
yields and the \Mtilde distributions.

The $\PQb$ tagging~\cite{CMS-BTV-16-002} efficiency scale factor
uncertainties for AK4 jets are measured in multijets and \ttjets samples,
separately for $\PQb$ quark jets and for light-quark and gluon jets.
These uncertainties affect only the signal event yields for the low-mass
analysis. For the high-mass analysis, the $\PQb$ tagging efficiency scale
factor uncertainty for subjets is measured using multijet and \ttjets samples and
is applied to the subjets for $\PH$ and $\PZ$ jets.
These affect both the event yields and the \Mtilde distributions.

For the top quark jet tagging used in the high-mass search,
the scale factor is measured using semileptonic \ttjets events,
with a boosted top quark jet, using the method
from Ref.~\cite{CMS-PAS-JME-15-002}. The scale factor with
uncertainty is $1.07^{+0.05}_{-0.03}$ for each top quark jet, and
affects both the event yields and the \Mtilde distributions.

The PDF uncertainties were evaluated with
the {PDF4LHC} procedure~\cite{Butterworth:2015oua}, by reweighting the simulated events using
the eigenvectors of the {NNPDF3.0} PDF set; it was found to change the overall event
yields by 0.5--1\%.
The renormalization and factorization scale uncertainties were estimated by doubling and halving the nominal values used in the simulations. These were found to
affect the high-mass search event yields for both the signal and
the \ttjets background.

The uncertainty in the measurement of the integrated luminosity amounts
to 2.5\%~\cite{CMS-PAS-LUM-17-001} and affects the overall event yields of
all simulated processes.
The uncertainty associated with the mismodeling of pileup
is evaluated based on a 4.6\% variation on the $\Pp\Pp$ total
inelastic cross section~\cite{Sirunyan:2018nqx}
and affects the
overall simulated event yields for both analyses.

The systematic uncertainties in the background estimation for the low-mass
search are taken into account by inflating the uncertainty in
the slope parameter of the transfer functions.
This includes an uncertainty component justified from the difference
observed when the weights are computed on light-flavor versus
heavy-flavor jets and an uncertainty component coming from the
uncertainties in the fit arising from the $\PQb$ tagging efficiency.
Based on the observed differences, the slope parameter uncertainty is increased by a factor
of four for the \TwoMOneL to \ThreeM region transfer function and by a factor of three for
the \ThreeM to \ThreeT region transfer function.
For the high-mass search, the uncertainty in the background prediction includes the systematic uncertainty in the double ratio used to constrain the QCD multijet event yields in the different control and the signal regions (Section~\ref{sec:fh-boosted-bkgd}). While the initial uncertainty in this double ratio is assigned to be 60\%,
the postfit uncertainty ranges from 23--25\%, depending on signal model.

The \ttjets cross section uncertainty is $^{+6.1}_{-5.5}\%$ at
next-to-next-to-leading order~\cite{Czakon:2013goa}, while that in
the \wjets cross section is $3.8\%$~\cite{Melnikov:2006kv,Li:2012wna}.
These uncertainties affect only
the high-mass search where both these backgrounds are taken from simulations.
These uncertainties are propagated to the estimation of
the QCD multijet background from the data, where the non-QCD multijet
background components are subtracted from the data. The uncertainties also
include \Mtilde-dependent modeling uncertainties for the various
background components.
\section{Search results}\label{sec:combination}
The low-mass search is sensitive to masses from 0.6 to 1.2\TeV,
while the high-mass
search is sensitive to masses from 0.7 to 2.6\TeV, but with
its main sensitivity
starting around 1\TeV.
The limits presented here correspond to a confidence level of 95\%.
They are based on whichever of the two searches for each considered
mass has the best estimated expected sensitivity for each
decay channel (\tH, \tZ, and the sum), production mode, and fractional
width. The limits from each search considered separately are presented
for completeness in Appendix~\ref{sec:supplementary}.

Upper limits are set on the cross sections for the
two production modes ($\Tbq$ and $\Ttq$) with
the two decay modes ($\tH$ and $\tZ$) as well as their
sum ($\tH$+$\tZ$), for four fractional width ($\GoM$) values.
The individual limits for the $\tH$ decay mode neglect potential contributions
from the $\tZ$ and $\bW$ decay modes and, similarly, the
individual limits for the $\tZ$ decay mode neglect potential contributions
from the $\tH$ and $\bW$ decay modes.
The ($\tH$+$\tZ$) sum limits are computed assuming the relative cross sections for
the two channels calculated in the
respective model for a particular width based
on Ref.~\cite{Carvalho:2018jkq}.
For \Tbq, this corresponds to the \PQT singlet model
with $\kw=\kh=\kz$ and
for \Ttq, to
the (\TB) doublet model
with $\kh=\kz$
and $\kw=0$, where
\kw, \kh, and \kz denote the \PQT coupling parameter to $\bW$, $\tH$, and $\tZ$, respectively.
Only potential contributions from \bW are neglected for the ($\tH$+$\tZ$) experimental limits.
The calculated cross sections for \tH and \tZ are approximately the same
but unequal. This is due both to mass effects explicit in the decay width expressions
of Ref.~\cite{AguilarSaavedra:2009es} and to additional amplitude contributions
associated with the $2 \to 4$ finite width calculation arising
from \PQT quark-mediated Feynman diagrams relevant to the large-width regime, such
as the t-channel exchange of a \PQT quark.
The fractional width values
include 10, 20, and 30\%, and the narrow-width case.
Given the estimated effective Gaussian mass resolution of about 5\%, the experimental limits set using
the narrow-width simulated signal samples are applicable to (Breit--Wigner) fractional widths of up to about 5\%.
The experimental upper limits on cross sections are generally more restrictive for smaller widths given the narrower line shapes.
The computed cross sections are found to depend approximately linearly on
the fractional width.

\begin{figure}[h]
  \centering
    \includegraphics[width=0.49\textwidth]{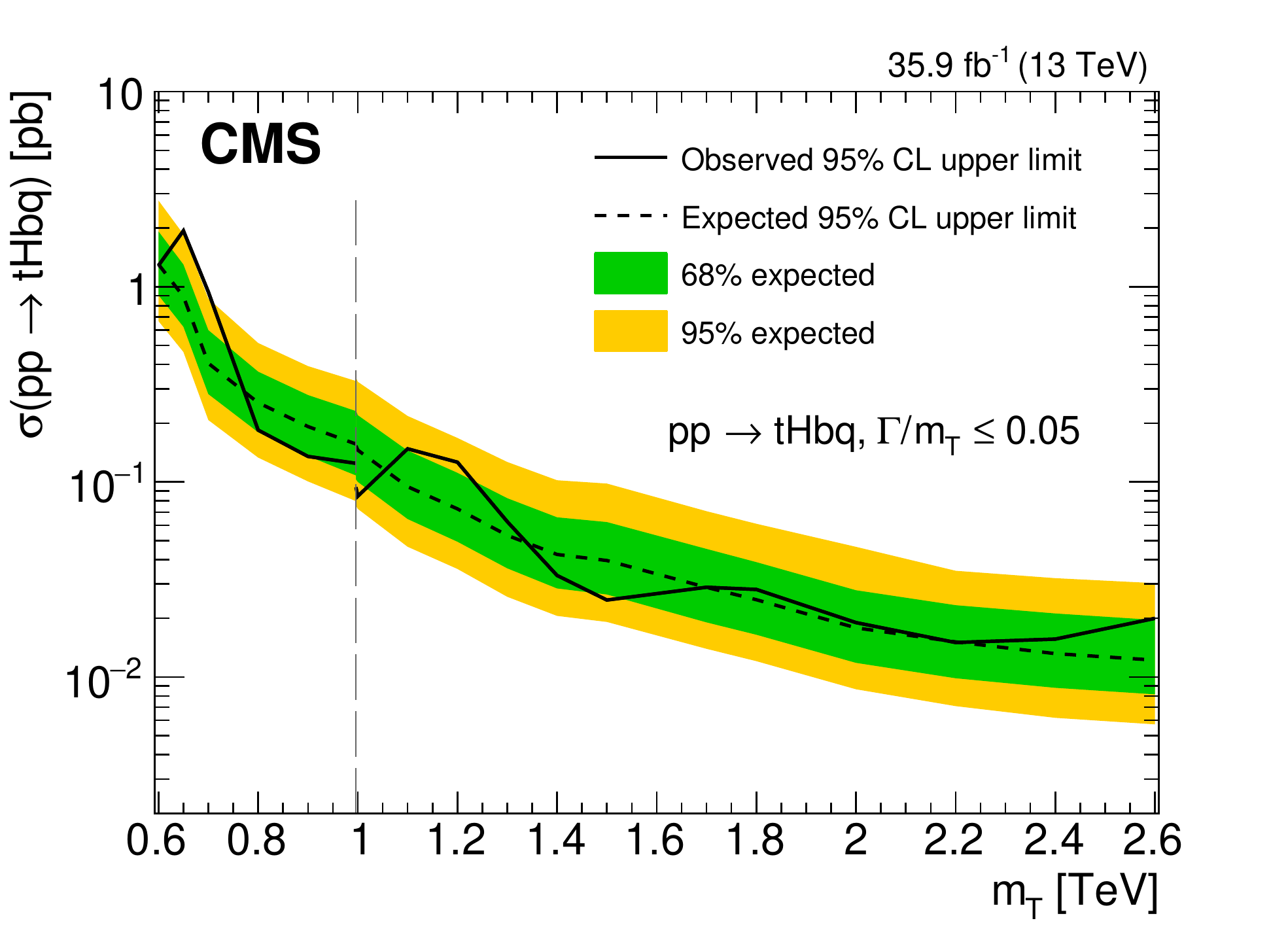}
    \includegraphics[width=0.49\textwidth]{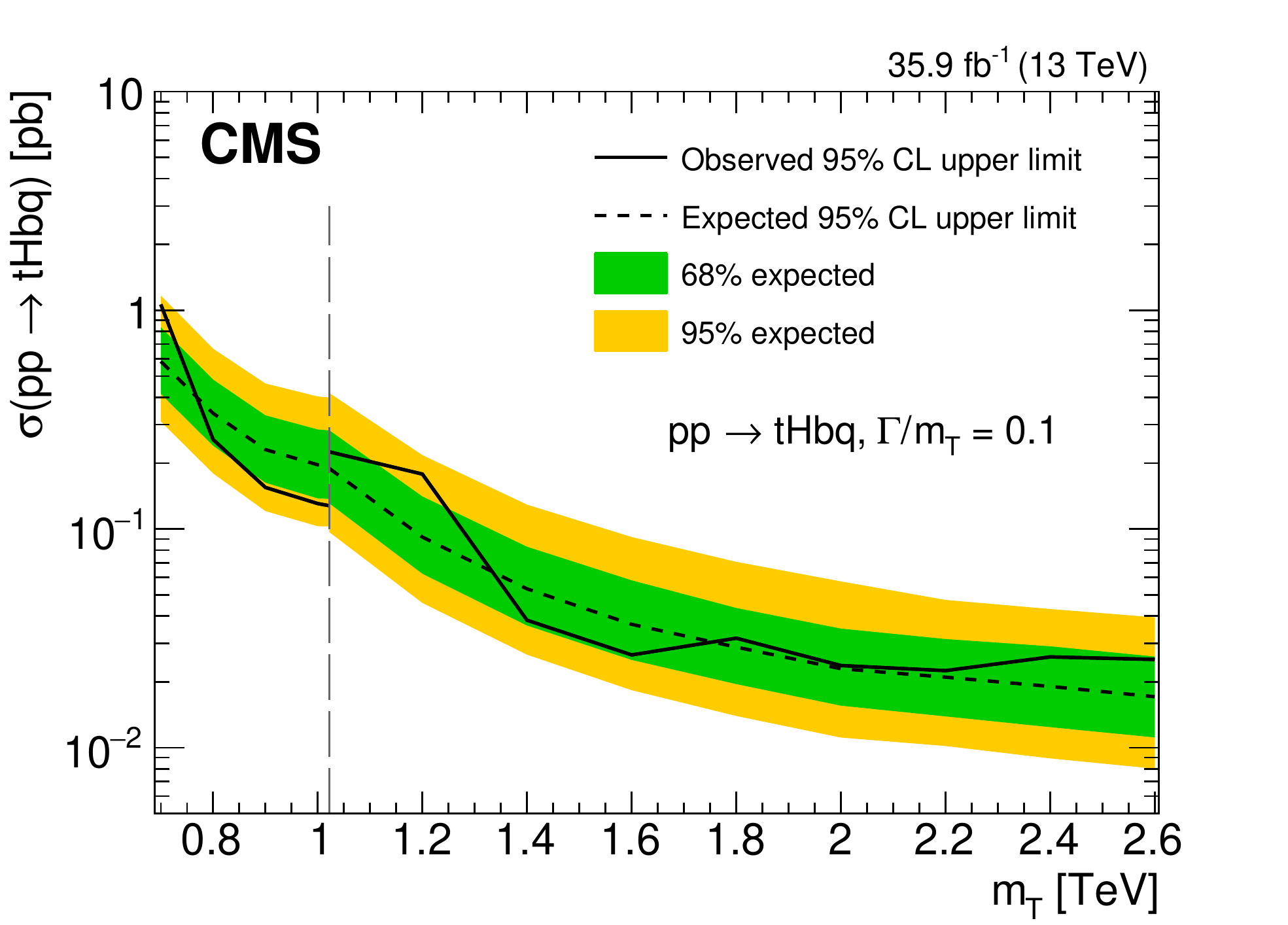}
    \includegraphics[width=0.49\textwidth]{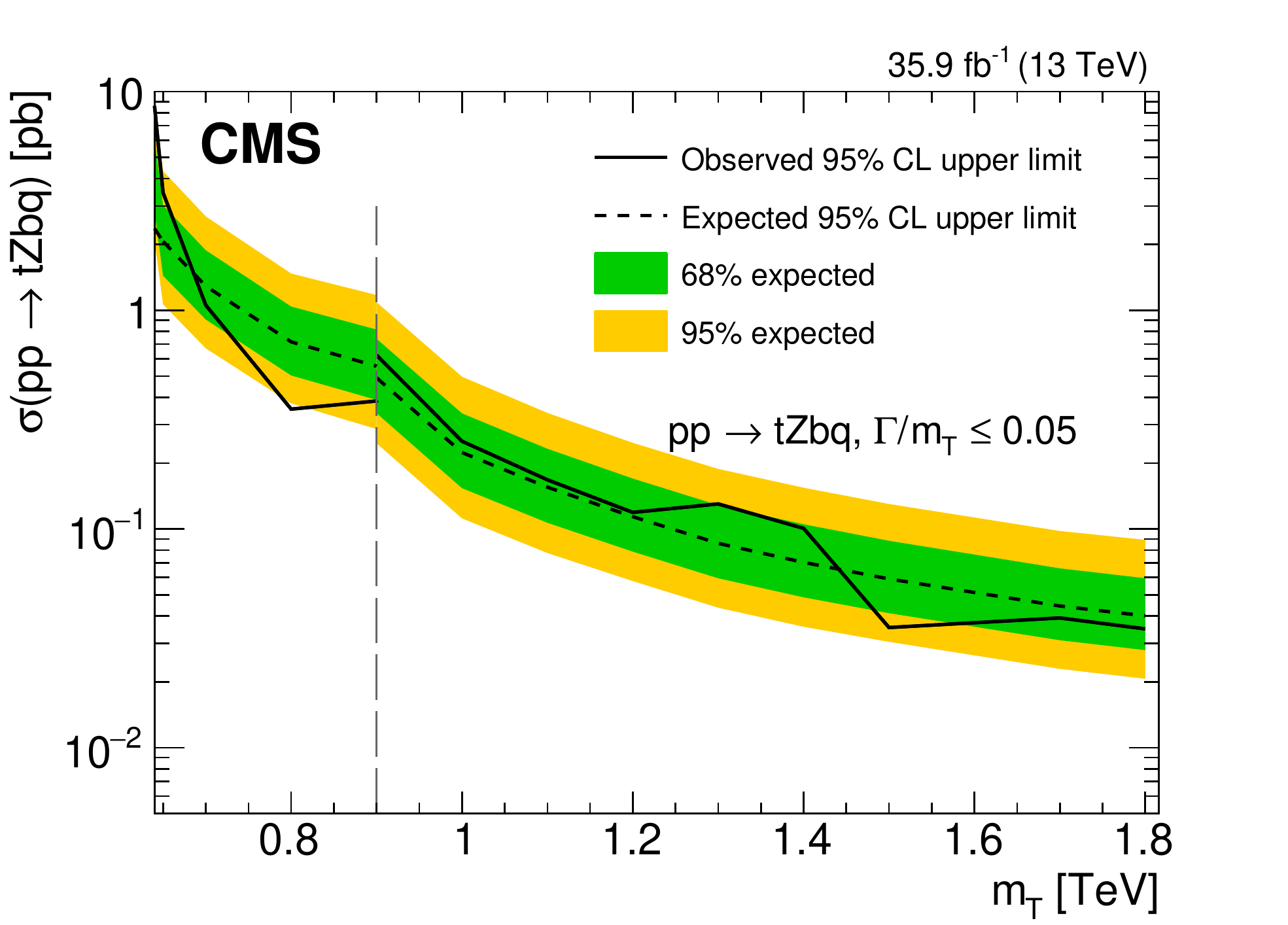}
    \includegraphics[width=0.49\textwidth]{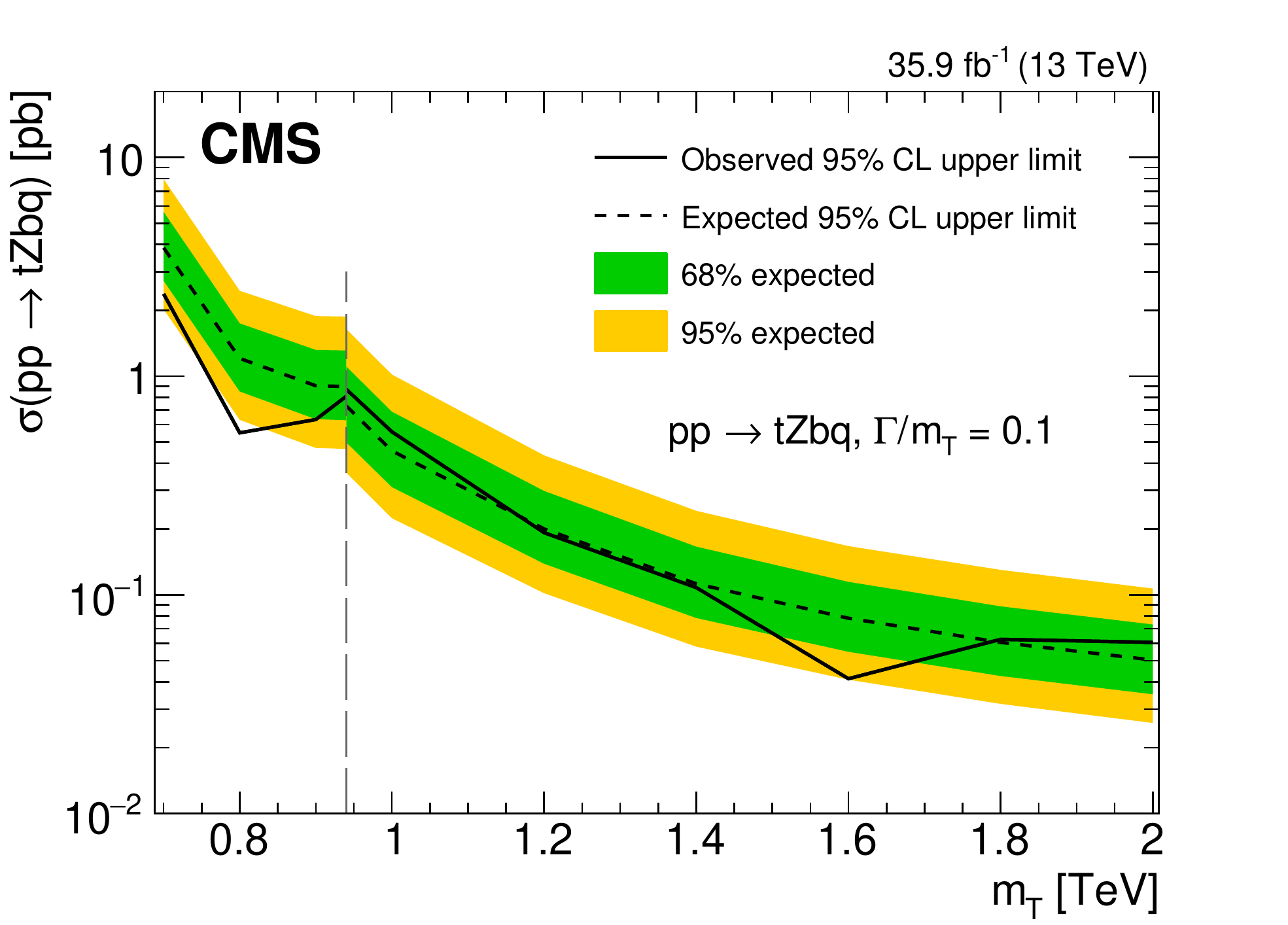}
    \includegraphics[width=0.49\textwidth]{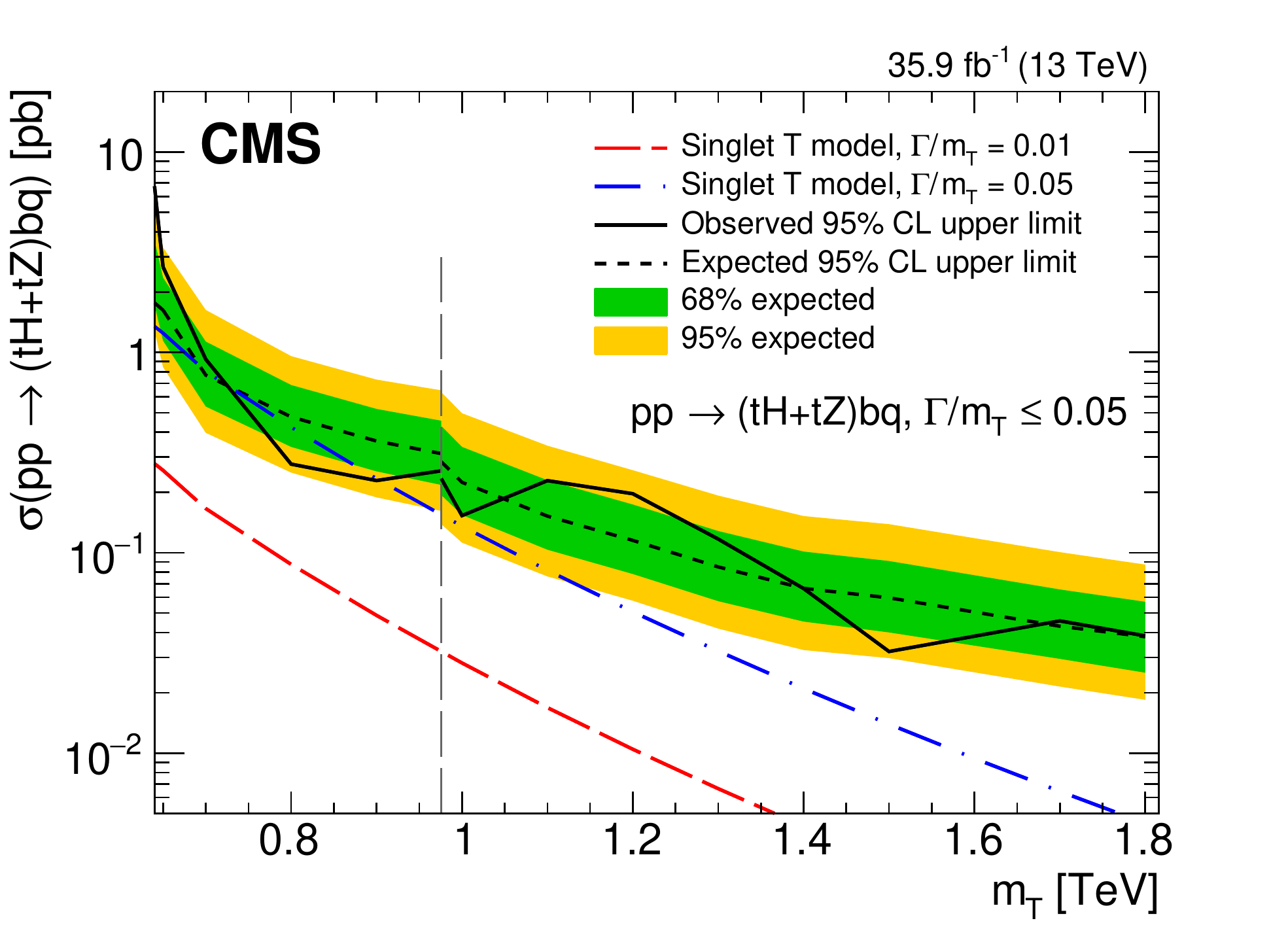}
    \includegraphics[width=0.49\textwidth]{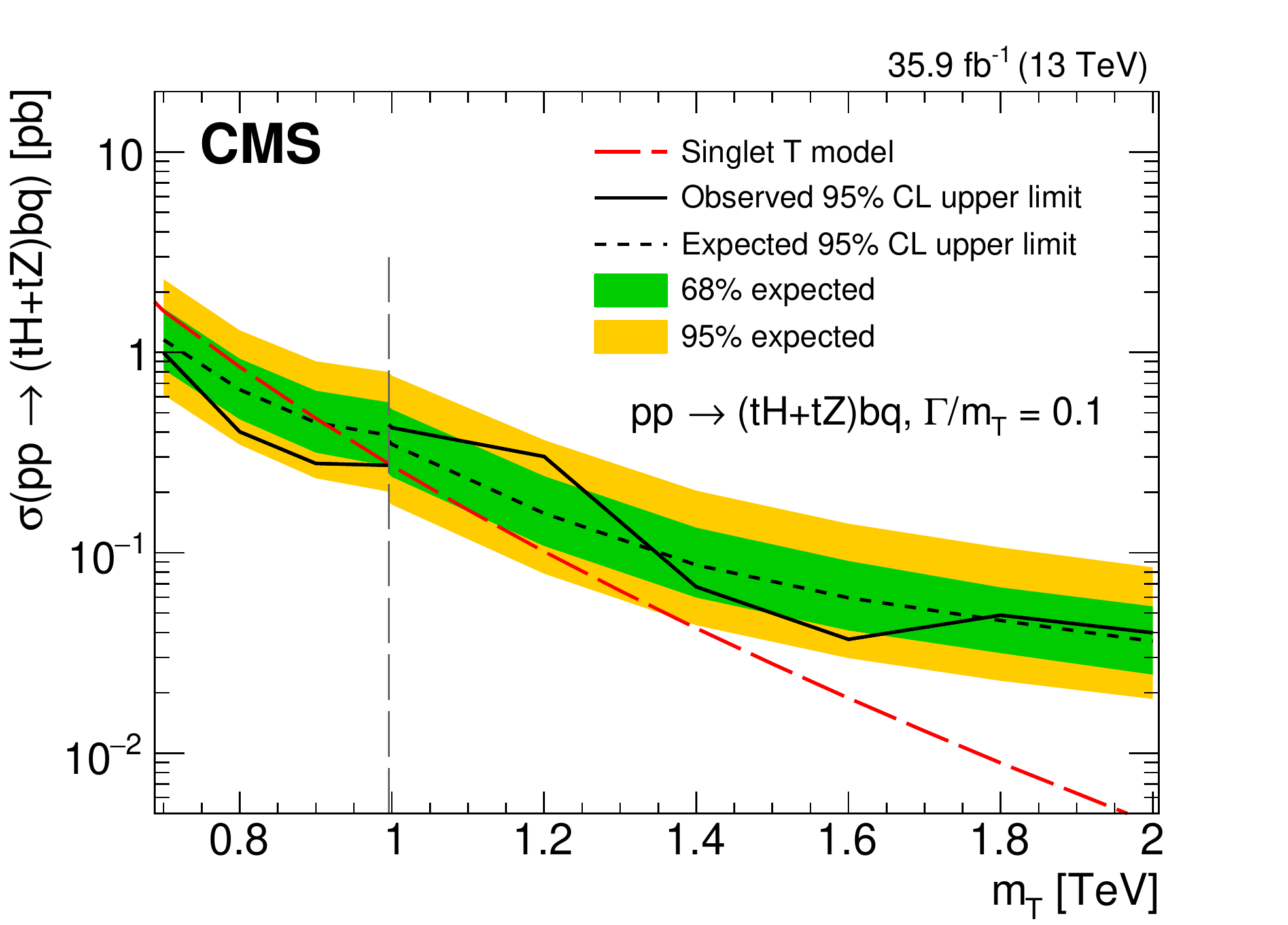}
    \caption{The observed and median expected upper limits at 95\%~\CL on the cross sections for production
associated with a bottom quark for the \tHbq (upper row) and \tZbq (middle row) channels, and their sum, \tHZbq (lower row), for
different assumed values of the \PQT quark mass.
The inner (green) band and the outer (yellow) band indicate the regions containing 68 and 95\%, respectively,
of the distribution of limits expected under the background-only hypothesis.
The left column is for a narrow fractional width ($\GoM \le 0.05$) and the right column is
for a fractional width of $\GoM = 0.1$.
The vertical dashed lines are
the crossover points in sensitivity that indicate the mass intervals used for presenting
the low-mass and high-mass search results.
The dashed red and dot-dashed blue curves are for the \PQT singlet model.
Given the specified width, the couplings are implicit in the model. Two curves
corresponding to $\GoM = 0.05$ (dot-dashed blue)
and $\GoM = 0.01$ (dashed red) are shown for the narrow fractional width.}
    \label{fig:massLim_Tbq1}

\end{figure}

\begin{figure}[h]
  \centering
    \includegraphics[width=0.49\textwidth]{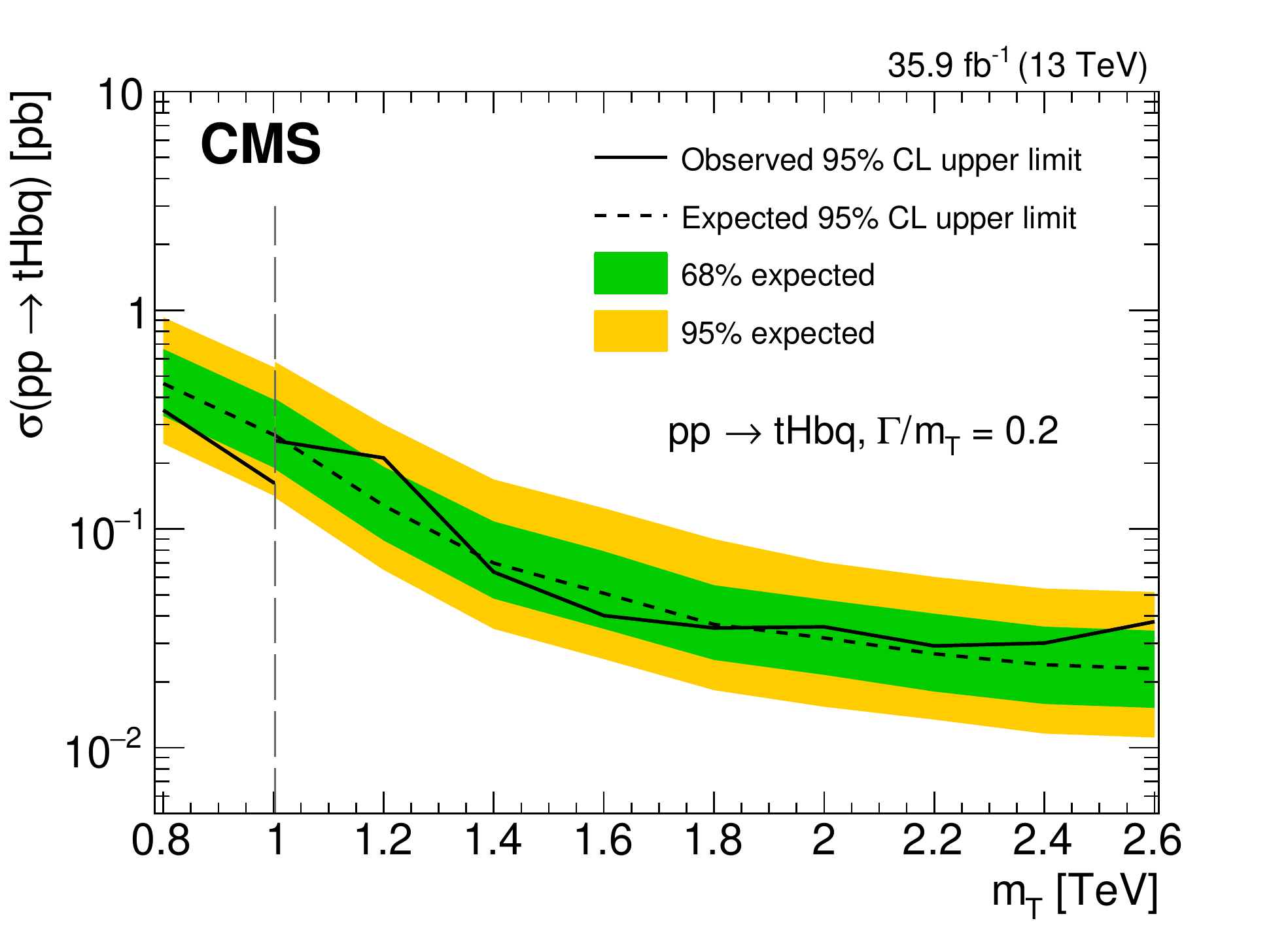}
    \includegraphics[width=0.49\textwidth]{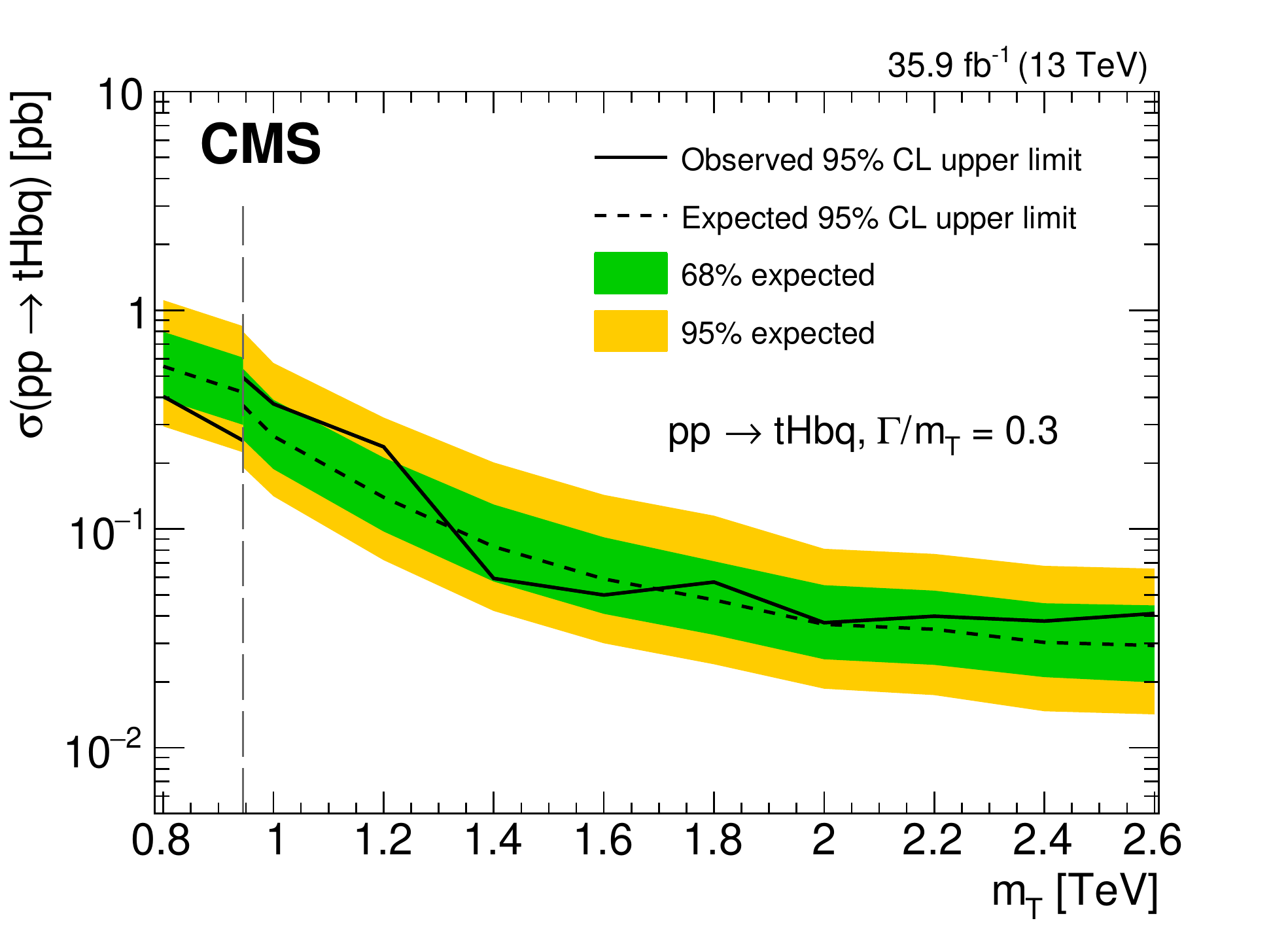}
    \includegraphics[width=0.49\textwidth]{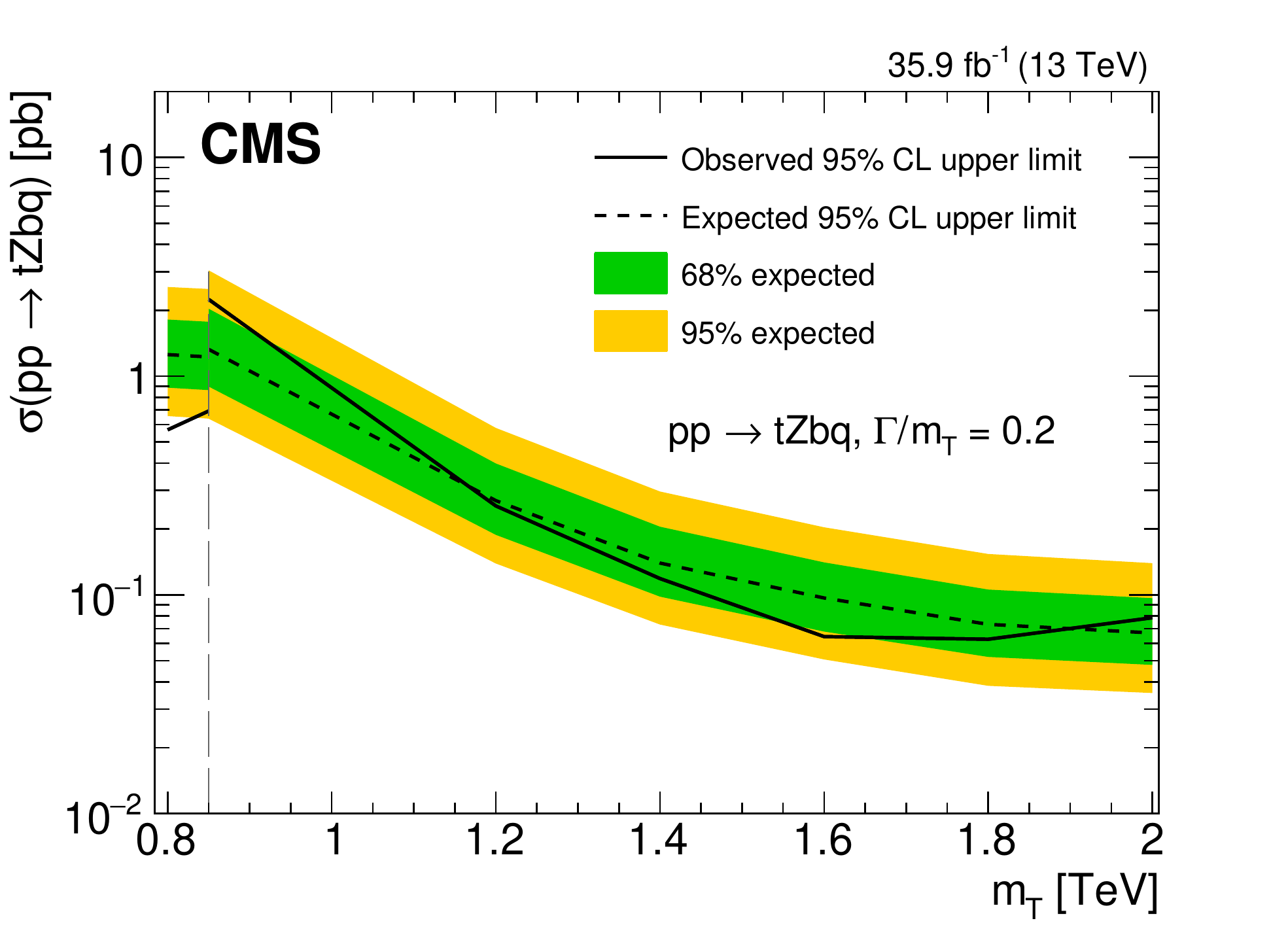}
    \includegraphics[width=0.49\textwidth]{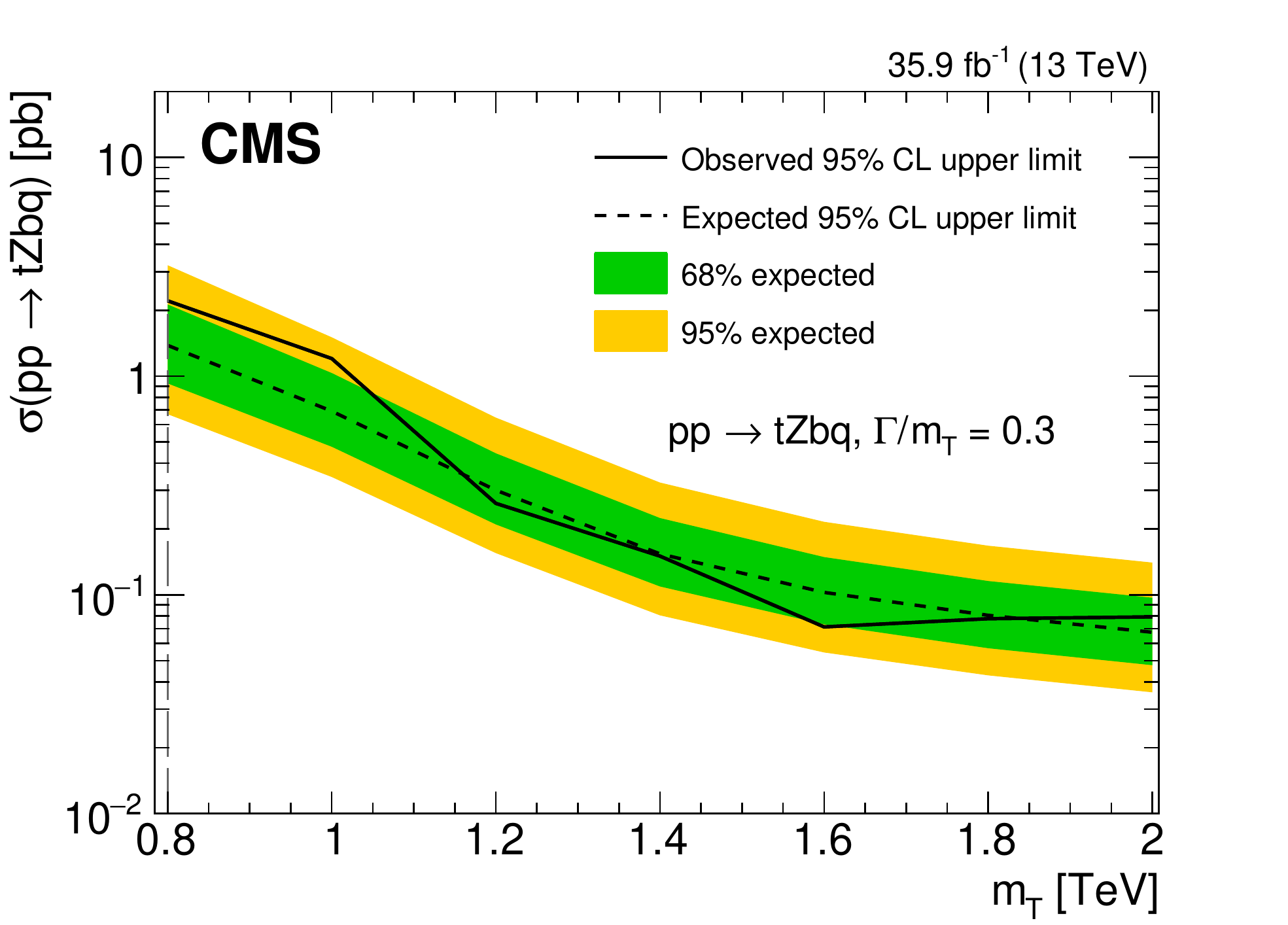}
    \includegraphics[width=0.49\textwidth]{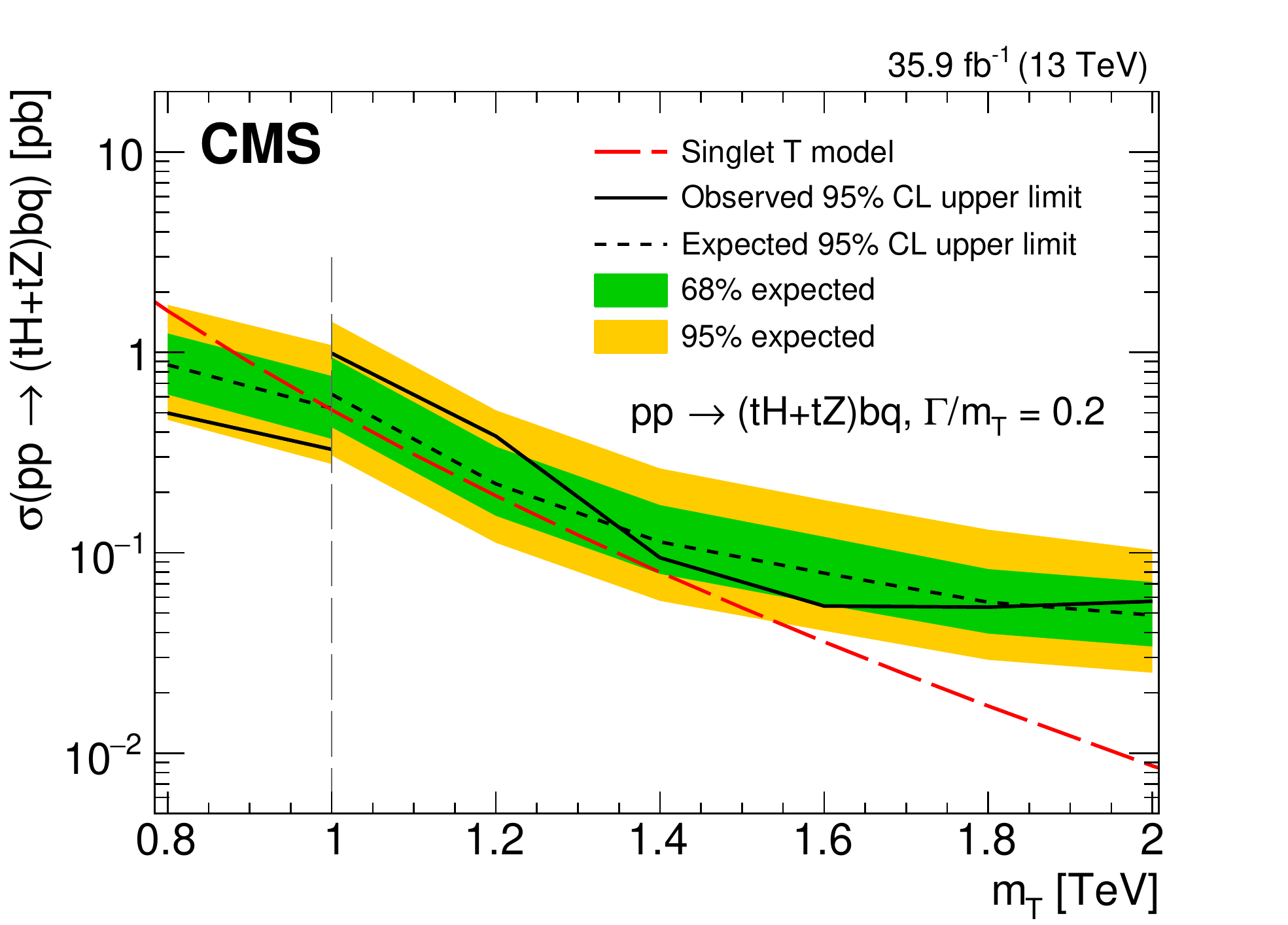}
    \includegraphics[width=0.49\textwidth]{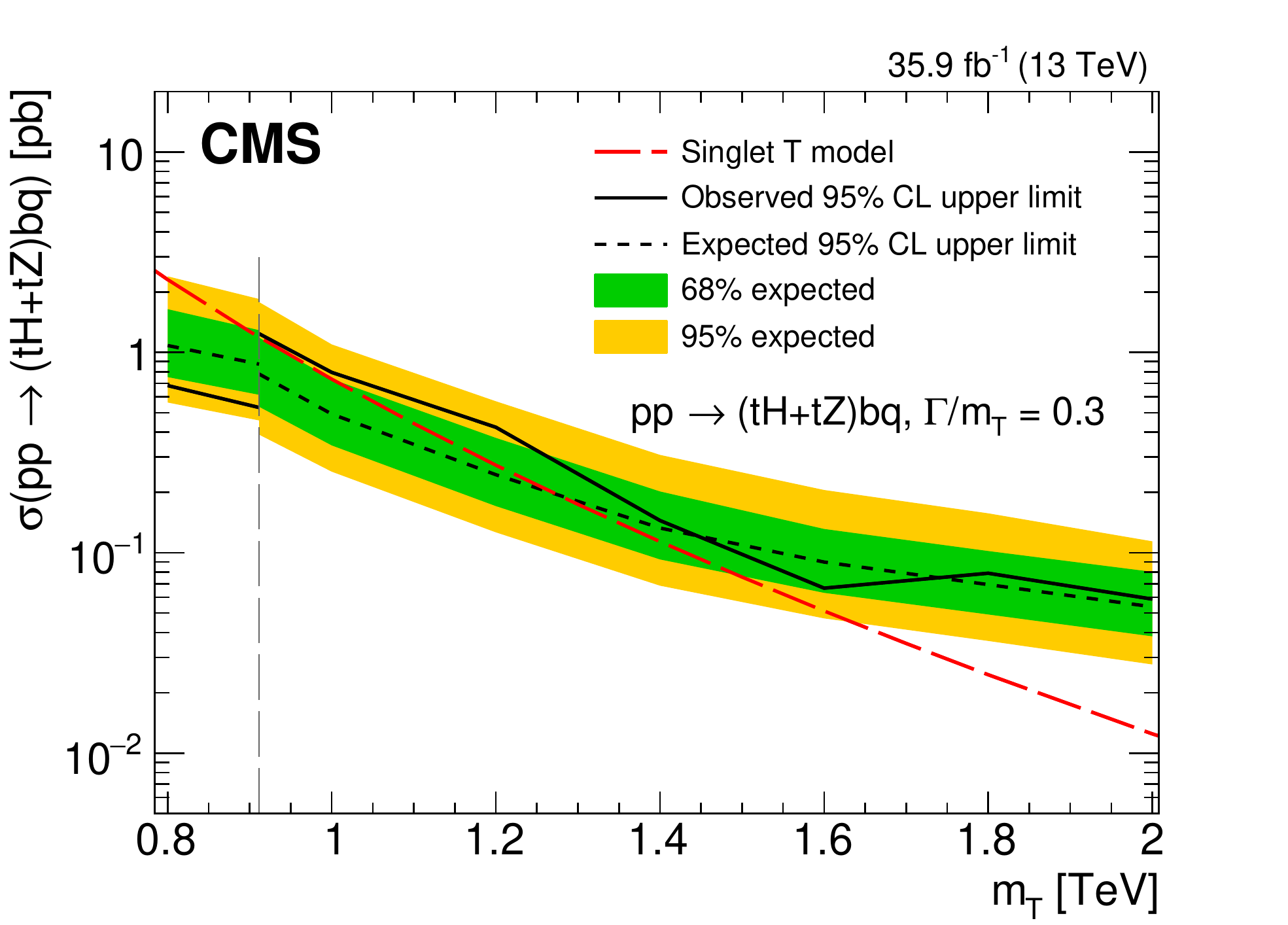}
    \caption{The observed and median expected upper limits at 95\%~\CL on the cross sections for production
associated with a bottom quark for the \tHbq (upper row) and \tZbq (middle row) channels,
and their sum, \tHZbq (lower row), for
different assumed values of the \PQT quark mass.
The inner (green) band and the outer (yellow) band indicate the regions containing 68 and 95\%, respectively,
of the distribution of limits expected under the background-only hypothesis.
The left column is for a fractional width of 20\% and the right column is for a fractional width of 30\%.
The vertical dashed lines are
the crossover points in sensitivity that indicate the mass intervals used for presenting
the low-mass and high-mass search results.
The dashed red curves are for the \PQT singlet model. Given the specified width, the couplings are implicit in the model.}
    \label{fig:massLim_Tbq2}

\end{figure}

\begin{figure}[!htb]
  \centering
    \includegraphics[width=0.49\textwidth]{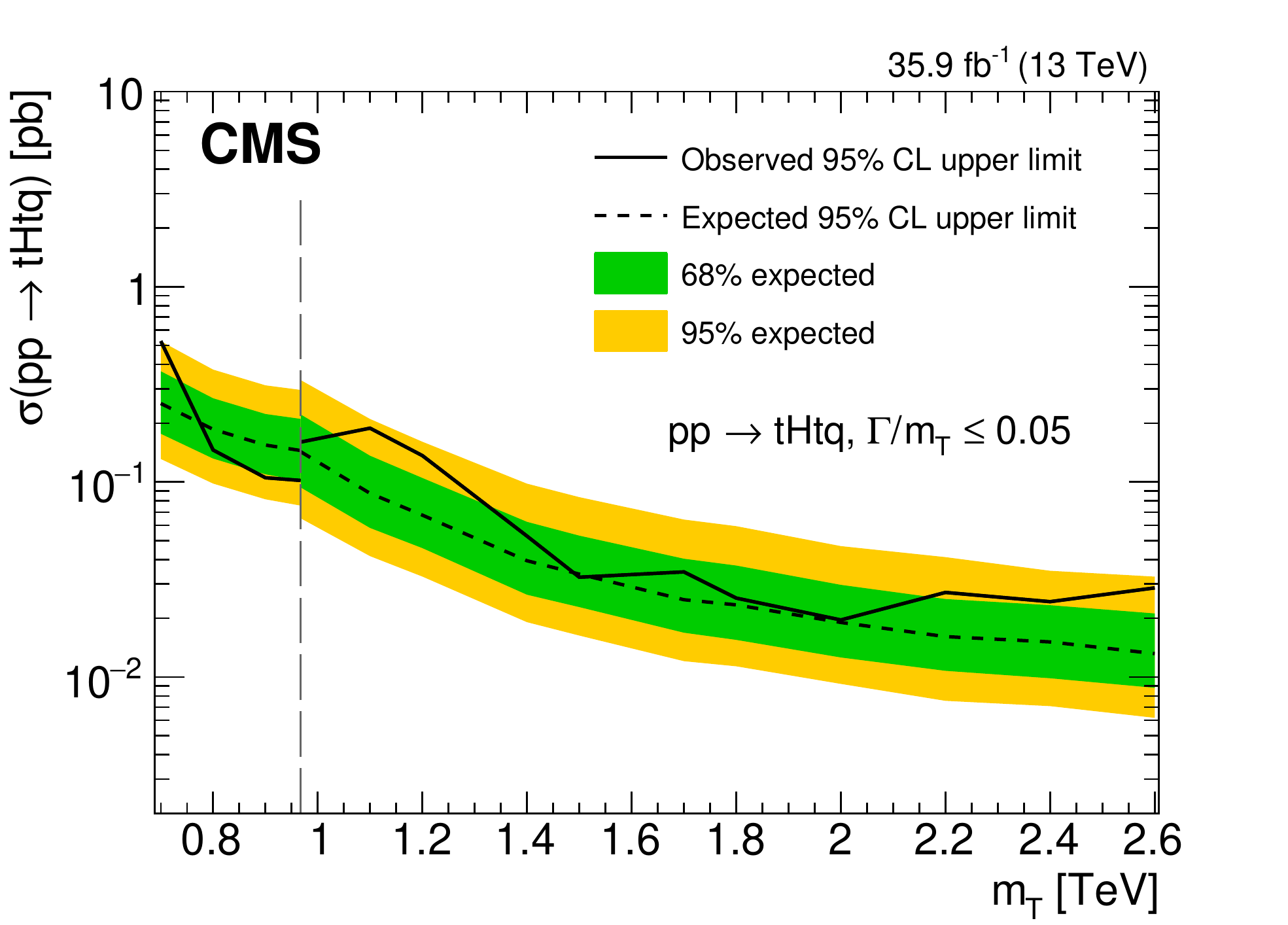}
    \includegraphics[width=0.49\textwidth]{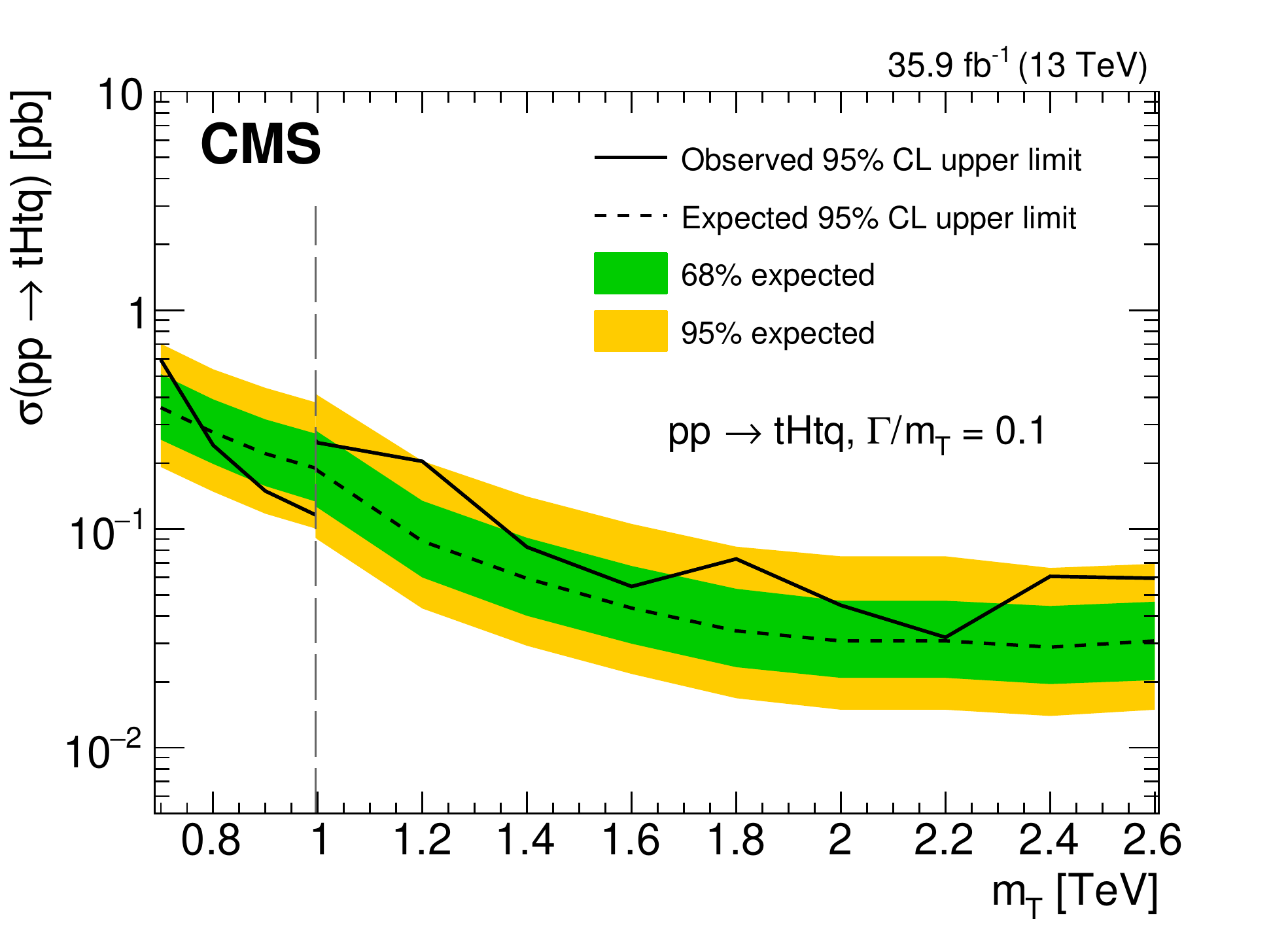}
    \includegraphics[width=0.49\textwidth]{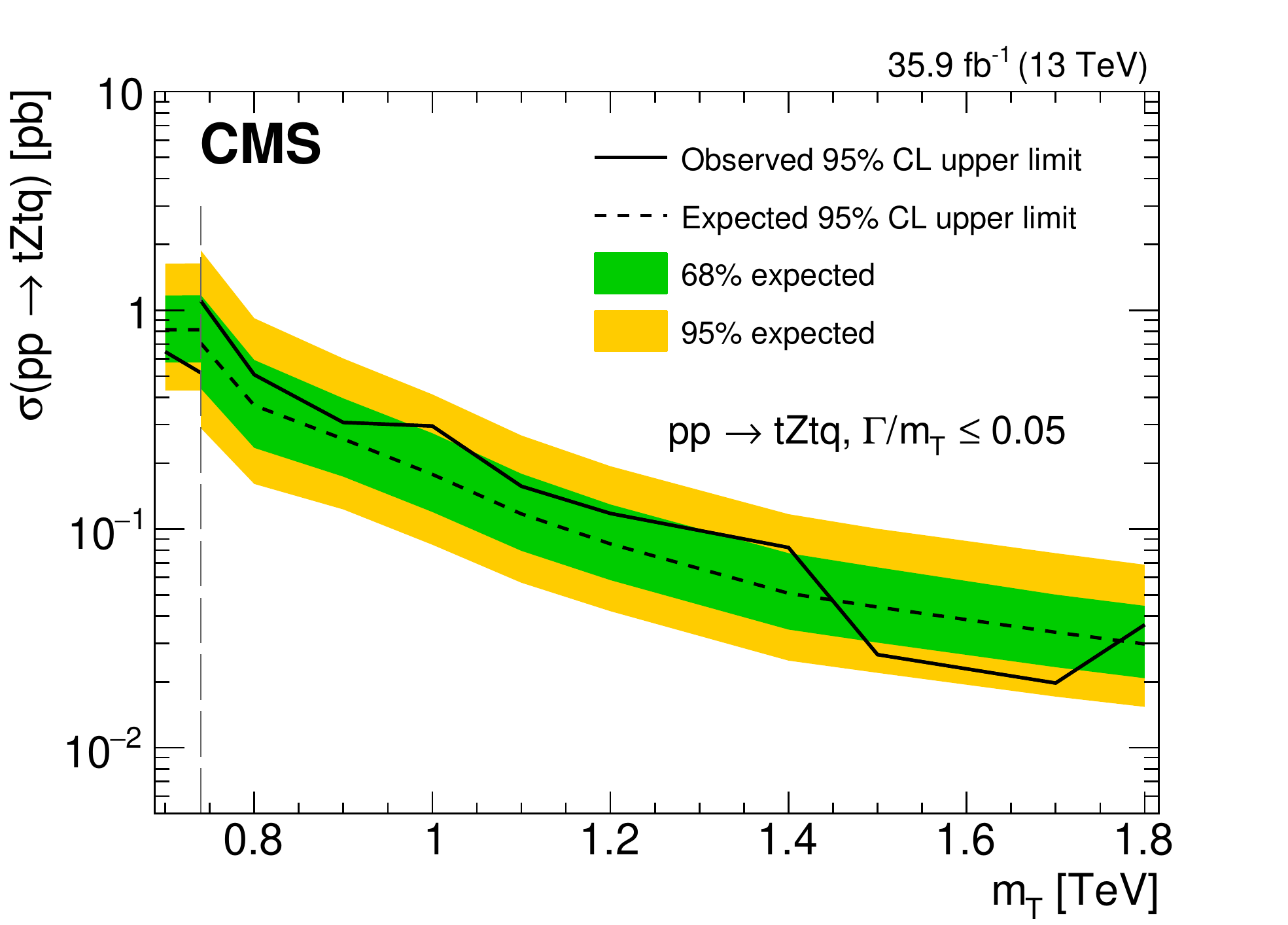}
    \includegraphics[width=0.49\textwidth]{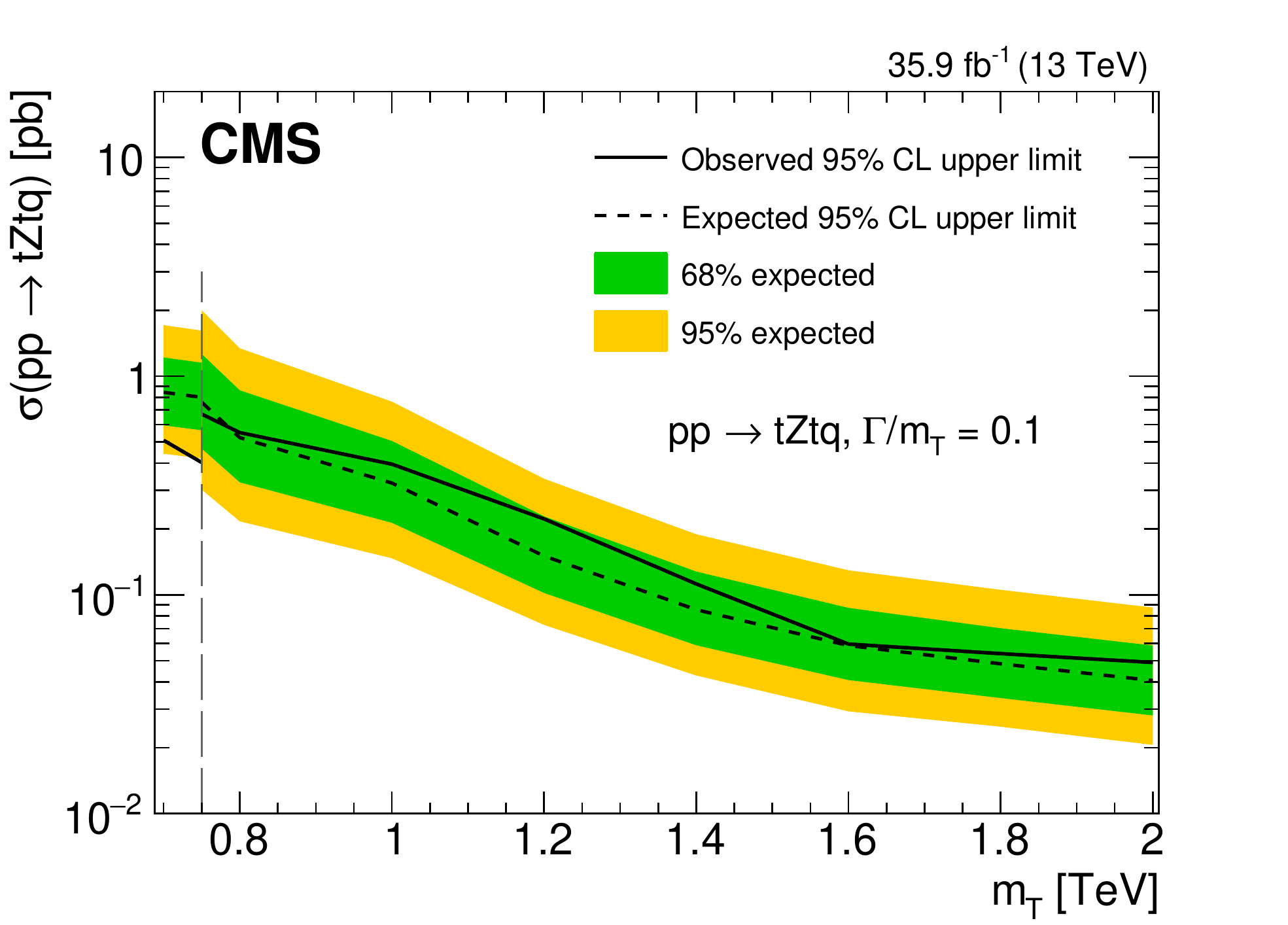}
    \includegraphics[width=0.49\textwidth]{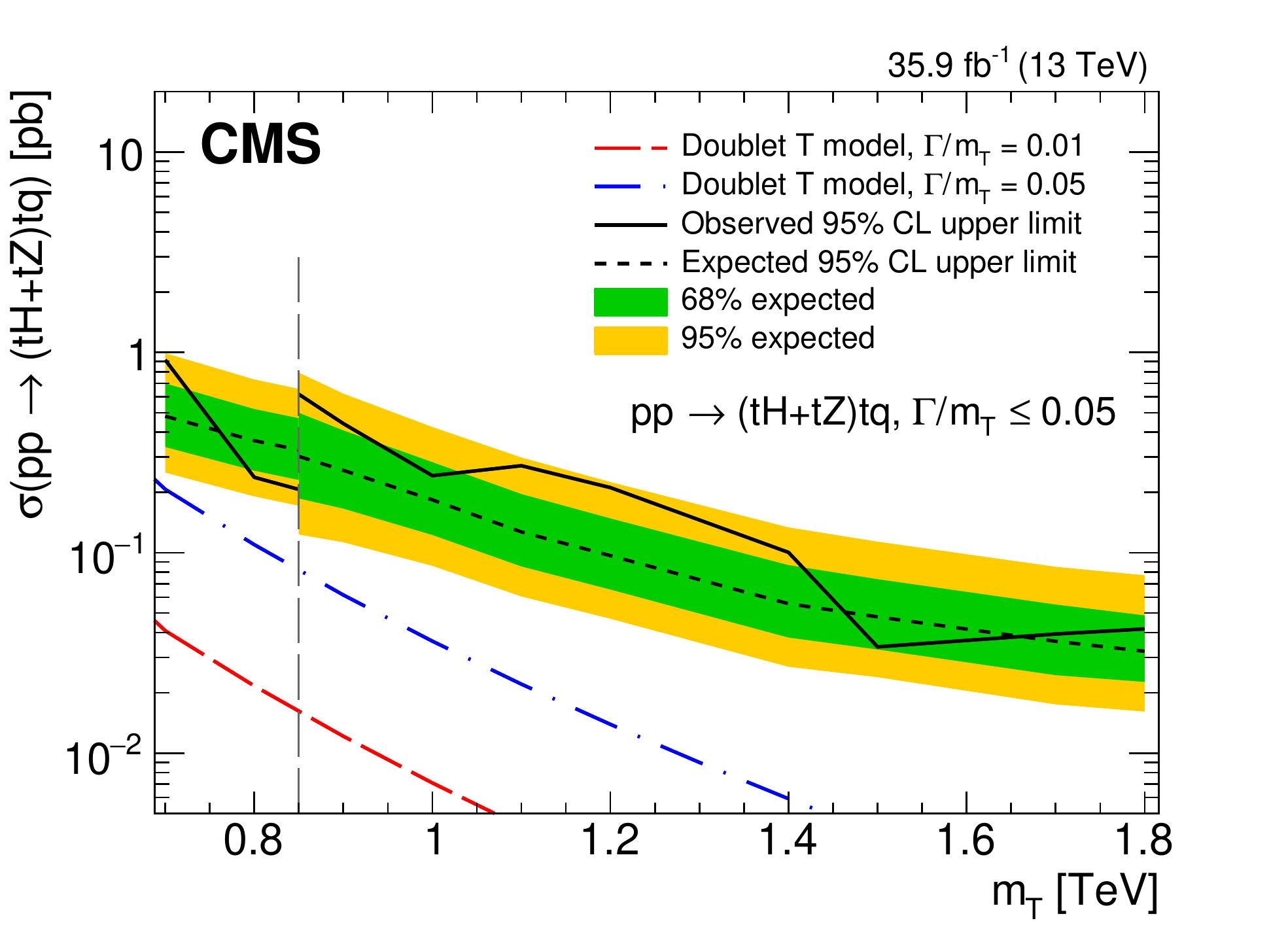}
    \includegraphics[width=0.49\textwidth]{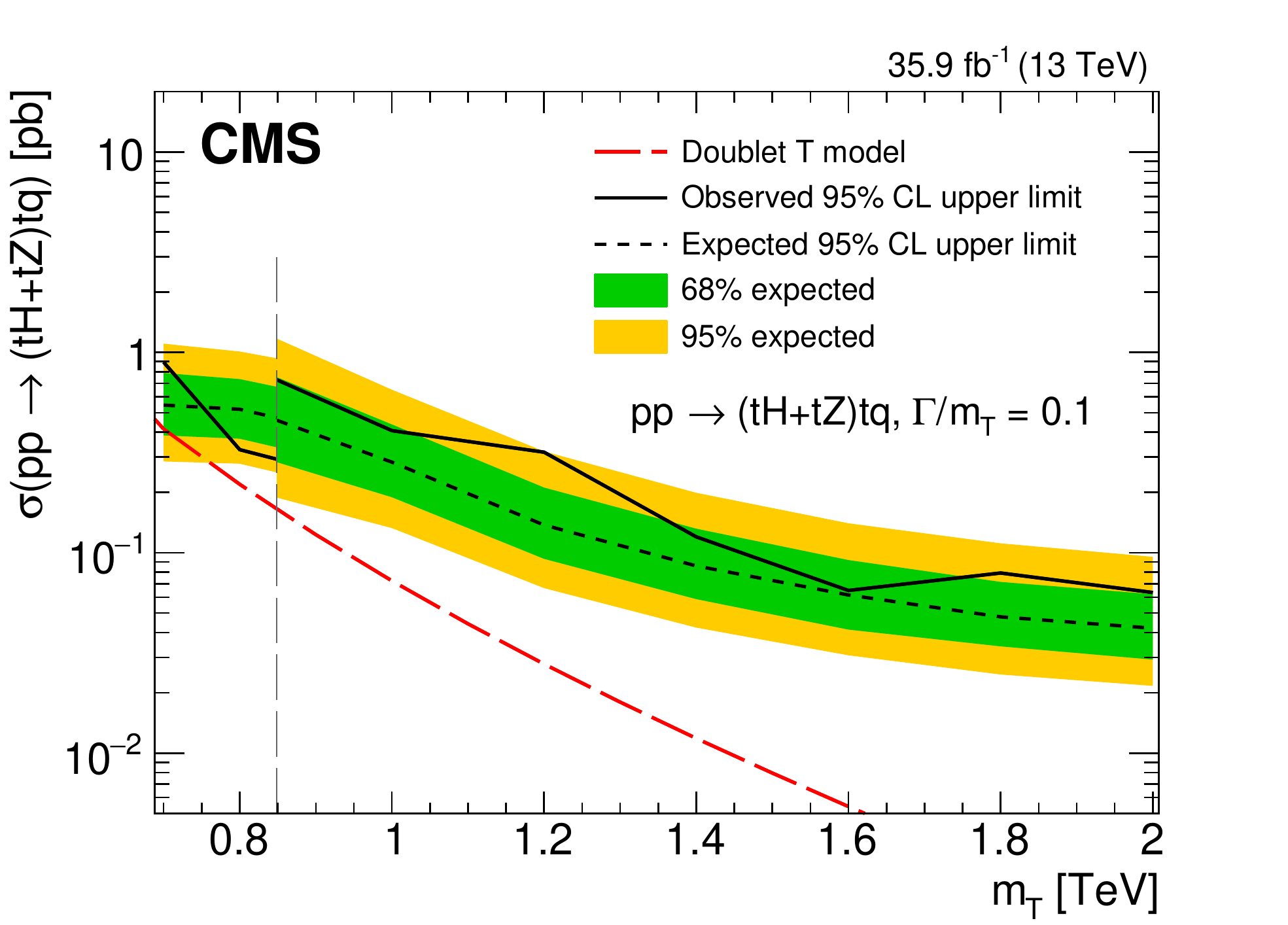}
    \caption{The observed and median expected upper limits at 95\%~\CL on the cross sections for production
associated with a top quark for the \tHtq (upper row) and \tZtq (middle row) channels, and their sum, \tHZtq (lower row), for
different assumed values of the \PQT quark mass.
The inner (green) band and the outer (yellow) band indicate the regions containing 68 and 95\%, respectively,
of the distribution of limits expected under the background-only hypothesis.
The left column is for a narrow fractional width ($\GoM \le 0.05$) and the right column is for a fractional width of $\GoM = 0.1$.
The vertical dashed lines are
the crossover points in sensitivity that indicate the mass intervals used for presenting
the low-mass and high-mass search results.
The dashed red and dot-dashed blue curves are for the (\TB) doublet model.
Given the specified width, the couplings are implicit in the model.
Two curves corresponding to $\GoM = 0.05$ (dot-dashed blue)
and $\GoM = 0.01$ (dashed red) are shown for the narrow fractional width.}
    \label{fig:massLim_Ttq1}

\end{figure}

\begin{figure}[!htb]
  \centering
    \includegraphics[width=0.49\textwidth]{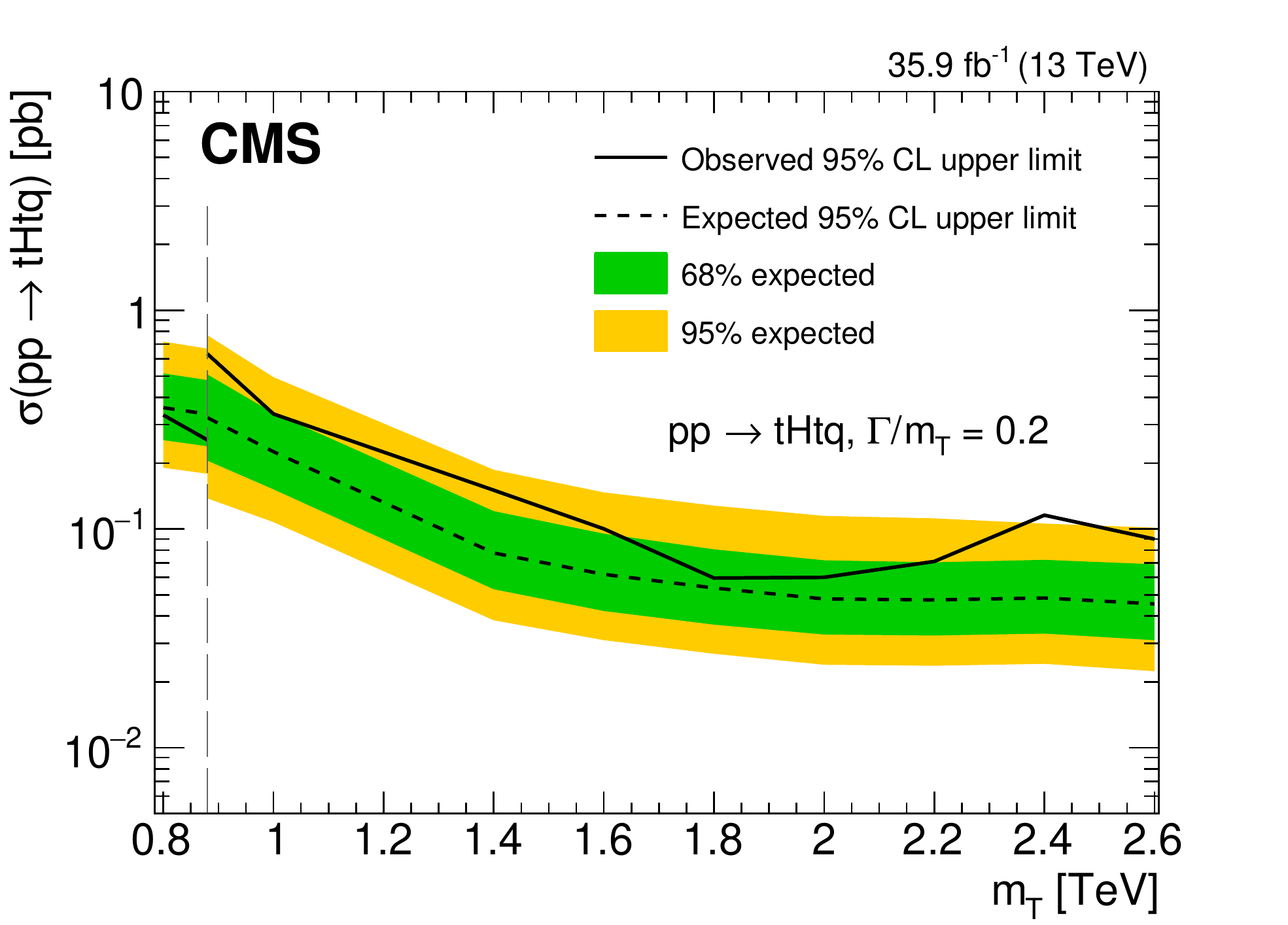}
    \includegraphics[width=0.49\textwidth]{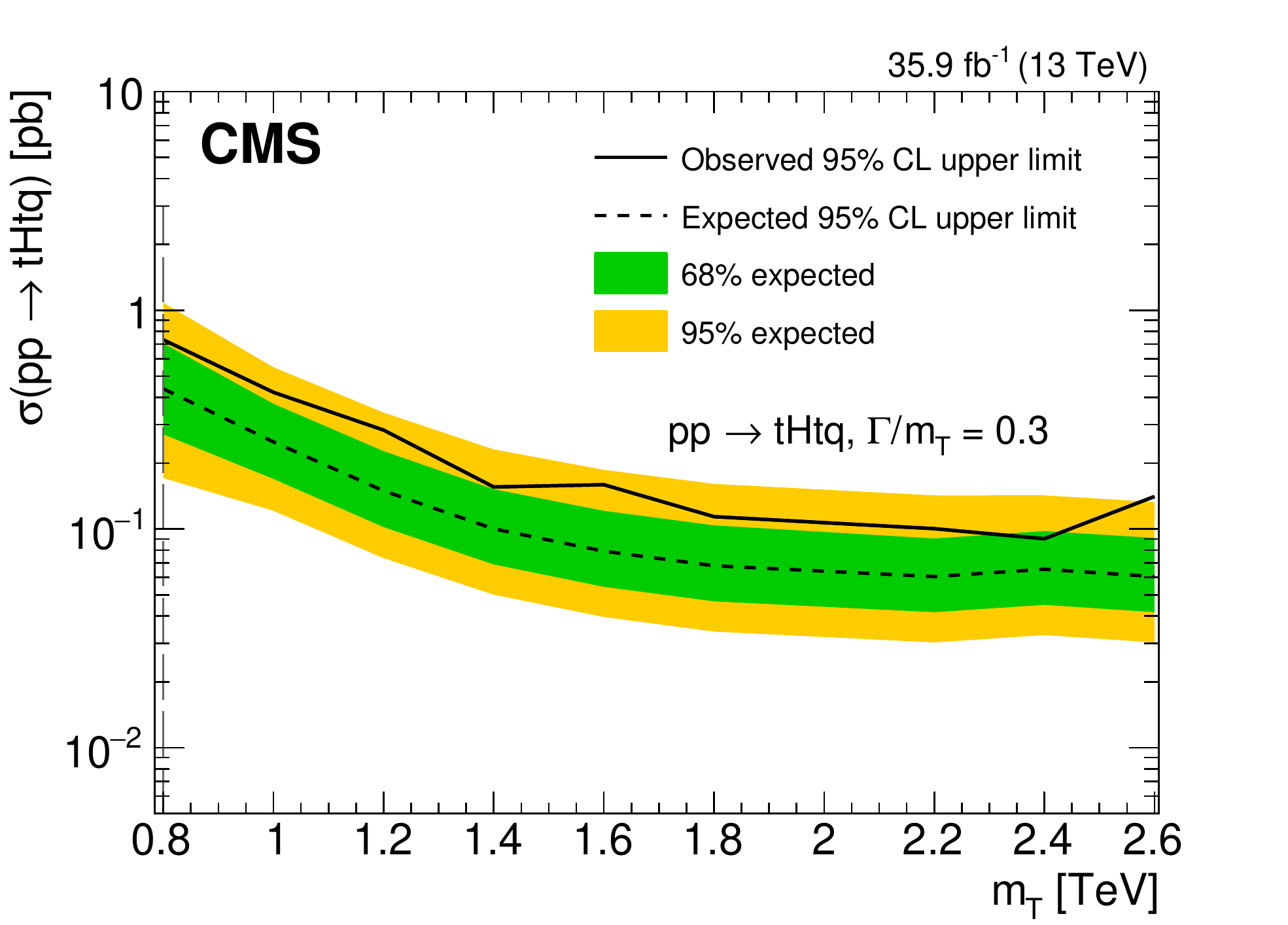}
    \includegraphics[width=0.49\textwidth]{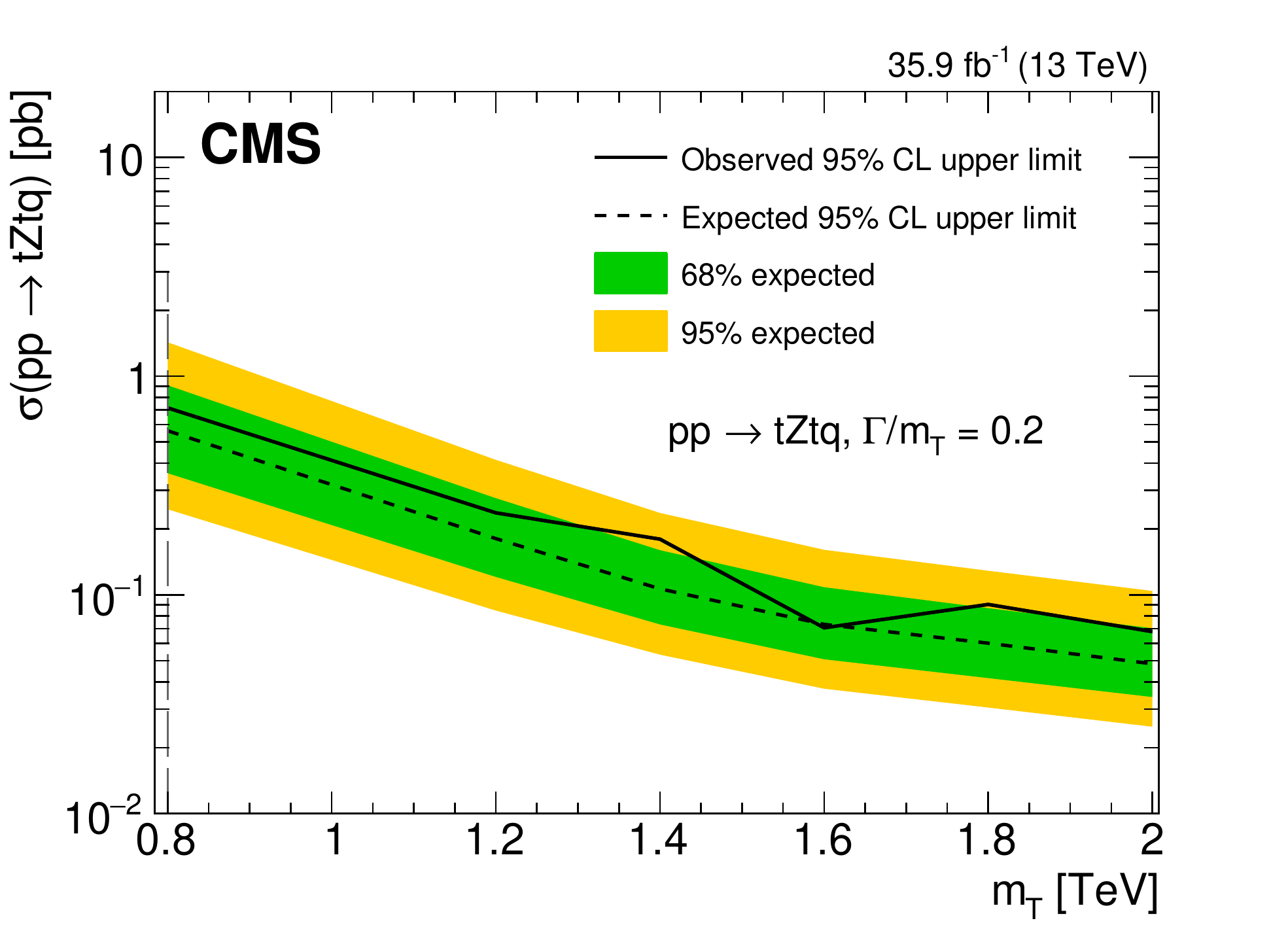}
    \includegraphics[width=0.49\textwidth]{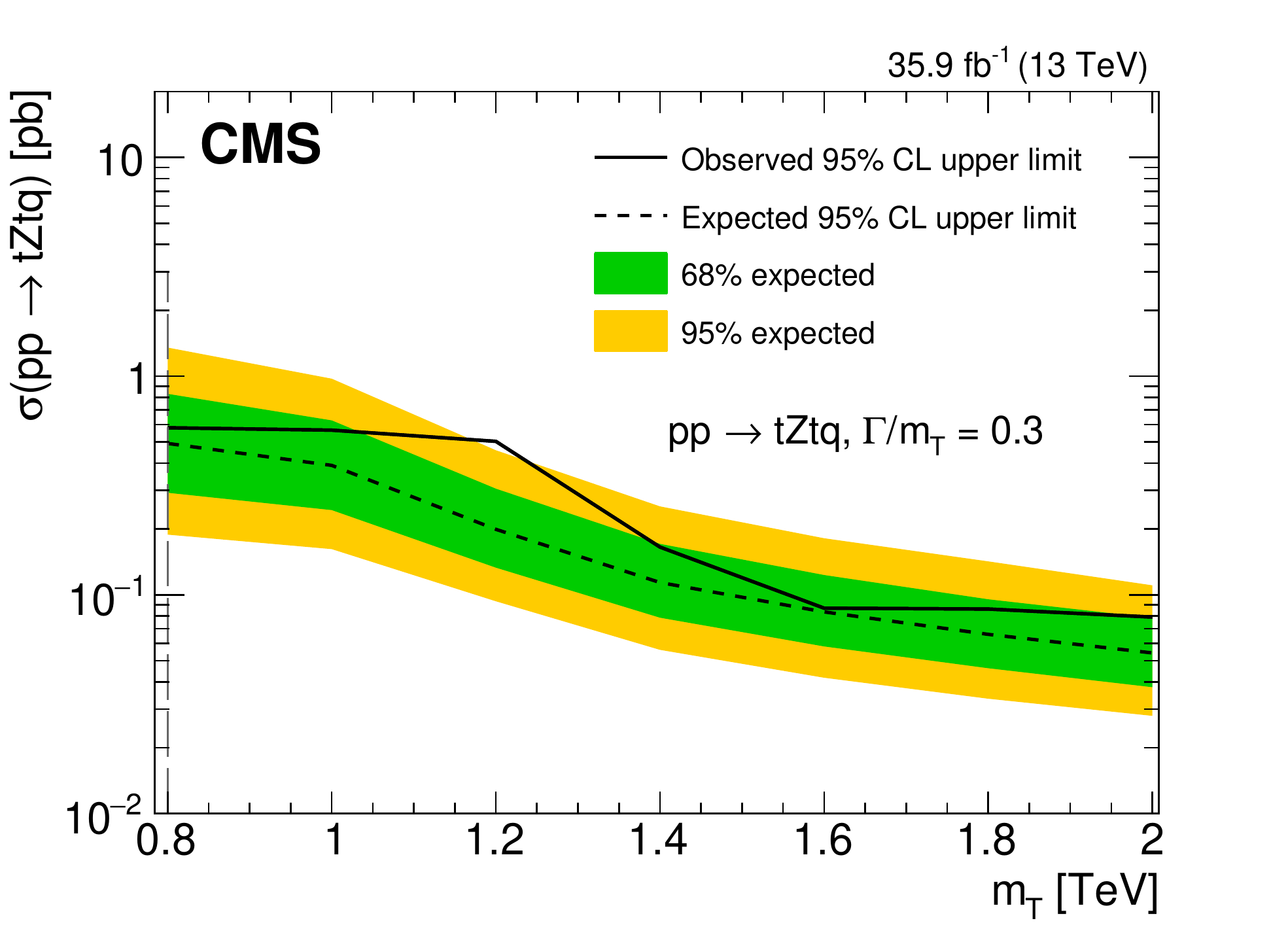}
    \includegraphics[width=0.49\textwidth]{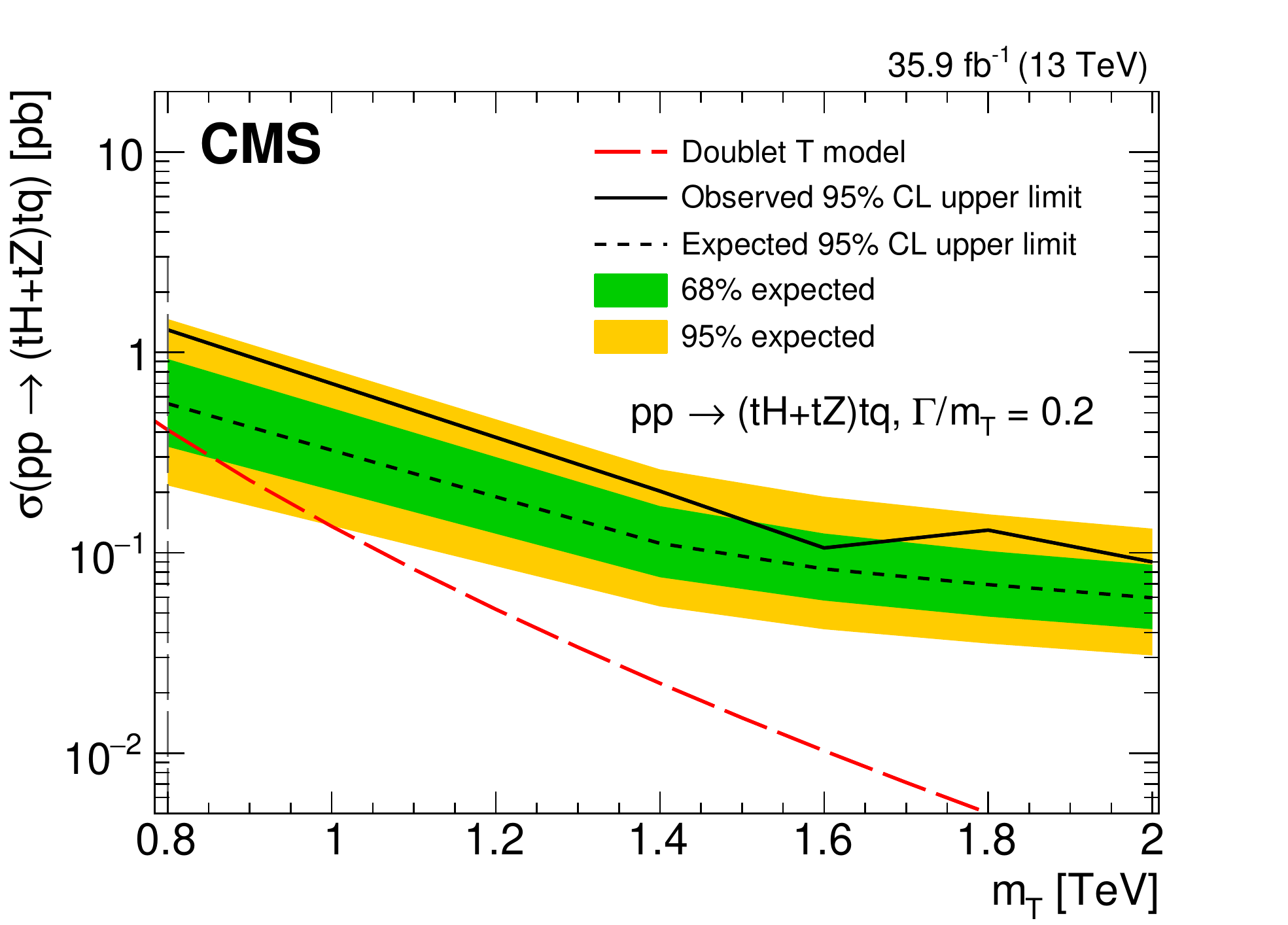}
    \includegraphics[width=0.49\textwidth]{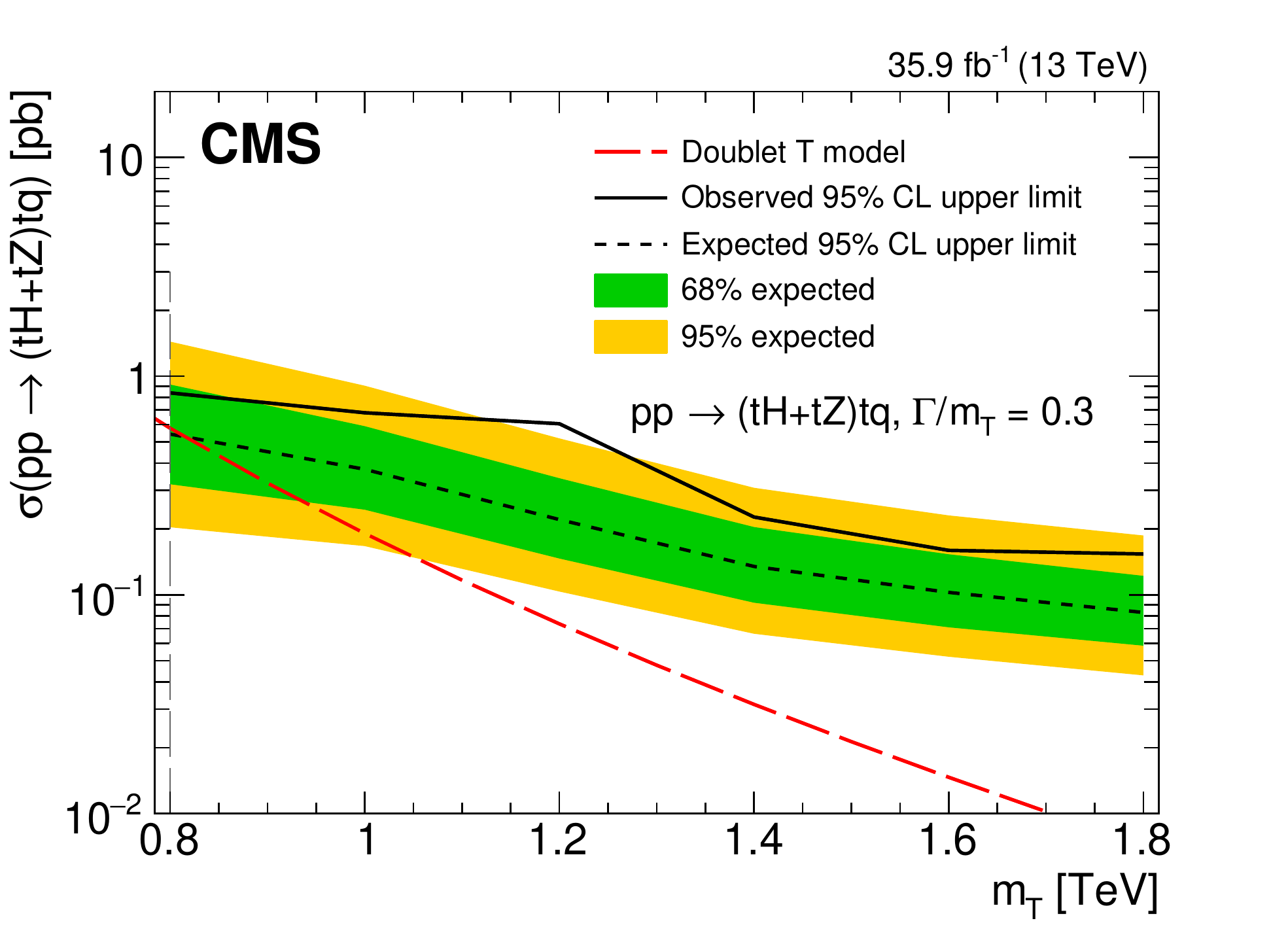}
    \caption{The observed and median expected upper limits at 95\%~\CL on the cross sections for production
associated with a top quark for the \tHtq (upper row) and \tZtq (middle row) channels, and their sum, \tHZtq (lower row), for
different assumed values of the \PQT quark mass.
The inner (green) band and the outer (yellow) band indicate the regions containing 68 and 95\%, respectively,
of the distribution of limits expected under the background-only hypothesis.
The left column is for a fractional width of 20\% and the right column is for a fractional width of 30\%.
The vertical dashed lines are
the crossover points in sensitivity that indicate the mass intervals used for presenting
the low-mass and high-mass search results.
The dashed red curves are for the (\TB) doublet model. Given the specified width, the couplings are implicit in the model.}
    \label{fig:massLim_Ttq2}

\end{figure}

Figure~\ref{fig:massLim_Tbq1} shows the cross section upper limits for
\Tbqonly production for the \tHbq and \tZbq channels, and their sum.
The figure includes results for a narrow fractional width,
corresponding to $\GoM \le 0.05$, and for a fractional width of 10\%.
Results for the same quantities are shown in Fig.~\ref{fig:massLim_Tbq2} for
fractional widths of 20 and 30\%.
Similar results for \Ttqonly production for the \tHtq and \tZtq channels, and their sum
are shown in
Fig.~\ref{fig:massLim_Ttq1} for narrow fractional width and
for 10\% fractional width,
and the corresponding results for fractional
widths of 20 and 30\% are shown in Fig.~\ref{fig:massLim_Ttq2}.
Superimposed on these results
are the expected cross sections for the \PQT singlet model and
for the (\TB) doublet model.
The \tH and \tZ branching fractions for a narrow width
are both approximately 25\% for the \PQT singlet model
and 50\% for the (\TB) doublet model.

The presented results are evaluated using left-handed chirality
for the \PQT in the \Tbq cases and right-handed
chirality for the \Ttq cases.
Studies with the opposite chirality
for narrow width \tprimetotH and \tprimetotZ have shown similar
sensitivity with the differences being small for the low-mass search and
at most 10\% for the high-mass search.

The results of this search show that the observed limits are consistent with the expected
limits arising from the background-only hypothesis.
Depending on the fractional width, it can be seen that this search
has expected sensitivity
for \PQT masses within the \PQT singlet model up to 1.28\TeV (for \tHZbq with 30\% fractional width).
For \PQT masses below 1\TeV, the models of the associated production with
a bottom quark are strongly constrained by the observed limits from
the low-mass search signature, which are generally more
stringent than expected above 0.75\TeV;
for the \PQT singlet model masses in the range 0.70 to 1.00\TeV
are excluded at 95\% \CL for some fractional widths between 5 and 30\%.
For the \PQT quark masses above 1.00\TeV, the observed limits are above model predictions,
and so no exclusion at 95\% \CL is possible with this data set in this
mass range. The models with the associated production with a top quark
have lower cross sections with a median expected sensitivity for \PQT quark masses within
the (\TB) doublet model of 0.82\TeV for the largest fractional
width of 30\%. However, for this model, no range of masses is excluded at 95\% \CL for any
of the masses and fractional widths considered here.

The presented experimental upper limits on the cross sections are of general
interest in more model independent approaches and demonstrate
the great potential of electroweak single production searches to
test vector-like quark production at mass scales far beyond
those accessible with pair production.
\clearpage
\section{Summary}\label{sec:summary}
A search for a vector-like top quark partner \PQT in the electroweak
single production mode with fully hadronic final states has been performed
using $\Pp\Pp$ collision events at $\sqrt{s} = 13\TeV$ collected by the CMS
experiment in 2016.
The data sample corresponds to an integrated luminosity of 35.9\fbinv.
The \PQT quarks  are assumed to couple only to the
standard model third-generation quarks.
The decay channels exploited are \tprimetotH and \tprimetotZ with the hadronic
decay of the top quark and primarily the $\bbbar$ decay
of the Higgs and \PZ bosons.
This search is designed to be sensitive to \PQT quark fractional widths
of up to 30\% and a wide range of masses.
The background is
mostly due to standard model \ttjets and QCD multijet events with some
contributions from \wjets processes.
No significant excess of data above the standard model background is observed
and upper limits at 95\% confidence level
are set on  $\sigma \:\mathcal{B}(\tprimetotH)$
and $\sigma \:\mathcal{B}(\tprimetotZ)$, which vary
between 2\unit{pb} and 20\unit{fb} for \PQT masses ranging
from 0.6 to 2.6\TeV in the \Tbqonly and \Ttqonly production channels.
Results from combining the two decay channels
assuming equal couplings are also reported.
Compared with prior electroweak single production searches,
this search is significantly more sensitive for \tprimetotH.
The search gives the first constraints using this production mode
on \tprimetotZ for hadronic decays of the \PZ boson. These results
are competitive with those from searches for \tprimetotZ using other \PZ decay modes.
The combined \tprimetotH and \tprimetotZ results for associated production with
a bottom quark lead to constraints
on \PQT quarks in the \PQT singlet model for masses below 1.00\TeV.
The expected sensitivity for this model extends to
1.28\TeV (for 30\% fractional width), which is
comparable to the mass reach of the most stringent pair production searches.

\begin{acknowledgments}
We congratulate our colleagues in the CERN accelerator departments for the excellent performance of the LHC and thank the technical and administrative staffs at CERN and at other CMS institutes for their contributions to the success of the CMS effort. In addition, we gratefully acknowledge the computing centers and personnel of the Worldwide LHC Computing Grid for delivering so effectively the computing infrastructure essential to our analyses. Finally, we acknowledge the enduring support for the construction and operation of the LHC and the CMS detector provided by the following funding agencies: BMBWF and FWF (Austria); FNRS and FWO (Belgium); CNPq, CAPES, FAPERJ, FAPERGS, and FAPESP (Brazil); MES (Bulgaria); CERN; CAS, MoST, and NSFC (China); COLCIENCIAS (Colombia); MSES and CSF (Croatia); RPF (Cyprus); SENESCYT (Ecuador); MoER, ERC IUT, PUT and ERDF (Estonia); Academy of Finland, MEC, and HIP (Finland); CEA and CNRS/IN2P3 (France); BMBF, DFG, and HGF (Germany); GSRT (Greece); NKFIA (Hungary); DAE and DST (India); IPM (Iran); SFI (Ireland); INFN (Italy); MSIP and NRF (Republic of Korea); MES (Latvia); LAS (Lithuania); MOE and UM (Malaysia); BUAP, CINVESTAV, CONACYT, LNS, SEP, and UASLP-FAI (Mexico); MOS (Montenegro); MBIE (New Zealand); PAEC (Pakistan); MSHE and NSC (Poland); FCT (Portugal); JINR (Dubna); MON, RosAtom, RAS, RFBR, and NRC KI (Russia); MESTD (Serbia); SEIDI, CPAN, PCTI, and FEDER (Spain); MOSTR (Sri Lanka); Swiss Funding Agencies (Switzerland); MST (Taipei); ThEPCenter, IPST, STAR, and NSTDA (Thailand); TUBITAK and TAEK (Turkey); NASU and SFFR (Ukraine); STFC (United Kingdom); DOE and NSF (USA).

\hyphenation{Rachada-pisek} Individuals have received support from the Marie-Curie program and the European Research Council and Horizon 2020 Grant, contract Nos.\ 675440, 752730, and 765710 (European Union); the Leventis Foundation; the A.P.\ Sloan Foundation; the Alexander von Humboldt Foundation; the Belgian Federal Science Policy Office; the Fonds pour la Formation \`a la Recherche dans l'Industrie et dans l'Agriculture (FRIA-Belgium); the Agentschap voor Innovatie door Wetenschap en Technologie (IWT-Belgium); the F.R.S.-FNRS and FWO (Belgium) under the ``Excellence of Science -- EOS" -- be.h project n.\ 30820817; the Beijing Municipal Science \& Technology Commission, No. Z181100004218003; the Ministry of Education, Youth and Sports (MEYS) of the Czech Republic; the Lend\"ulet (``Momentum") Program and the J\'anos Bolyai Research Scholarship of the Hungarian Academy of Sciences, the New National Excellence Program \'UNKP, the NKFIA research grants 123842, 123959, 124845, 124850, 125105, 128713, 128786, and 129058 (Hungary); the Council of Science and Industrial Research, India; the HOMING PLUS program of the Foundation for Polish Science, cofinanced from European Union, Regional Development Fund, the Mobility Plus program of the Ministry of Science and Higher Education, the National Science Center (Poland), contracts Harmonia 2014/14/M/ST2/00428, Opus 2014/13/B/ST2/02543, 2014/15/B/ST2/03998, and 2015/19/B/ST2/02861, Sonata-bis 2012/07/E/ST2/01406; the National Priorities Research Program by Qatar National Research Fund; the Ministry of Science and Education, grant no. 3.2989.2017 (Russia); the Programa Estatal de Fomento de la Investigaci{\'o}n Cient{\'i}fica y T{\'e}cnica de Excelencia Mar\'{\i}a de Maeztu, grant MDM-2015-0509 and the Programa Severo Ochoa del Principado de Asturias; the Thalis and Aristeia programs cofinanced by EU-ESF and the Greek NSRF; the Rachadapisek Sompot Fund for Postdoctoral Fellowship, Chulalongkorn University and the Chulalongkorn Academic into Its 2nd Century Project Advancement Project (Thailand); the Nvidia Corporation; the Welch Foundation, contract C-1845; and the Weston Havens Foundation (USA).
\end{acknowledgments}
\bibliography{auto_generated}
\clearpage
\appendix
\section{Low-mass and high-mass search limits}\label{sec:supplementary}
For information, we show here separately the limits obtained with
each of the two search signatures that were used to give
the final search limit results presented in
Figs.~\ref{fig:massLim_Tbq1}--\ref{fig:massLim_Ttq2} of the paper.

\begin{figure}[!htb]
  \centering
    \includegraphics[width=0.49\textwidth]{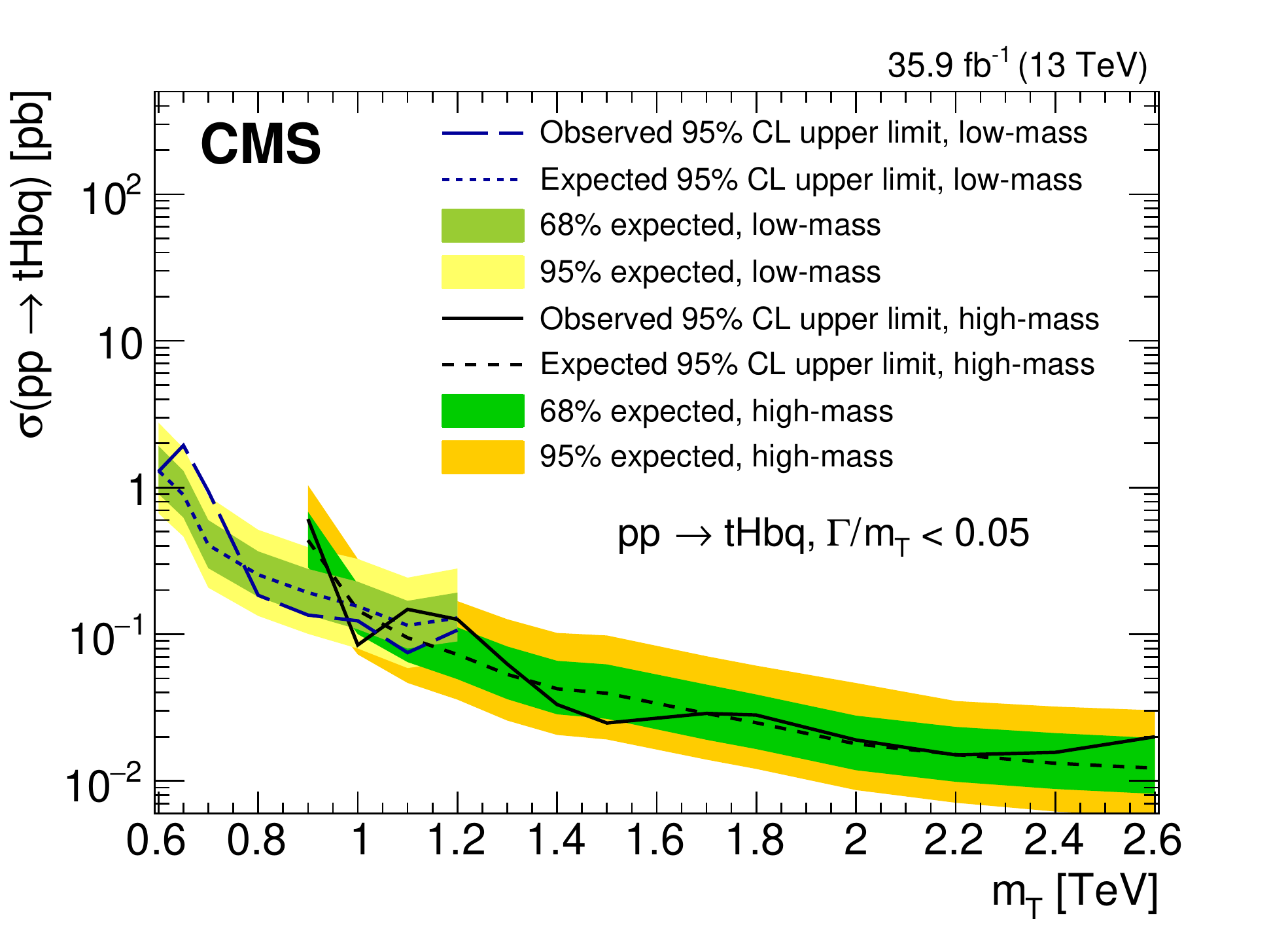}
    \includegraphics[width=0.49\textwidth]{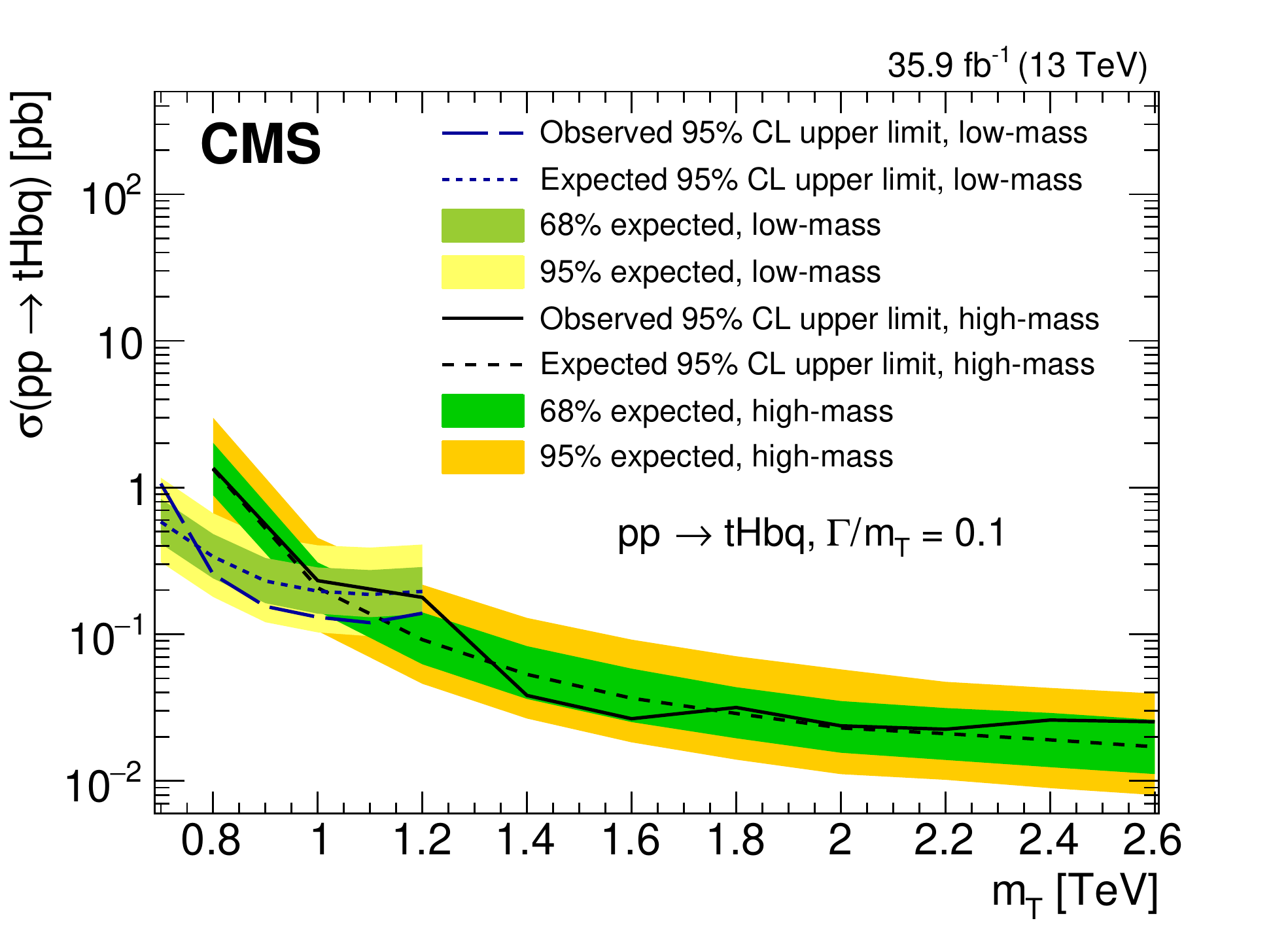}
    \includegraphics[width=0.49\textwidth]{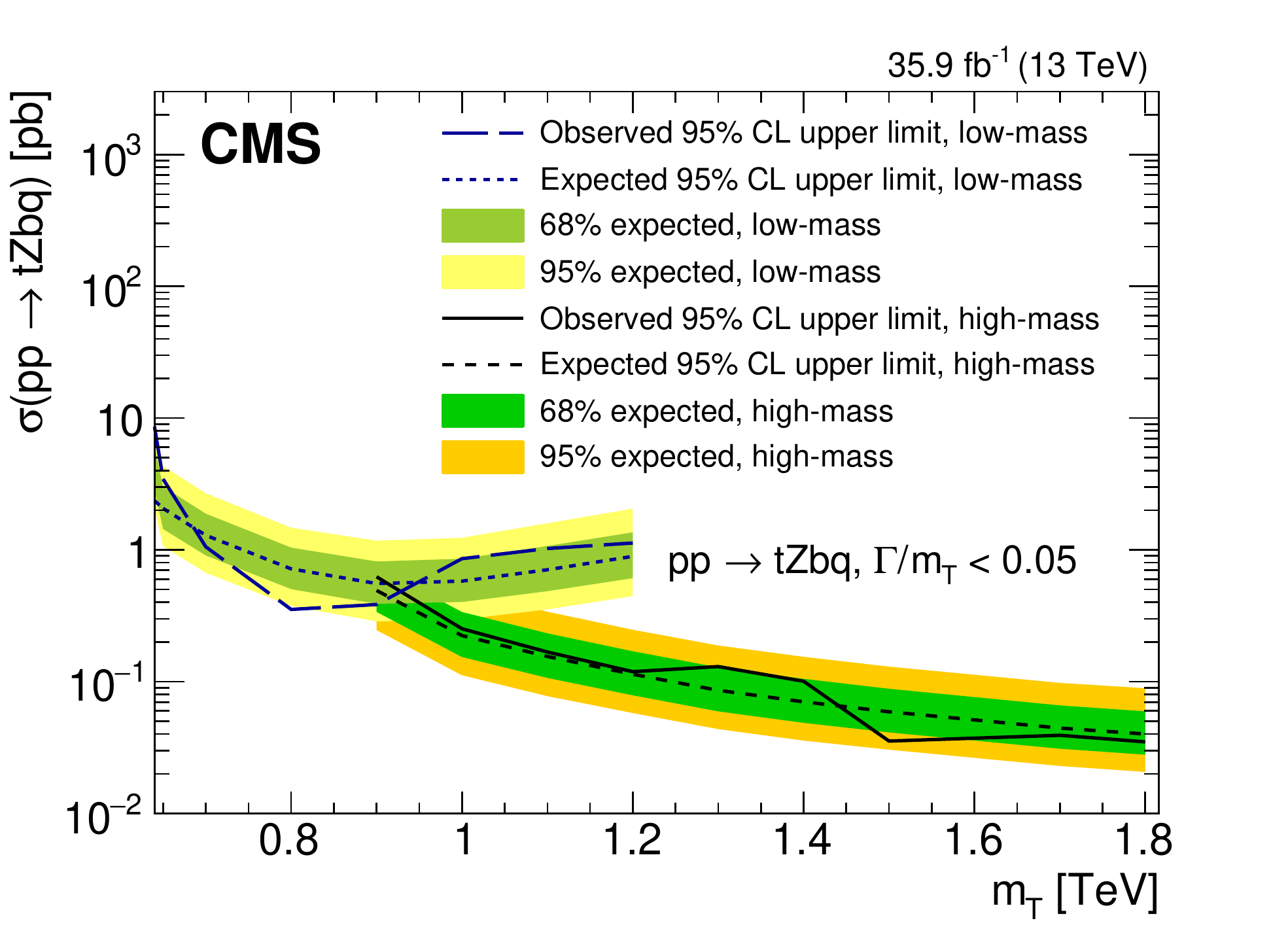}
    \includegraphics[width=0.49\textwidth]{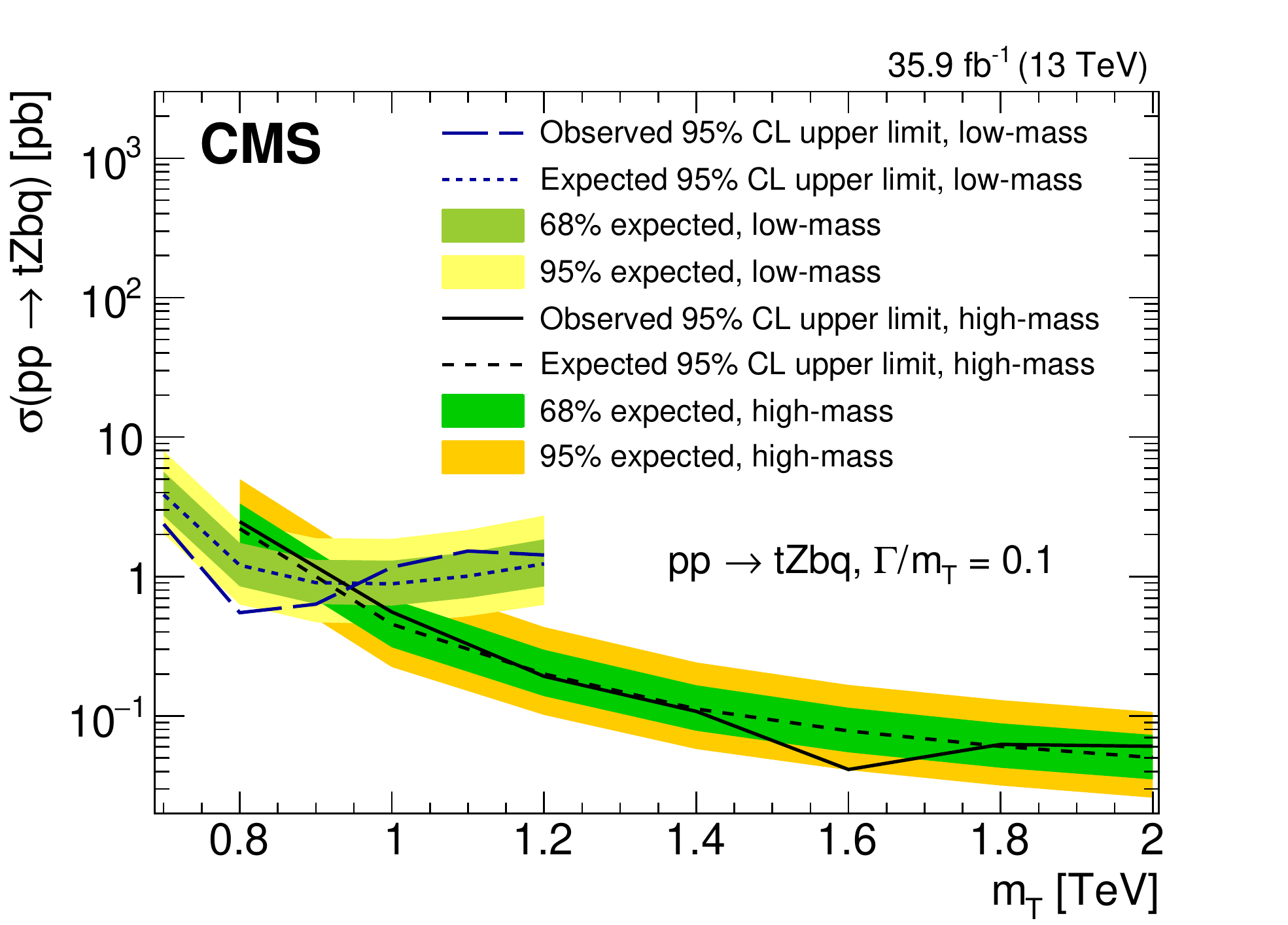}
    \includegraphics[width=0.49\textwidth]{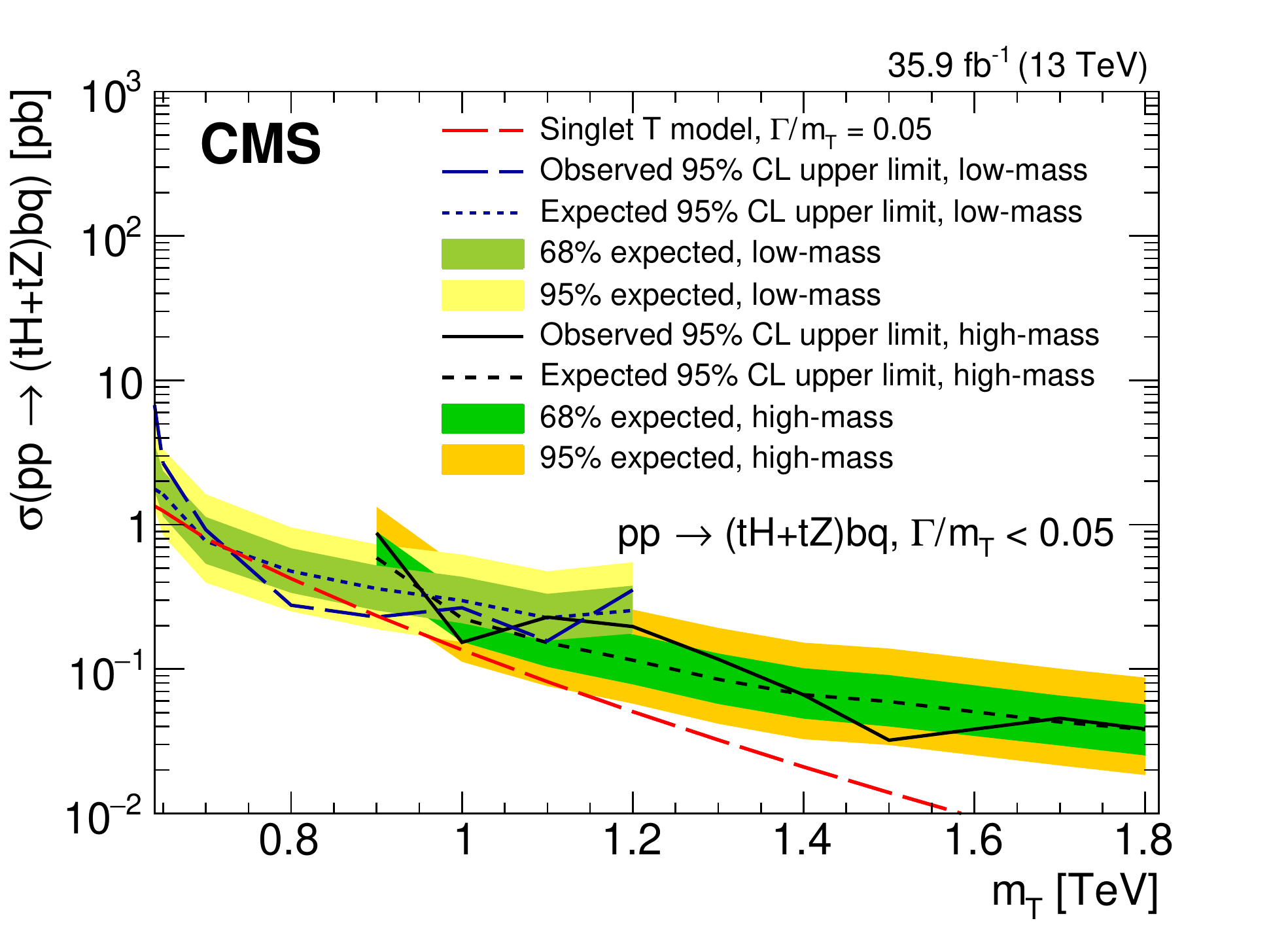}
    \includegraphics[width=0.49\textwidth]{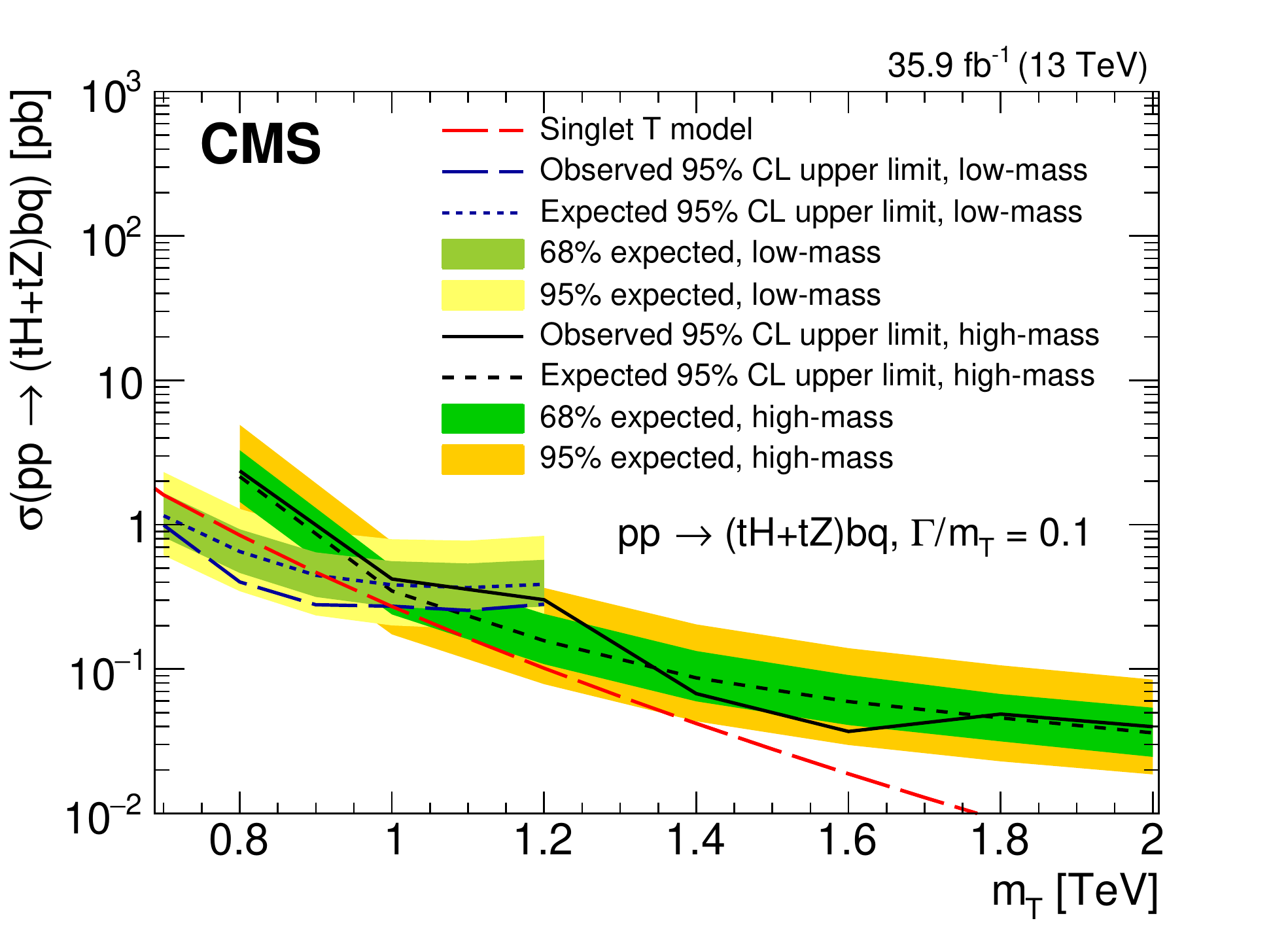}
    \caption{The observed and median expected upper limits at 95\%~\CL on the cross sections for production
associated with a bottom quark for the \tHbq (upper row) and \tZbq (middle row) channels, and their sum, \tHZbq (lower row), for
different assumed values of the \PQT quark mass.
The inner (green) bands and the outer (yellow) bands indicate the regions containing 68 and 95\%, respectively,
of the distribution of limits expected under the background-only hypothesis.
The left column is for a narrow fractional width ($\GoM \le 0.05$) and the right column is for a fractional width of $\GoM = 0.1$.
The dashed red curves are for the \PQT singlet model.
Given the specified width, the couplings are implicit in the model.}
    \label{fig2:massLim_Tbq1}

\end{figure}

\begin{figure}[!htb]
  \centering
    \includegraphics[width=0.49\textwidth]{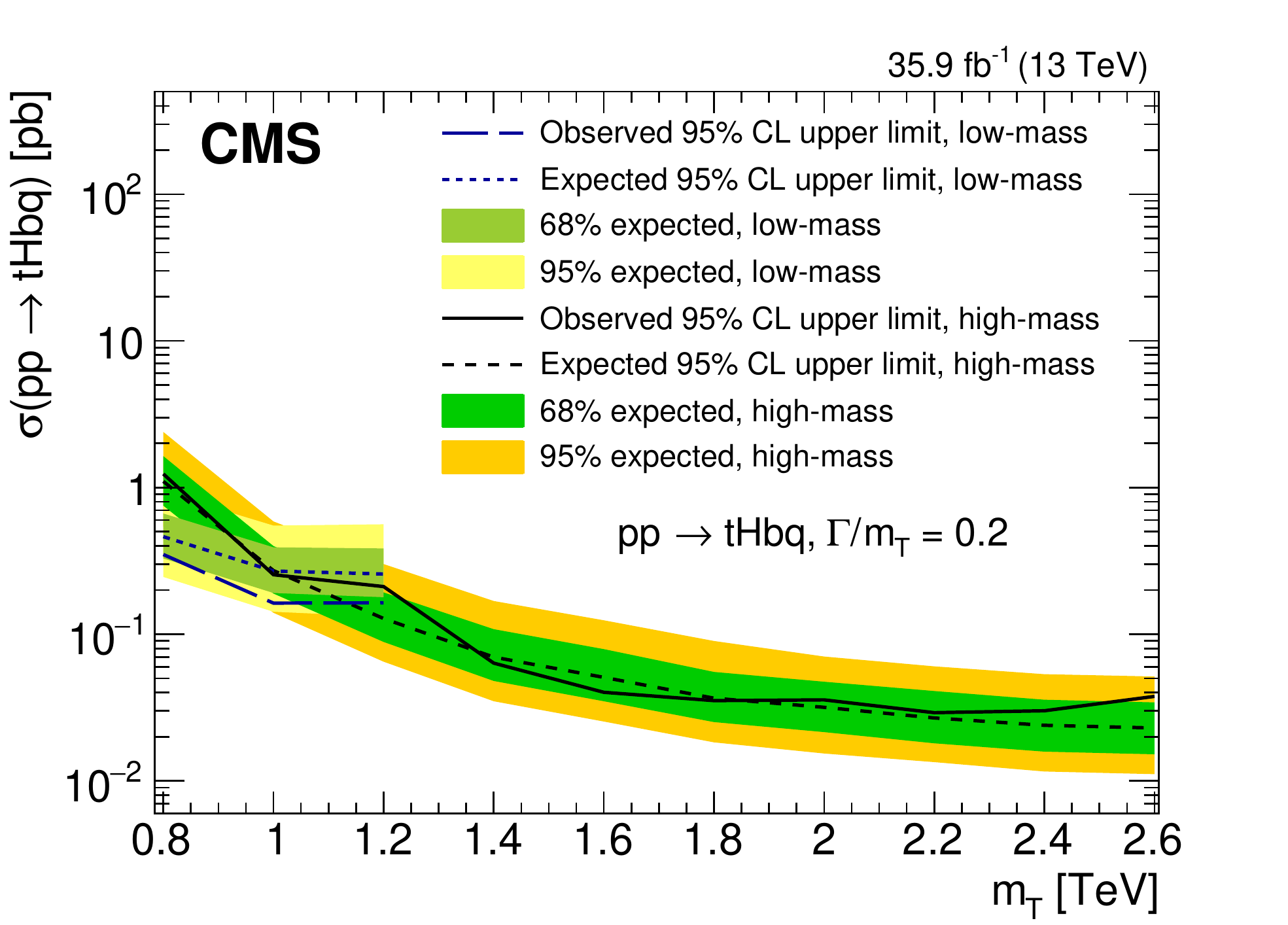}
    \includegraphics[width=0.49\textwidth]{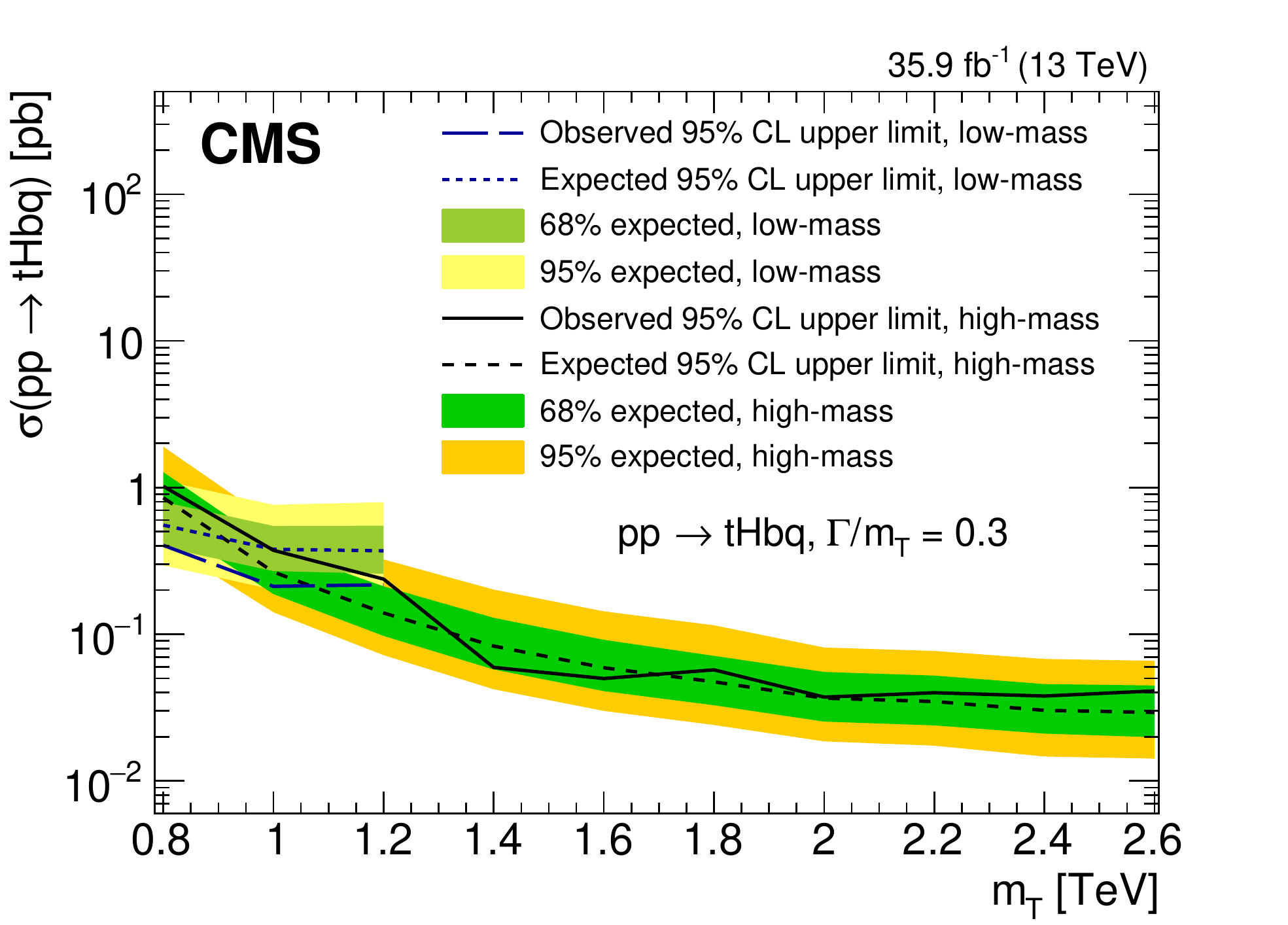}
    \includegraphics[width=0.49\textwidth]{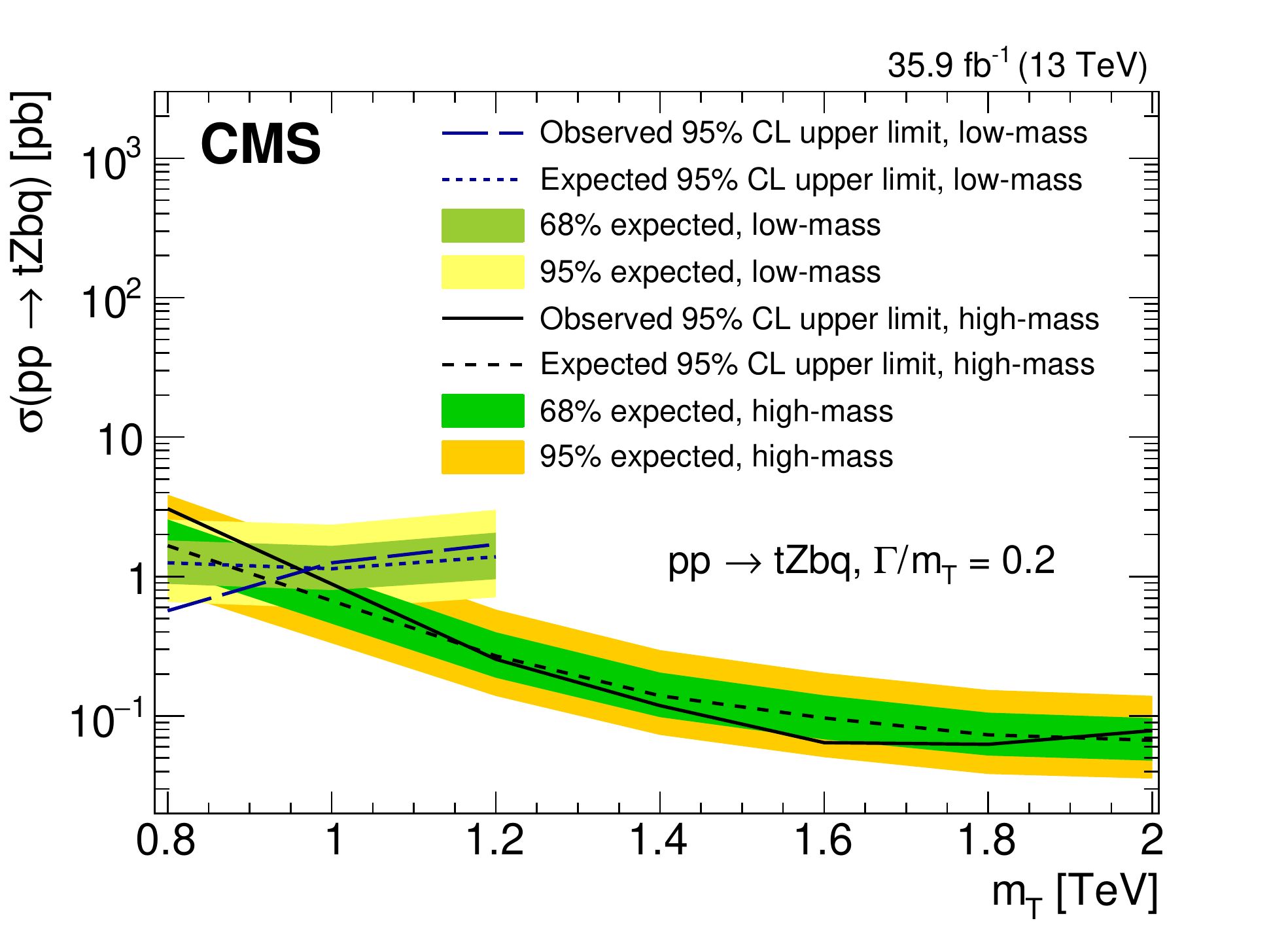}
    \includegraphics[width=0.49\textwidth]{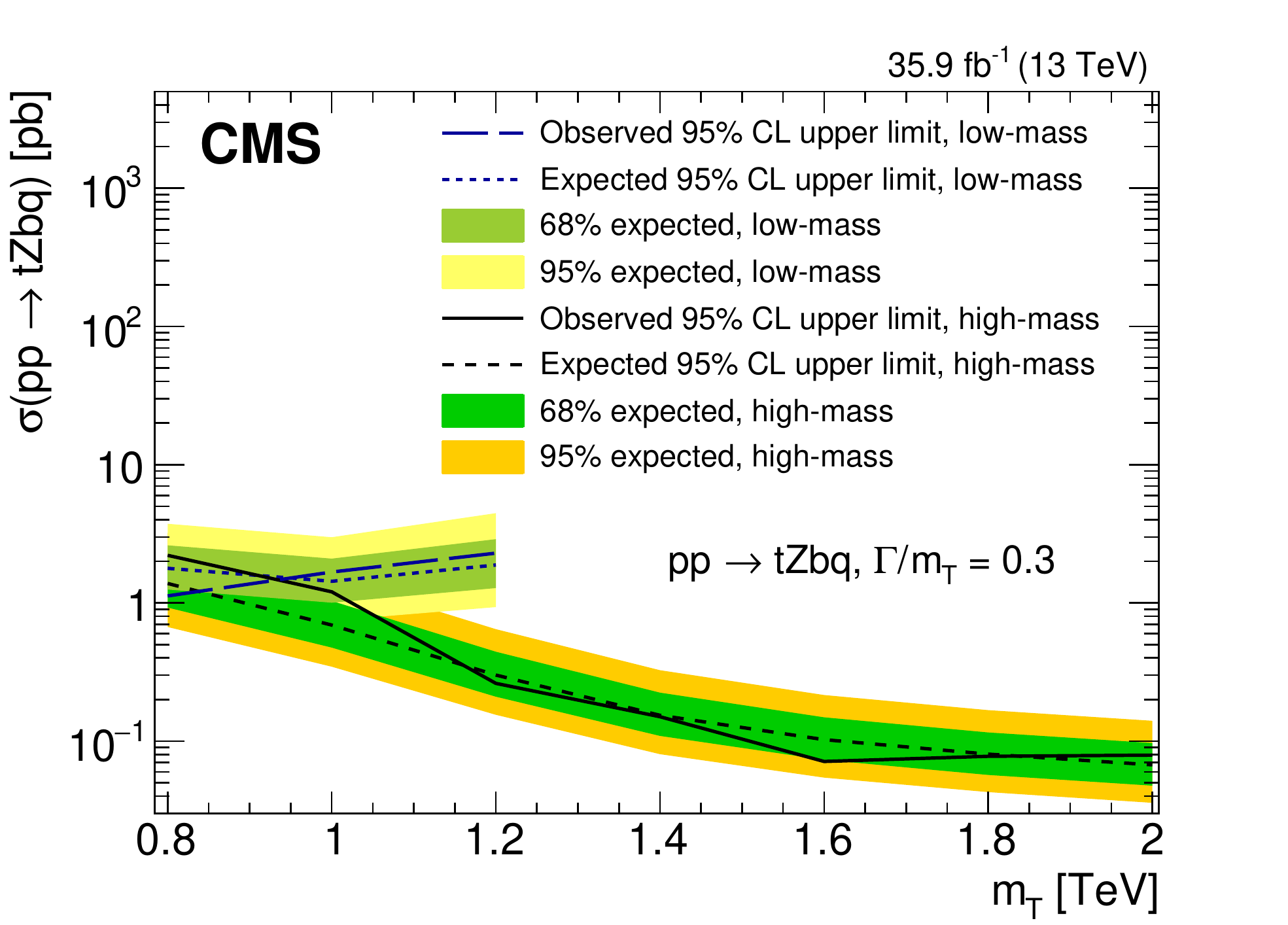}
    \includegraphics[width=0.49\textwidth]{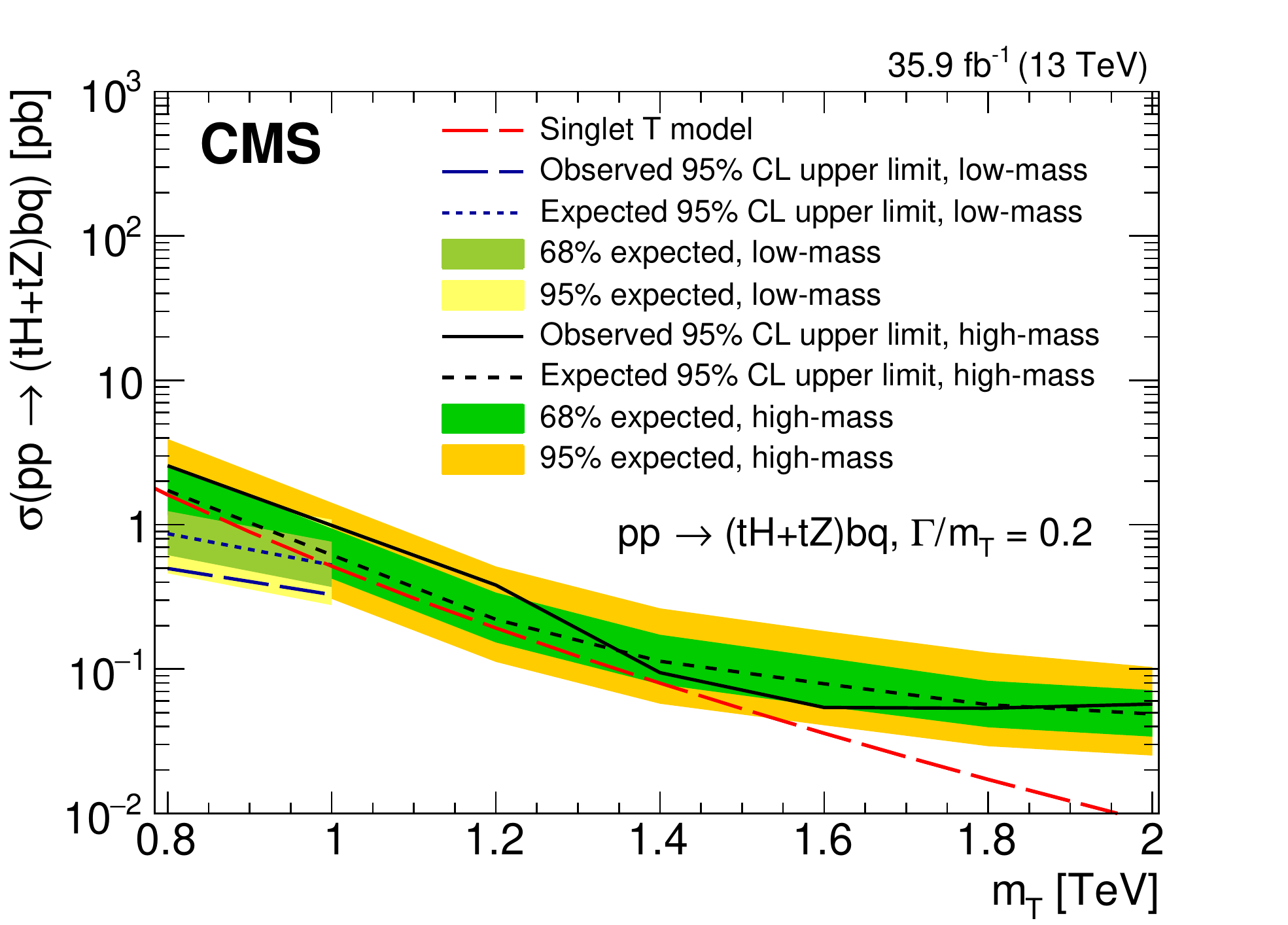}
    \includegraphics[width=0.49\textwidth]{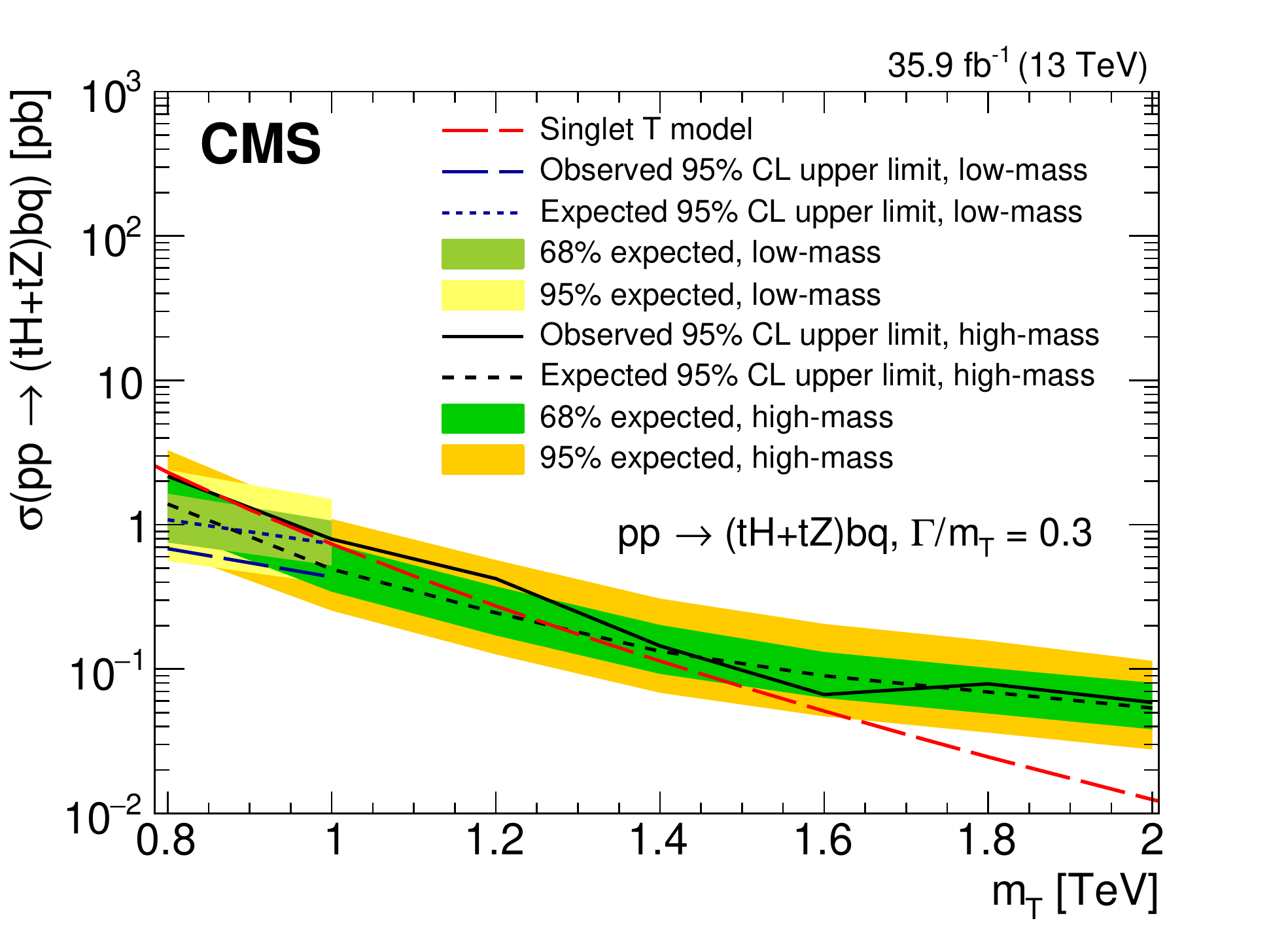}
    \caption{The observed and median expected upper limits at 95\%~\CL on the cross sections for production
associated with a bottom quark for the \tHbq (upper row) and \tZbq (middle row) channels, and their sum, \tHZbq (lower row), for
different assumed values of the \PQT quark mass.
The inner (green) bands and the outer (yellow) bands indicate the regions containing 68 and 95\%, respectively,
of the distribution of limits expected under the background-only hypothesis.
The left column is for a fractional width of 20\% and the right column is for a fractional width of 30\%.
The dashed red curves are for the \PQT singlet model.
Given the specified width, the couplings are implicit in the model.}
    \label{fig2:massLim_Tbq2}

\end{figure}

\begin{figure}[!htb]
  \centering
    \includegraphics[width=0.49\textwidth]{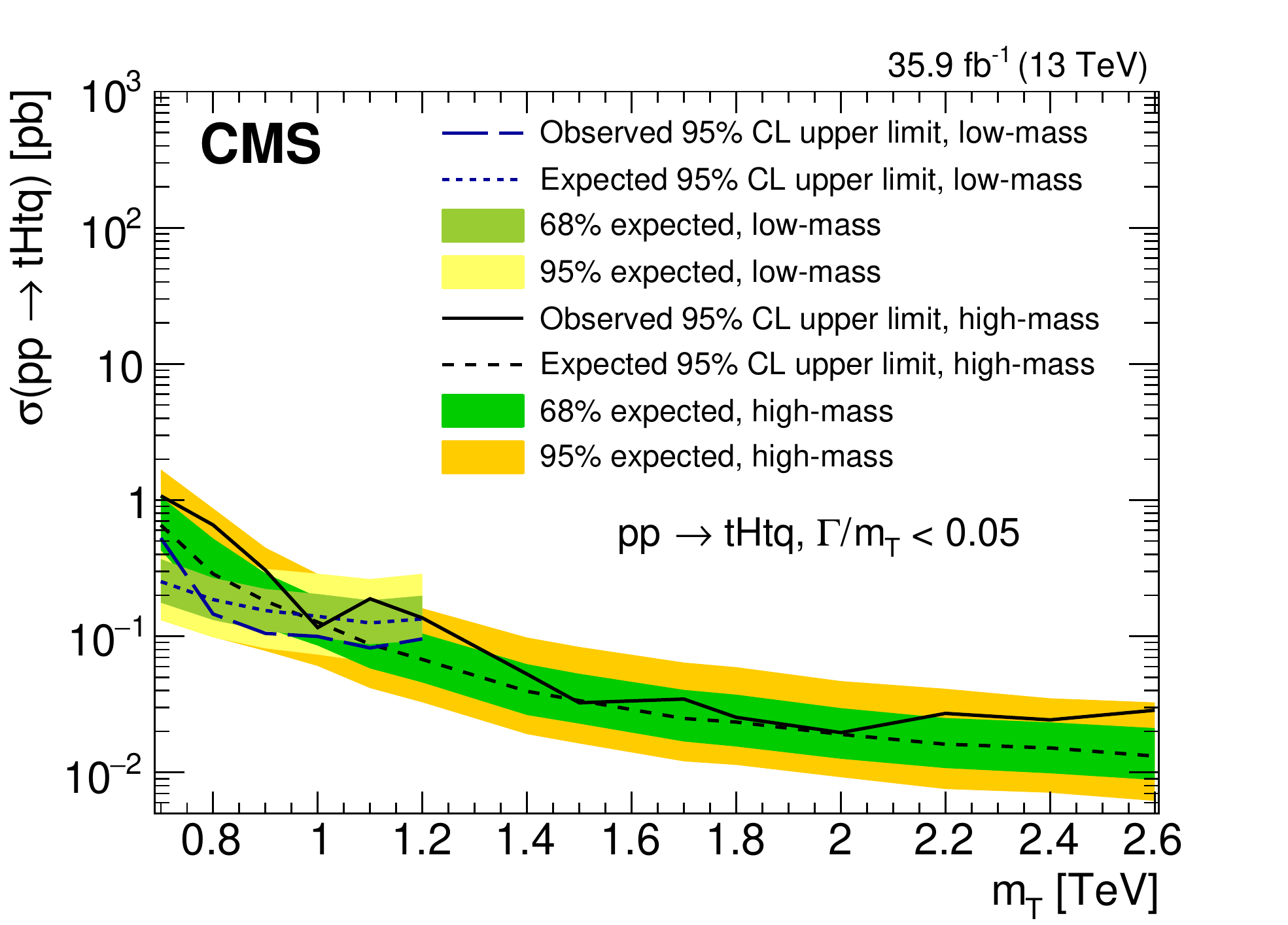}
    \includegraphics[width=0.49\textwidth]{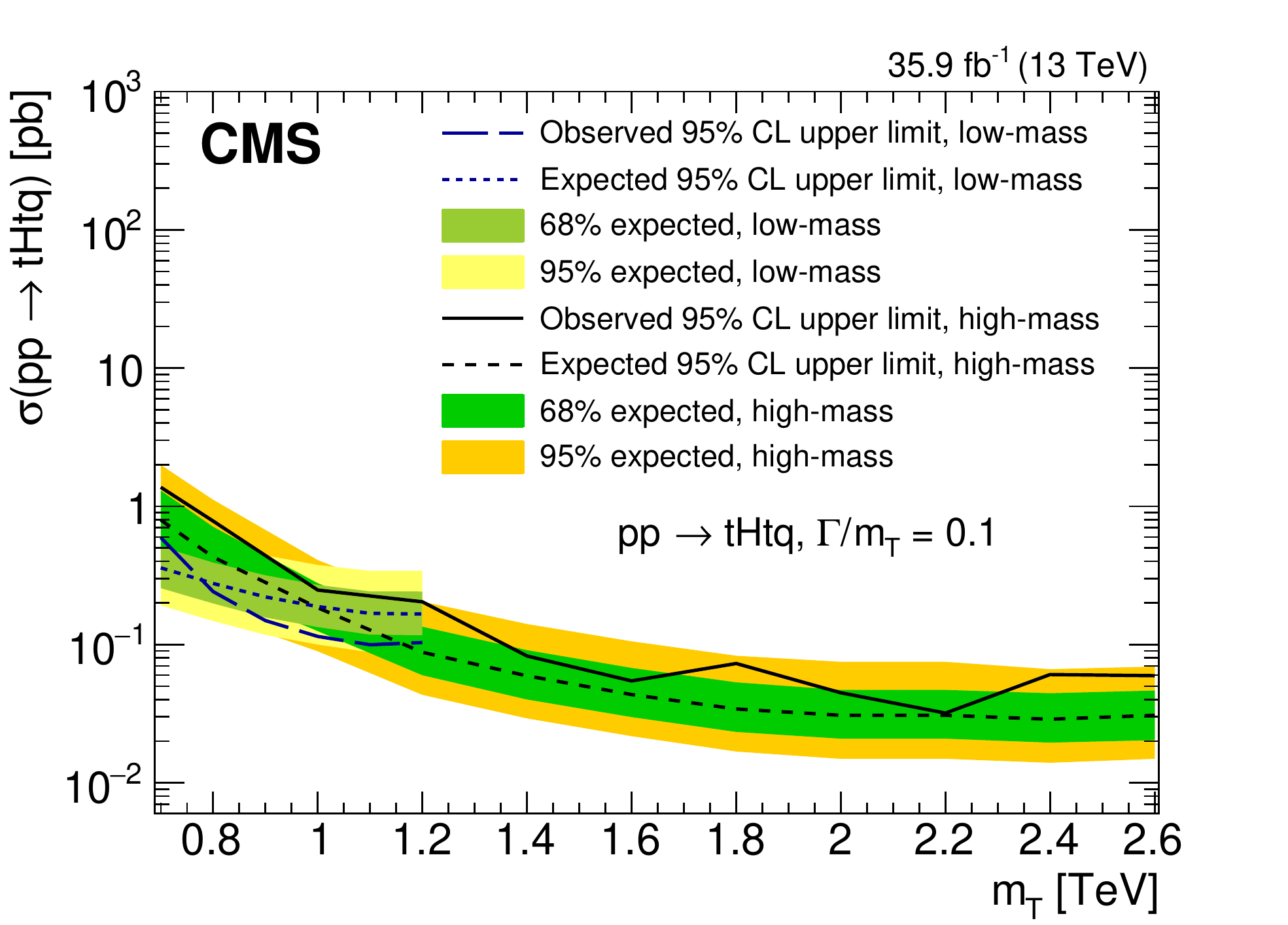}
    \includegraphics[width=0.49\textwidth]{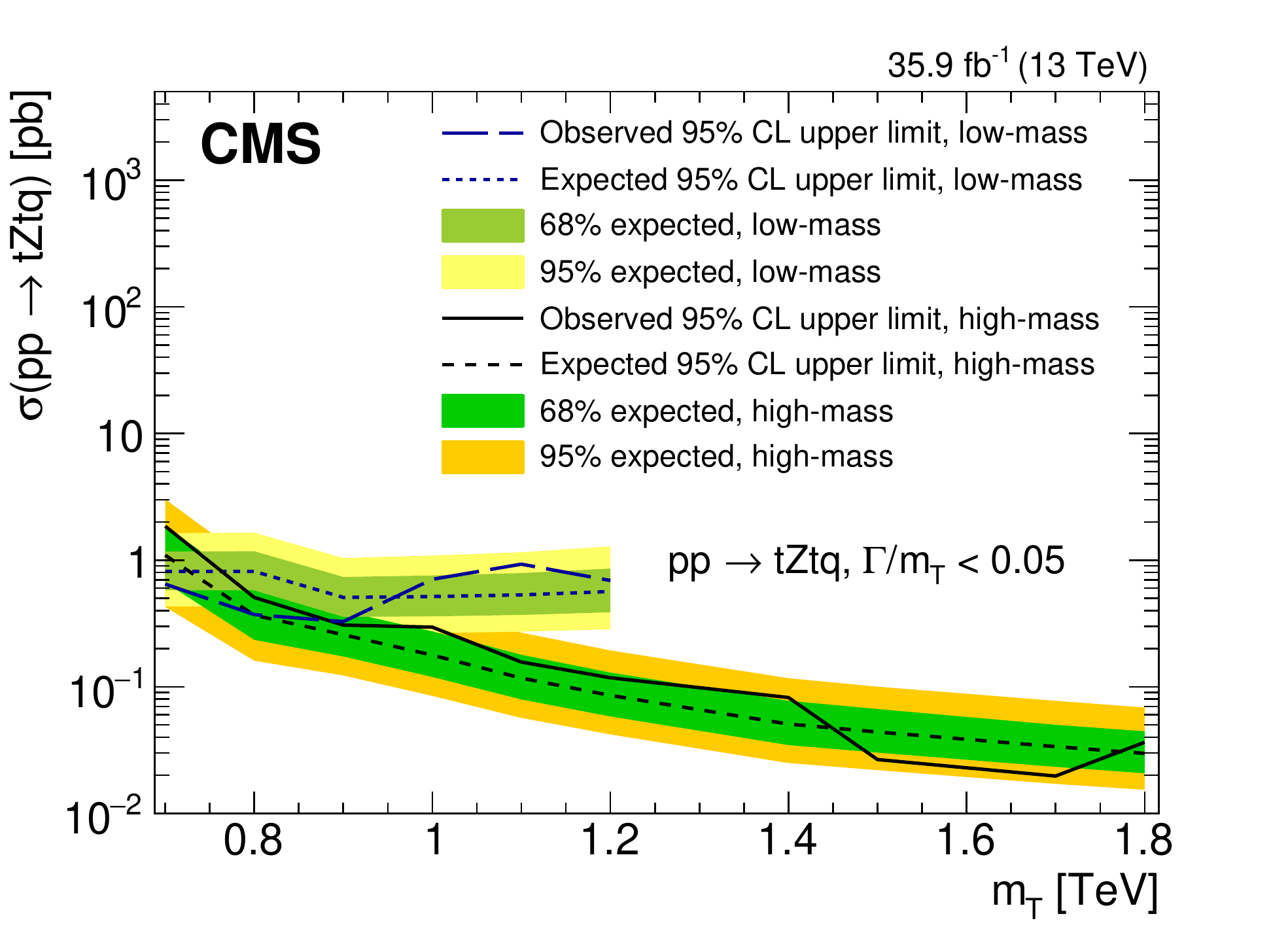}
    \includegraphics[width=0.49\textwidth]{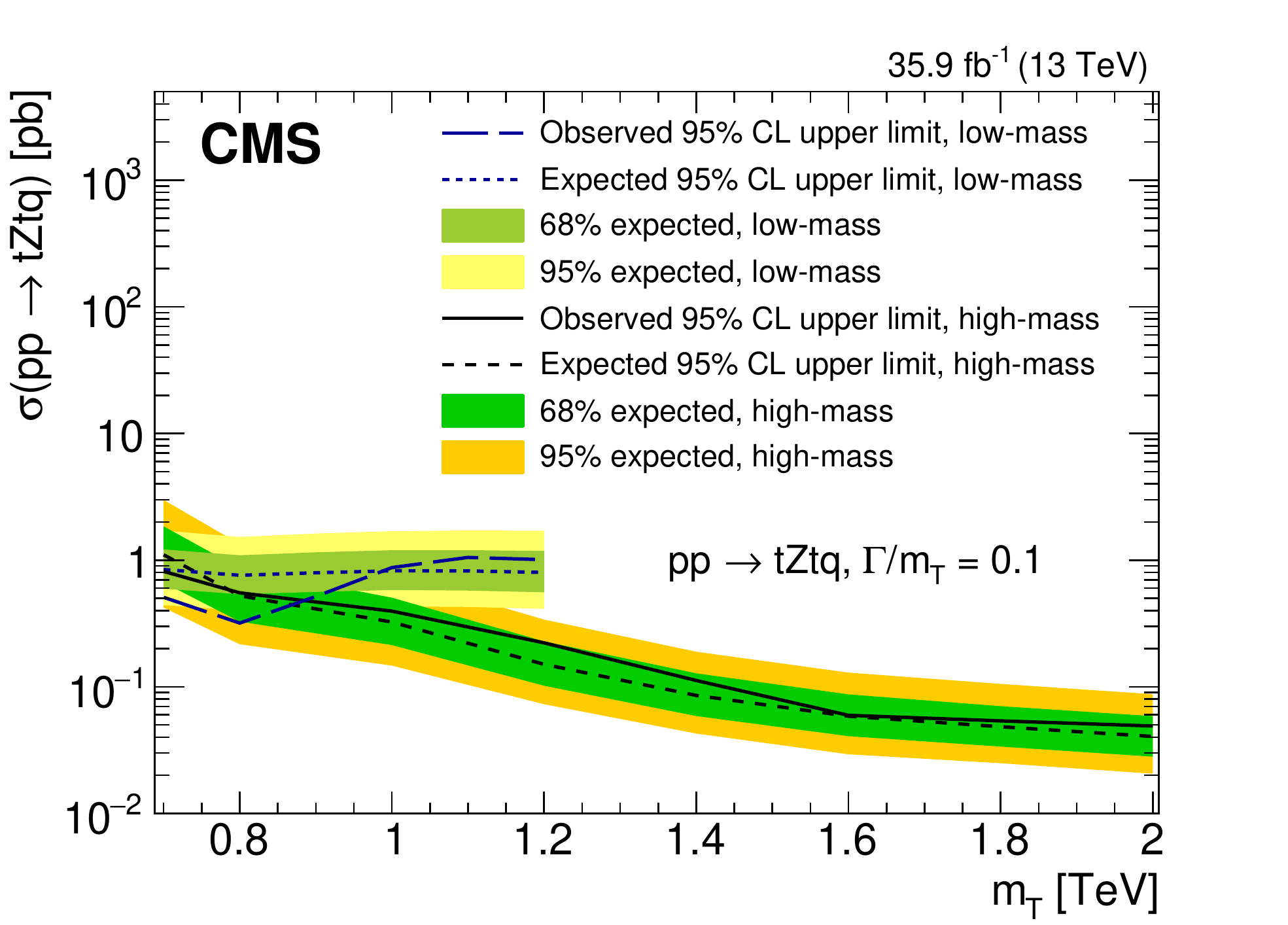}
    \includegraphics[width=0.49\textwidth]{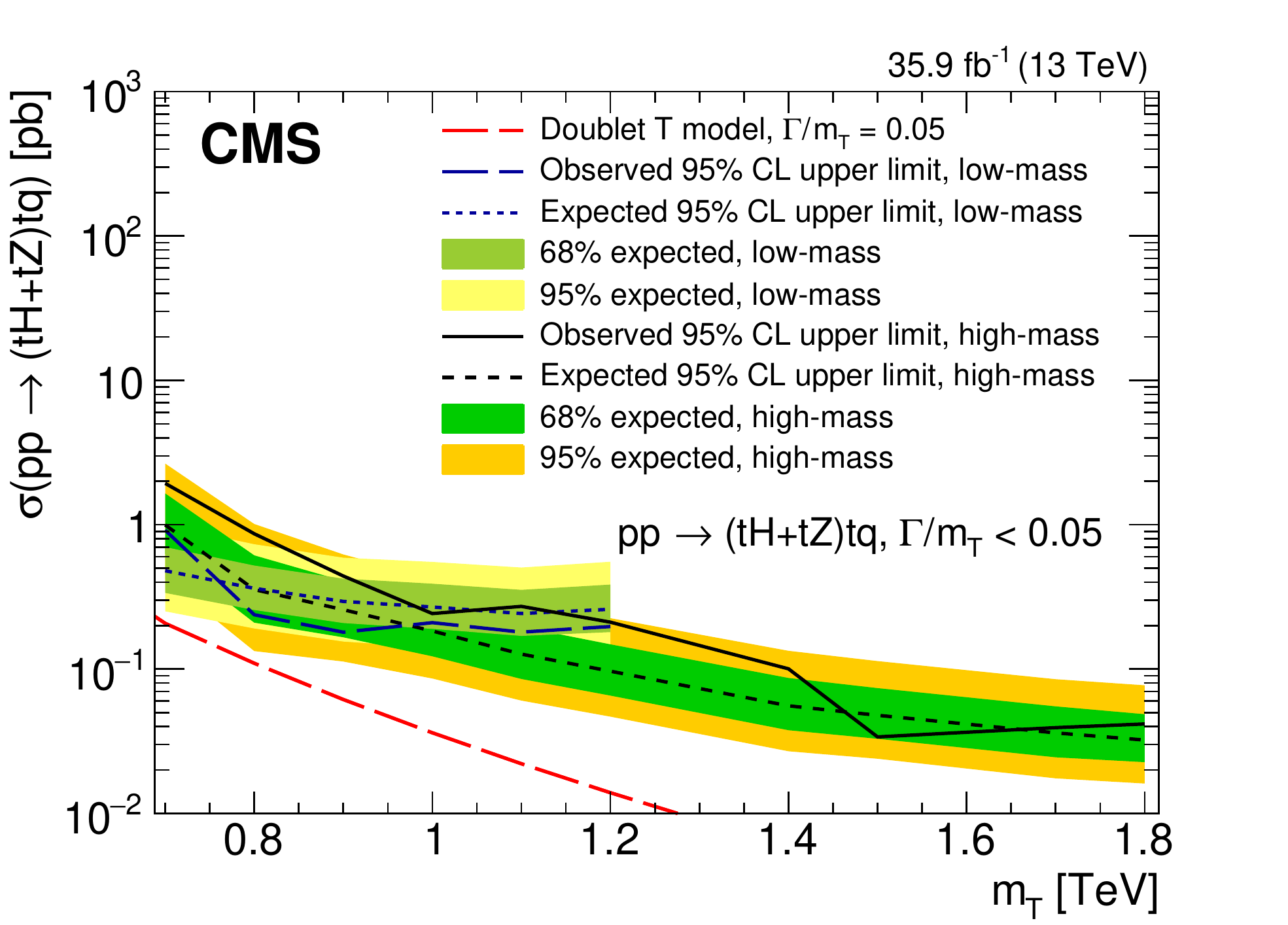}
    \includegraphics[width=0.49\textwidth]{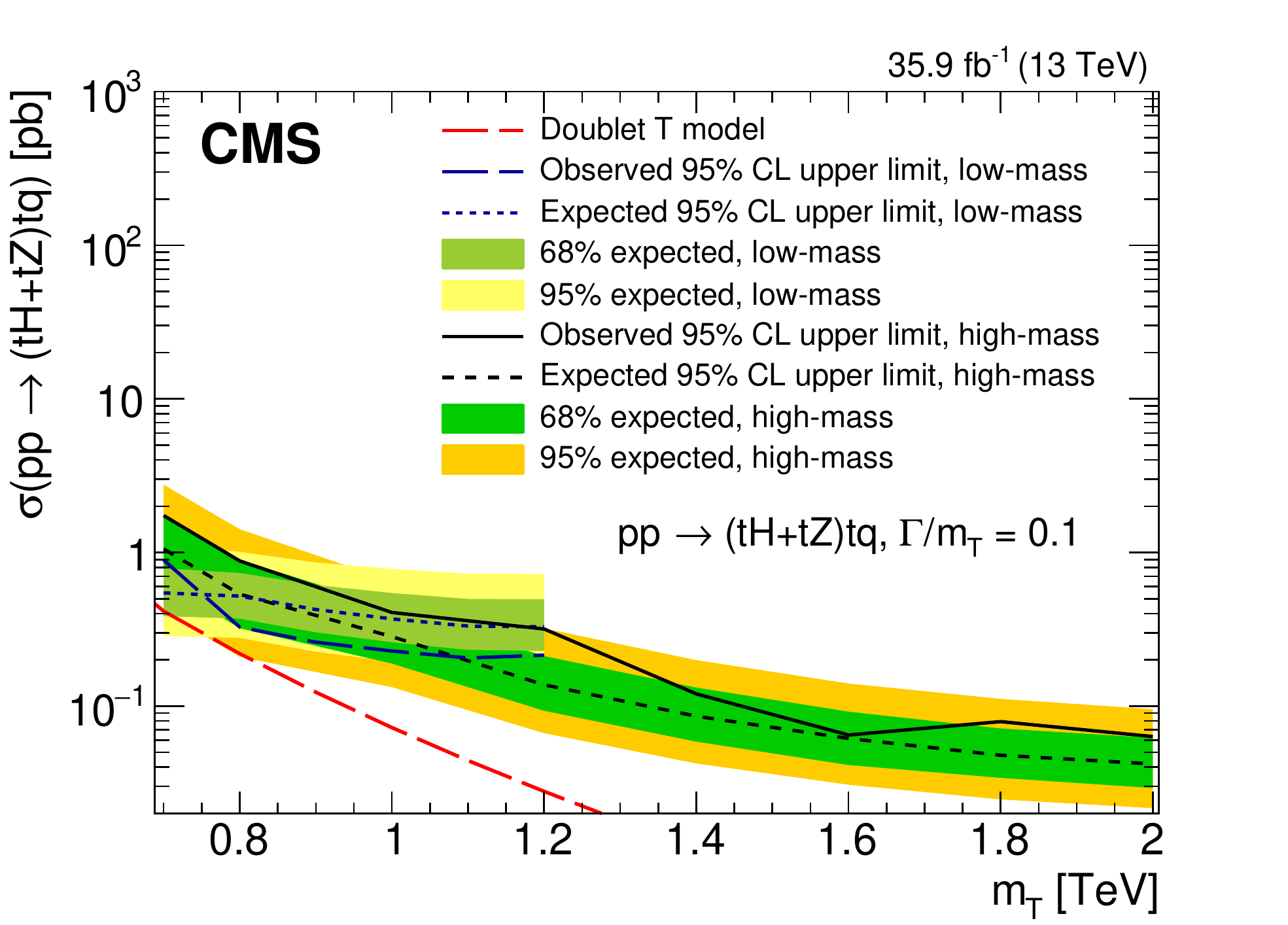}
    \caption{The observed and median expected upper limits at 95\%~\CL on the cross sections for production
associated with a top quark for the \tHtq (upper row) and \tZtq (middle row) channels, and their sum, \tHZtq (lower row), for
different assumed values of the \PQT quark mass.
The inner (green) bands and the outer (yellow) bands indicate the regions containing 68 and 95\%, respectively,
of the distribution of limits expected under the background-only hypothesis.
The left column is for a narrow fractional width ($\GoM \le 0.05$) and the right column is for a fractional width of $\GoM = 0.1$.
The dashed red curves are for the (\TB) doublet model.
Given the specified width, the couplings are implicit in the model.}
    \label{fig2:massLim_Ttq1}

\end{figure}

\begin{figure}[!htb]
  \centering
    \includegraphics[width=0.49\textwidth]{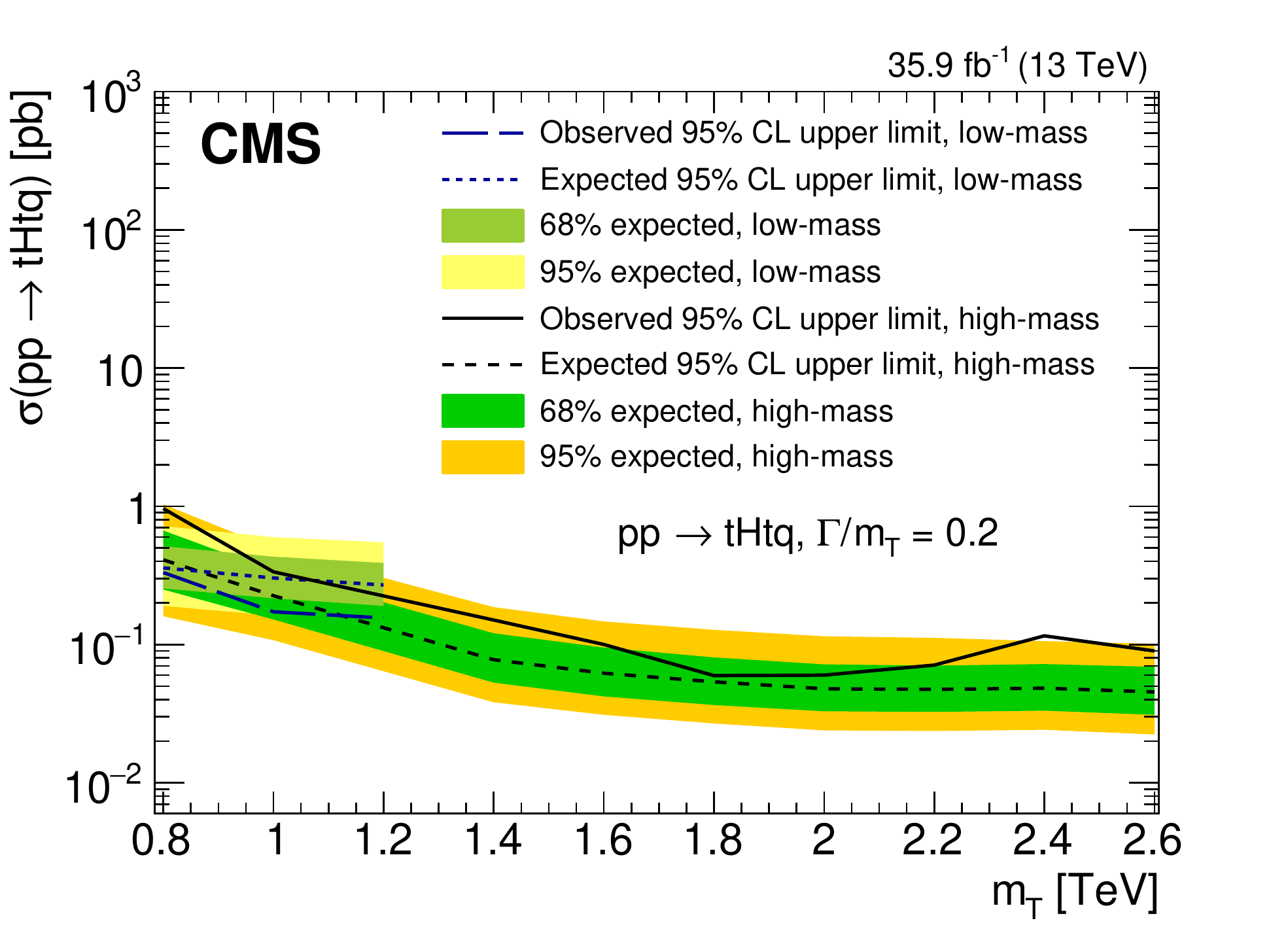}
    \includegraphics[width=0.49\textwidth]{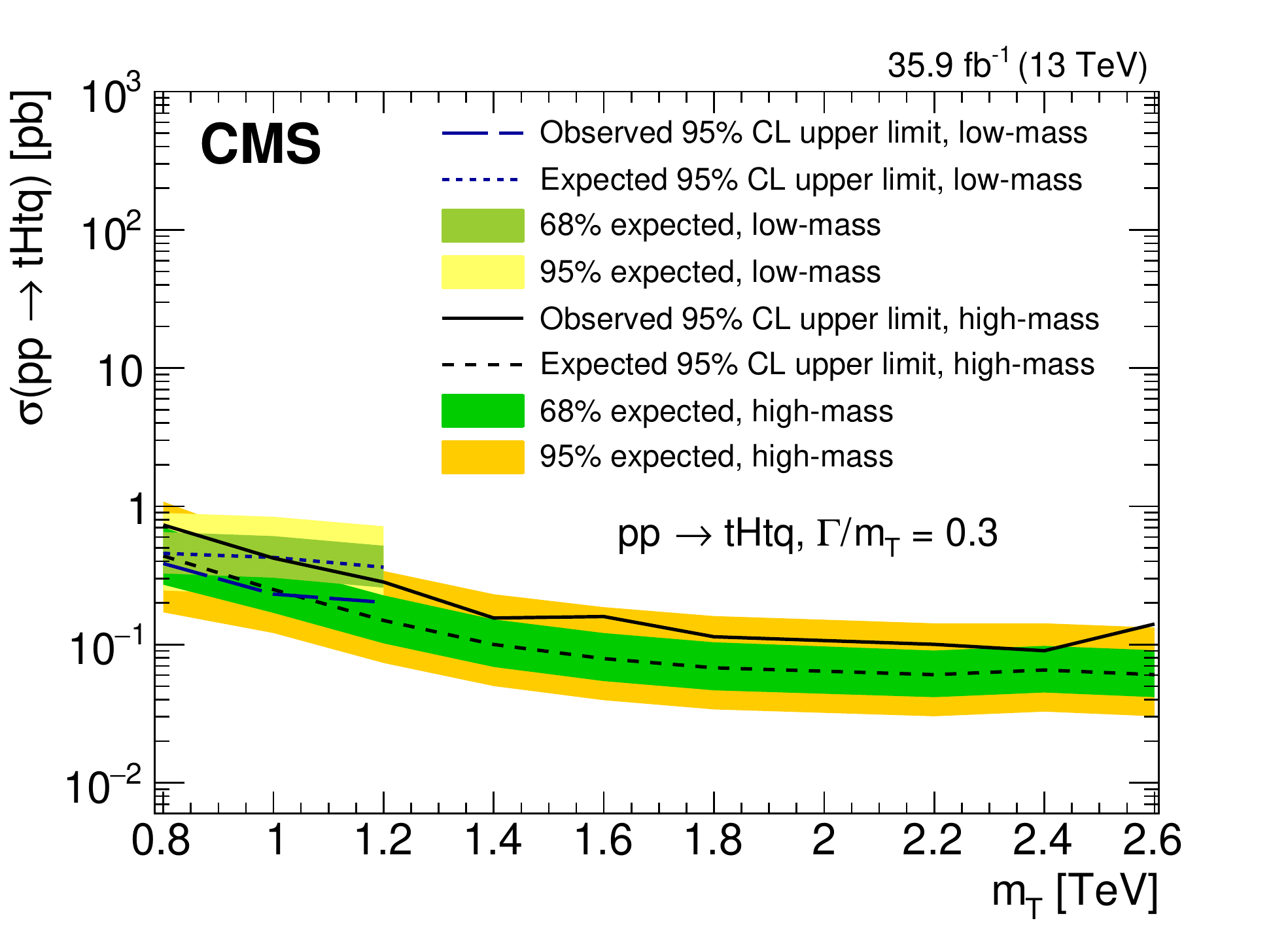}
    \includegraphics[width=0.49\textwidth]{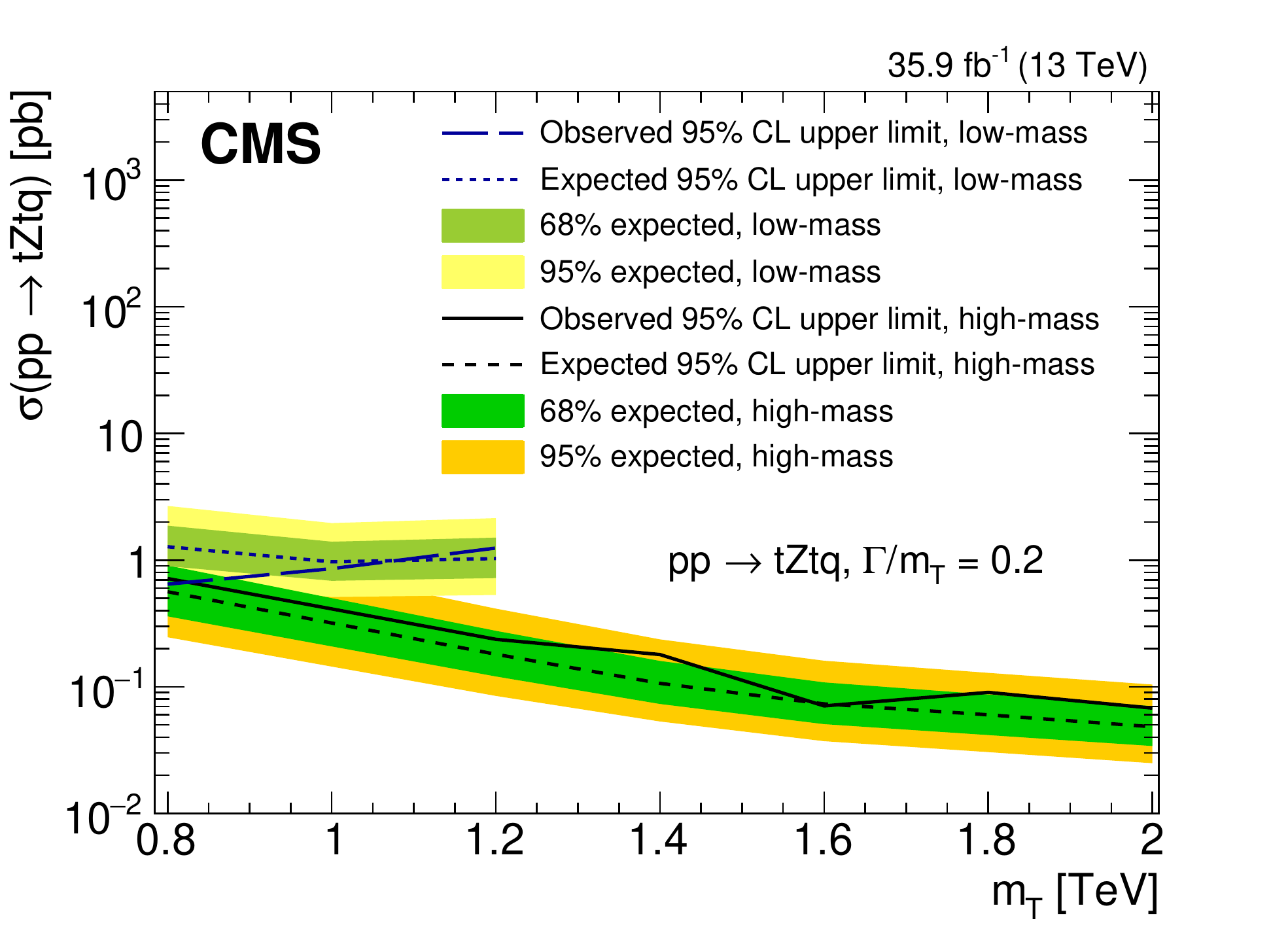}
    \includegraphics[width=0.49\textwidth]{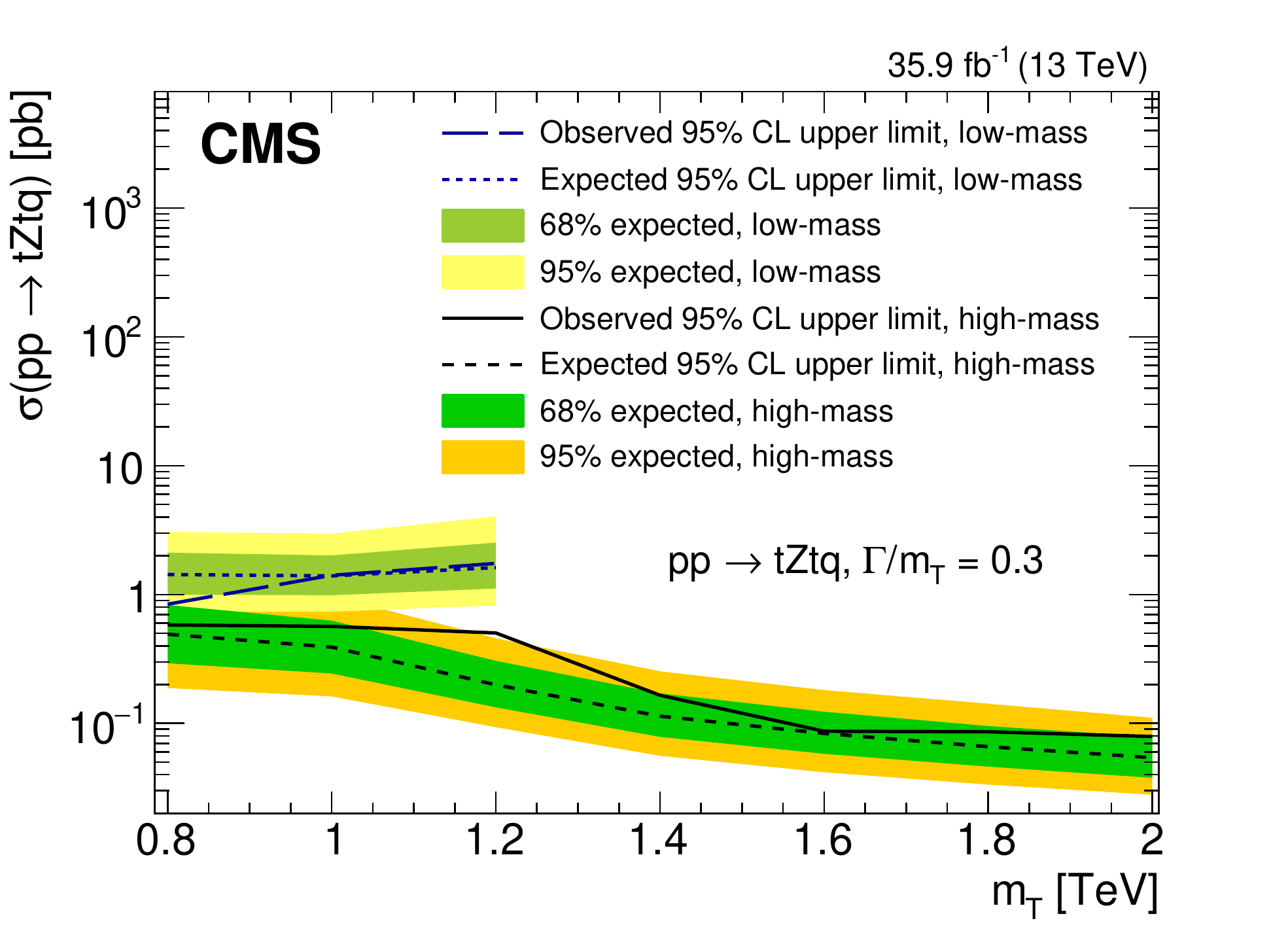}
    \includegraphics[width=0.49\textwidth]{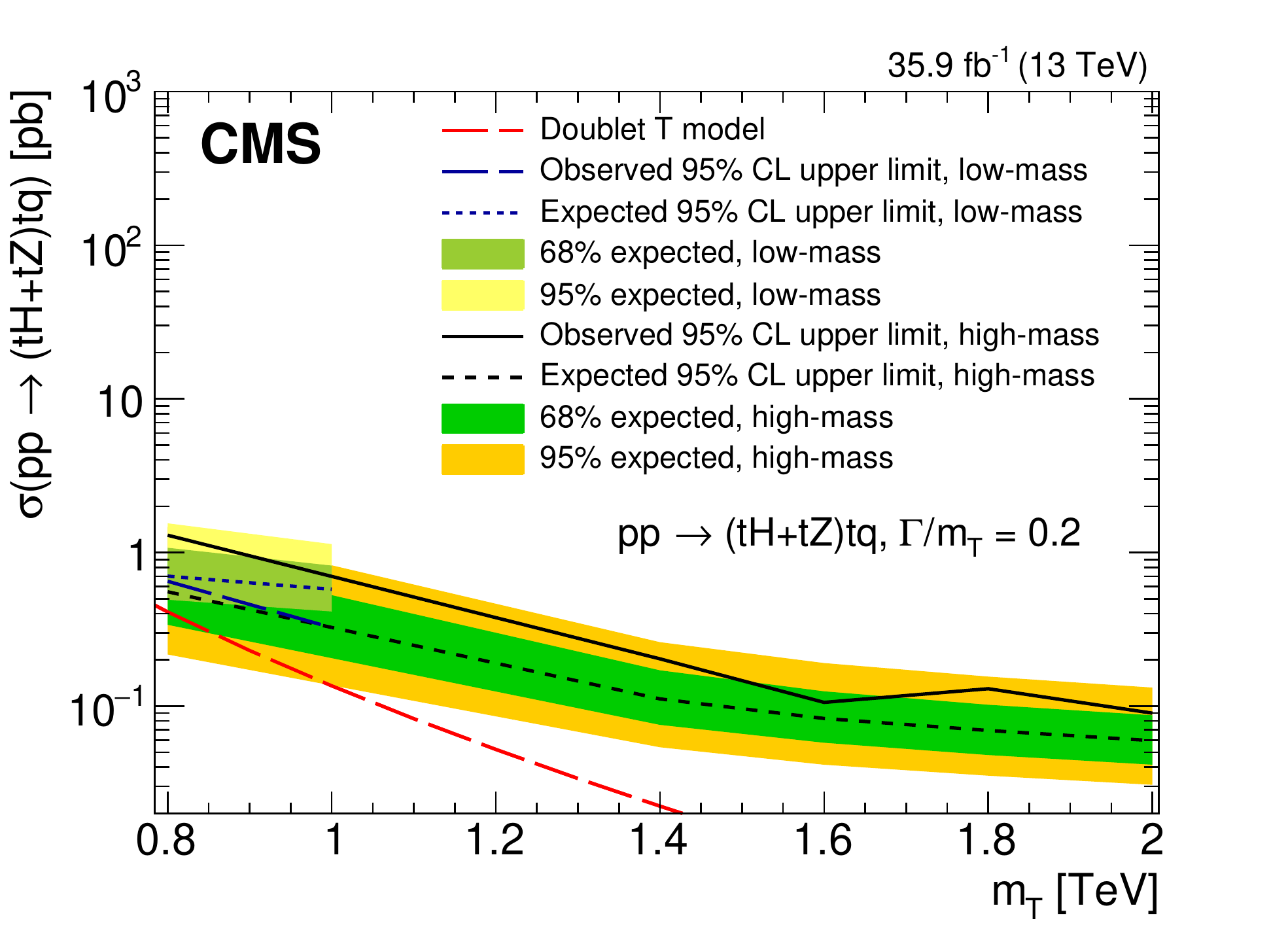}
    \includegraphics[width=0.49\textwidth]{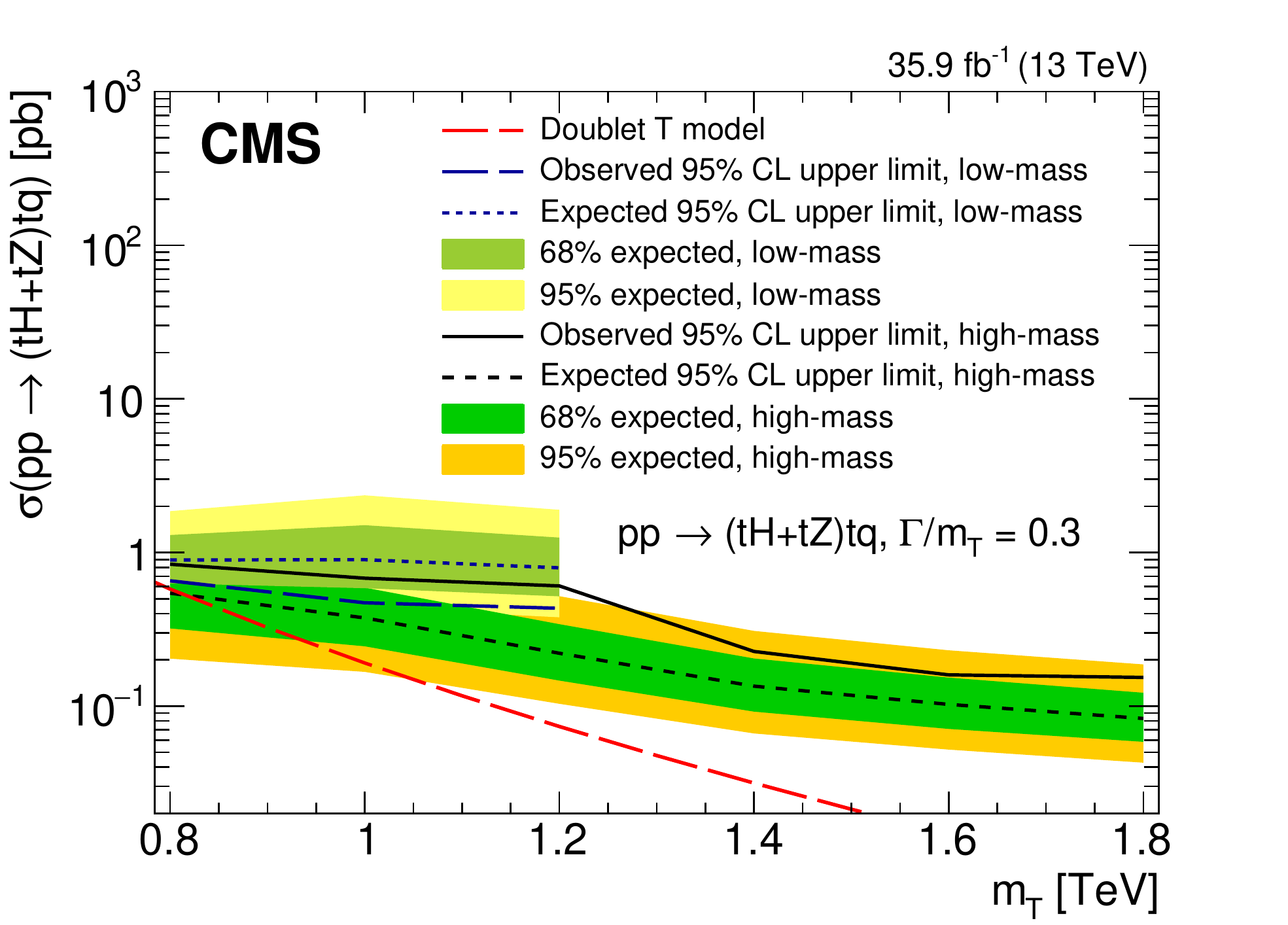}
    \caption{The observed and median expected upper limits at 95\%~\CL on the cross sections for production
associated with a top quark for the \tHtq (upper row) and \tZtq (middle row) channels, and their sum, \tHZtq (lower row), for
different assumed values of the \PQT quark mass.
The inner (green) bands and the outer (yellow) bands indicate the regions containing 68 and 95\%, respectively,
of the distribution of limits expected under the background-only hypothesis.
The left column is for a fractional width of 20\% and the right column is for a fractional width of 30\%.
The dashed red curves are for the (\TB) doublet model.
Given the specified width, the couplings are implicit in the model.}
    \label{fig2:massLim_Ttq2}

\end{figure}
\cleardoublepage \section{The CMS Collaboration \label{app:collab}}\begin{sloppypar}\hyphenpenalty=5000\widowpenalty=500\clubpenalty=5000\vskip\cmsinstskip
\textbf{Yerevan Physics Institute, Yerevan, Armenia}\\*[0pt]
A.M.~Sirunyan$^{\textrm{\dag}}$, A.~Tumasyan
\vskip\cmsinstskip
\textbf{Institut f\"{u}r Hochenergiephysik, Wien, Austria}\\*[0pt]
W.~Adam, F.~Ambrogi, T.~Bergauer, J.~Brandstetter, M.~Dragicevic, J.~Er\"{o}, A.~Escalante~Del~Valle, M.~Flechl, R.~Fr\"{u}hwirth\cmsAuthorMark{1}, M.~Jeitler\cmsAuthorMark{1}, N.~Krammer, I.~Kr\"{a}tschmer, D.~Liko, T.~Madlener, I.~Mikulec, N.~Rad, J.~Schieck\cmsAuthorMark{1}, R.~Sch\"{o}fbeck, M.~Spanring, D.~Spitzbart, W.~Waltenberger, C.-E.~Wulz\cmsAuthorMark{1}, M.~Zarucki
% ...
\vskip\cmsinstskip
\dag: Deceased\\
1:  Also at Vienna University of Technology, Vienna, Austria\\
% ...
\end{sloppypar}
\end{document}